\documentclass[10pt,aps,prd,superscriptaddress,nofootinbib,amsmath,amssymb,floatfix,showkeys,notitlepage,longbibliography,twocolumn]{revtex4-1}
\usepackage[T1]{fontenc}
\usepackage{mathptmx}
\usepackage{graphicx}
\usepackage{amsmath}
\usepackage{amssymb}
\usepackage{bm}
\usepackage{array}
\usepackage{booktabs}
\usepackage{orcidlink}
\usepackage{hyperref}
\usepackage{placeins}

\hypersetup{hidelinks}
\allowdisplaybreaks
\newcommand{\hyp}[3]{{}_2F_1\!\left(#1;#2;#3\right)}
\newcommand{\ee}{\mathrm{e}}
\newcommand{\pan}[2]{\begin{minipage}[t]{0.49\columnwidth}\centering\includegraphics[width=\linewidth]{#1}\\[-1pt]{\small #2}\end{minipage}}
\begin{document}

\title{Effective Optical Geometry and Holographic Phase Structure of a Magnetic AdS Black Hole}

\author{Mohammad Ali S. Afshar\,\orcidlink{0009-0001-3133-5992}}
\email{m.a.s.afshar@gmail.com}
\affiliation{Department of Physics, Faculty of Basic Sciences,
University of Mazandaran\\
P. O. Box 47416-95447, Babolsar, Iran}
\affiliation{School of Physics, Damghan University, P. O. Box 3671641167, Damghan, Iran}
\affiliation{Center for Theoretical Physics, Khazar University, 41 Mehseti Street, Baku, AZ1096, Azerbaijan}
\author{Mohammad Reza Alipour\,\orcidlink{0000-0001-8074-7865}}
\email{mohamad.alipour.1994@gmail.com, mr.alipour@stu.umz.ac.ir}
\affiliation{School of Physics, Damghan University, P. O. Box 3671641167, Damghan, Iran}
\affiliation{Center for Theoretical Physics, Khazar University, 41 Mehseti Street, Baku, AZ1096, Azerbaijan}
\author{Saeed Noori Gashti\,\orcidlink{0000-0001-7844-2640}}
\email{sn.gashti@du.ac.ir, saeed.noorigashti70@gmail.com}
\affiliation{School of Physics, Damghan University, P. O. Box 3671641167, Damghan, Iran}
\affiliation{Center for Theoretical Physics, Khazar University, 41 Mehseti Street, Baku, AZ1096, Azerbaijan}
\author{Jafar Sadeghi\,\orcidlink{0000-0003-4549-1766}}
\email{pouriya@ipm.ir}
\affiliation{Department of Physics, Faculty of Basic Sciences,
University of Mazandaran\\
P. O. Box 47416-95447, Babolsar, Iran}
\affiliation{School of Physics, Damghan University, P. O. Box 3671641167, Damghan, Iran}

\begin{abstract}
We study circular null orbits, thermodynamic topology, and the holographic thermodynamic description of a four-dimensional magnetically charged AdS black hole in nonlinear electrodynamics. The background orbit structure is classified using Duan's $\phi$-mapping construction and compared with the extraordinary electromagnetic characteristics. The latter define an effective optical metric with an additional radial factor that modifies the orbit condition and introduces explicit dependence on the AdS radius. For four representative configurations, the local indices of the five isolated orbits are unchanged while their radii shift by up to $5.19\%$. The background mass thresholds vary smoothly across $\sigma=1/4$ when the total mass is used, whereas the fixed-$m_0$ parametrization is singular there. Regularity at the center requires $\sigma\geq1$ and $m_0=0$. The thermodynamic analysis separates local stability indices from equilibrium coexistence: stable Gibbs-free-energy branches cross only in model-dependent parameter intervals. The boundary canonical ensemble exhibits critical behavior consistent with mean-field exponents, and the examined grand-canonical ensemble has a Hawking--Page crossing. In the mixed ensemble, the free-energy branches have a finite endpoint discontinuity, but the sampled competing states do not satisfy the full local-stability criterion.
\end{abstract}

\maketitle

\section{Introduction}\label{sec:intro}
Topological methods provide a way to classify black-hole properties that is less sensitive to the details of a particular solution. The approach adopted here is based on Duan's $\phi$-mapping theory \cite{ref001}: a two-dimensional vector field is constructed from a scalar function, and its zero points carry integer winding numbers. Their sum, the total topological charge (TTC), remains unchanged under smooth deformations, provided the relevant boundary conditions are preserved, and can therefore be used to distinguish topological classes. This framework has been applied to light rings and photon spheres \cite{ref058,ref059,ref060,ref061} and to black-hole thermodynamics \cite{ref072,ref073}, and has since been explored in a wide range of gravitational models \cite{ref002,ref003,ref004,ref005,ref006,ref007,ref008,ref009,ref010,ref011,ref012,ref013,ref014,ref015,ref016,ref017,ref018,ref019,ref020,ref021,ref022,ref023,ref024,ref025,ref026,ref027,ref028,ref029,ref030,ref031,ref032,ref033,ref034,ref035,ref036,ref037,ref038,ref039,ref040,ref041,ref042,ref043,ref044,ref045,ref046,ref047,ref048,ref049,ref050,ref051,ref052,ref053,ref054,ref055,ref056,ref057}.

For photon spheres, Cunha and collaborators established topological results for light rings around ultracompact objects and black holes, including the existence and stability properties of the relevant circular null orbits \cite{ref058,ref059,ref060}. Wei subsequently related these results to a topological charge and extended the analysis to asymptotically anti-de Sitter and de Sitter spacetimes \cite{ref061}. In our previous work, this correspondence was used to classify the spacetime around black holes and to organize parameter ranges by their horizon and orbit structure \cite{ref062,ref063,ref064,ref065,ref066,ref067,ref068}. We apply this construction to the circular null geodesics of the background metric and to the extraordinary electromagnetic characteristics. The distinction is relevant in nonlinear electrodynamics, where the propagation cone depends on derivatives of the electromagnetic Lagrangian \cite{refOptNovello,refOptCosta}.

A complementary line of research applies topological methods to black-hole thermodynamics. In the conventional approach, one studies the temperature as a function of the horizon radius and examines the Gibbs free energy to identify phase transitions; a crossing of the locally stable branches of a free-energy swallowtail, for example, identifies a first-order small--large black-hole transition \cite{ref069,ref070,ref071}. Wei, Liu and Mann showed that this information can also be encoded in the zeros of a vector field constructed from a generalized free energy, with the associated topological charges distinguishing stable and unstable branches \cite{ref072,ref073}. Subsequent studies have applied this method to a variety of black-hole solutions and thermodynamic ensembles. The present analysis uses the topological classification alongside the temperature and free-energy curves, so that the topological results can be checked against the conventional thermodynamic description.

The holographic description of black-hole thermodynamics provides a further connection between bulk gravitational physics and the thermodynamics of a dual conformal field theory (CFT). In extended black-hole thermodynamics, variations of the cosmological constant and Newton's constant can be translated into variations of boundary thermodynamic quantities, including the CFT central charge and its conjugate chemical potential \cite{refH1,refH3}. This dictionary has been used to study ensemble-dependent phase transitions and critical behavior for charged and rotating AdS black holes \cite{refH2,refH4}. Related work has examined holographic phase transitions in charged Gauss--Bonnet gravity \cite{ref071}, the relation between CFT thermodynamics and the weak gravity and weak cosmic censorship conjectures \cite{refH7}, and classical and topological CFT thermodynamics in Lifshitz and hyperscaling-violating geometries \cite{refH5}. Recent applications to Einstein--Maxwell--power--Yang--Mills AdS black holes further illustrate how nonlinear gauge sectors can influence the phase structure of the boundary theory \cite{refH6}. Together, these studies provide the context for examining the holographic thermodynamics of the magnetically charged nonlinear-electrodynamics black hole considered here.

The black hole studied in this work is the magnetically charged solution of Einstein gravity with a negative cosmological constant coupled to a two-parameter nonlinear electrodynamics, introduced by Kruglov \cite{ref089}. That work presented the metric and mass functions, the extended first law, the Smarr relation, the heat capacity and the Gibbs free energy, and estimated a critical pressure of approximately $0.003$ for $q=1$, $\beta=0.1$ and $\sigma=0.75$ from the Gibbs free energy, since the critical point is not available in closed analytic form. We study the dependence of the horizon and background-orbit structure, together with thermodynamic topology, on $\beta$, $\sigma$ and $q$. We also verify numerically the Smarr relation, including the $2\beta\mathcal{B}$ term established in Ref.~\cite{ref089}, and extend the holographic dictionary of Refs.~\cite{refH1,refH2,refH3} to this model.

The paper is organized as follows. Section~\ref{sec:method} summarizes the topological construction for photon spheres. Section~\ref{sec:model} introduces the model, discusses the relation between the integration constant in the metric and the total mass, and presents the background classification and the effective-optical comparison. Section~\ref{sec:thermo} gives the extended first law and Smarr relation and examines the thermodynamic topology and free energy. Section~\ref{sec:holo} constructs the holographic dictionary, checks the first law and Euler relation of the dual CFT, and studies the relevant ensembles, critical point and critical exponents. Section~\ref{sec:concl} presents our conclusions.

\section{Topological Description of Circular Null Orbits}\label{sec:method}
Photon spheres are linked to gravitational lensing and to the shadow of a black hole \cite{ref074,ref075,ref004,ref076,ref077,ref078,ref079,ref080}, and circular null orbits, their stability and the associated photon surfaces have been studied extensively \cite{ref081,ref082,ref083,ref084,ref085,ref086,ref087,ref088}. The construction used here is described in detail in Refs.~\cite{ref058,ref059,ref060,ref061,ref062,ref063,ref064}; only the relations needed below are recalled. The construction applies to a static metric on a nondegenerate domain. In the background analysis below, ``photon sphere'' denotes a circular null orbit of the spacetime metric; the effective electromagnetic orbits are treated in Sec.~\ref{sec:optical}. A static, spherically symmetric metric is written as
\begin{equation}\label{eq:metric}
ds^{2}=-f(r)\,dt^{2}+\frac{dr^{2}}{g(r)}+h(r)\left(d\theta^{2}+\sin^{2}\theta\,d\varphi^{2}\right).
\end{equation}
For null geodesics in the equatorial plane $\theta=\pi/2$ the Lagrangian is $\mathcal{L}=\tfrac{1}{2}g_{\mu\nu}\dot{x}^{\mu}\dot{x}^{\nu}=0$, with conserved energy $E=f\dot t$ and angular momentum $L=h\dot\varphi$. The condition $\mathcal{L}=0$ is equivalent to $\dot r^{2}+V_{\mathrm{eff}}(r)=0$ with
\begin{equation}\label{eq:Veff}
V_{\mathrm{eff}}(r)=g(r)\left(\frac{L^{2}}{h(r)}-\frac{E^{2}}{f(r)}\right).
\end{equation}
A circular null orbit satisfies $V_{\mathrm{eff}}=V_{\mathrm{eff}}'=0$. Because $V_{\mathrm{eff}}$ depends on the null geodesic's constants of motion $E$ and $L$, the function
\begin{equation}\label{eq:H}
\mathbf{H}(r,\theta)=\sqrt{\frac{-g_{tt}}{g_{\varphi\varphi}}}=\frac{1}{\sin\theta}\sqrt{\frac{f(r)}{h(r)}}
\end{equation}
is used instead, since it is built from the metric alone. Its extrema in $r$ at $\theta=\pi/2$ coincide with the circular photon orbits: a maximum corresponds to an unstable photon sphere (U-PS) and a minimum to a stable one (S-PS). In the $\theta$ direction, $\mathbf{H}$ always has a minimum at $\theta=\pi/2$. The vector field
\begin{equation}\label{eq:phi}
\phi=\left(\phi^{r},\phi^{\theta}\right)=\left(\sqrt{g}\,\partial_{r}\mathbf{H},\ \frac{\partial_{\theta}\mathbf{H}}{\sqrt{h}}\right)
\end{equation}
vanishes precisely at these orbits. With $\phi=\|\phi\|\,\ee^{i\Omega}$, the winding number of a zero point $z_i$ is
\begin{equation}\label{eq:winding}
\omega_{i}=\frac{1}{2\pi}\oint_{C_{i}}d\Omega ,\qquad \Omega=\arctan\frac{\phi^{\theta}}{\phi^{r}},
\end{equation}
where $C_i$ is a closed counterclockwise curve in the $(r,\theta)$ plane that encloses only $z_i$, and $\Omega$ is evaluated with a two-argument arctangent and unwrapped continuously along a parametrization of $C_i$. Equivalently, $\omega_i=\kappa_i\zeta_i$, where the Hopf index $\kappa_i\geq1$ is the number of times the normalized field $n^{a}=\phi^{a}/\|\phi\|$ covers the unit circle around $z_i$, and $\zeta_i=\pm1$ is the sign of the Jacobian of $\phi$ at the zero point. The associated topological current $j^{\mu}=(2\pi)^{-1}\varepsilon^{\mu\nu\rho}\varepsilon_{ab}\,\partial_{\nu}n^{a}\partial_{\rho}n^{b}$, with the indices running over $(t,r,\theta)$, is identically conserved, $\partial_{\mu}j^{\mu}=0$, and is supported on the zeros of $\phi$. The total charge is
\begin{equation}\label{eq:TTC}
Q=\sum_{i}\omega_{i}.
\end{equation}
Near a maximum of $\mathbf{H}$ in $r$ the field behaves as $(\phi^{r},\phi^{\theta})\propto(-x,\,y)$, a saddle with $\omega=-1$; near a minimum it behaves as $(x,y)$ with $\omega=+1$. For an exterior domain satisfying the horizon and asymptotic hypotheses of Refs.~\cite{ref058,ref059,ref060,ref061}, the background charge is $Q=-1$: the number of unstable photon spheres outside the horizon exceeds the number of stable ones by one. We distinguish singular horizonless configurations with a stable--unstable background pair (NS) from those without background circular null orbits (NS0). Both can have $Q=0$: a pair contributes $+1-1$, whereas an orbit-free domain has an empty sum. Singularity is determined from the central geometry and horizon structure, not from the total index. Their formation from regular initial data is a dynamical question related to weak cosmic censorship \cite{refCensWald}.

\section{The Four-Dimensional AdS Magnetic Black Hole}\label{sec:model}
\subsection{Solution and mass parameters}
The action is
\begin{equation}\label{eq:action}
I=\int d^{4}x\sqrt{-g}\left[\frac{R-2\Lambda}{16\pi G}+\mathcal{L}(\mathcal{F})\right],
\end{equation}
with $\Lambda=-3/l^{2}$ and the nonlinear electrodynamics Lagrangian \cite{ref089}
\begin{equation}\label{eq:lag}
\mathcal{L}(\mathcal{F})=-\frac{\mathcal{F}}{4\pi\,(1+2\beta\mathcal{F})^{\sigma}},\qquad \mathcal{F}=\frac{F_{\mu\nu}F^{\mu\nu}}{4}=\frac{B^{2}-E^{2}}{2}.
\end{equation}
The coupling $\beta>0$ has dimension $[L^{4}]$ in units $c=\hbar=1$ and $[L^{2}]$ in the geometrized units $G=c=1$ used below, in which the charge $q$ has the dimension of a length, and $\sigma>0$ is dimensionless. In the weak-field limit the Lagrangian reduces to that of Maxwell. For the magnetic field $B=q/r^{2}$ one finds, in units $G=1$,
\begin{equation}\label{eq:f}
f(r)=1-\frac{2m_{0}}{r}+\frac{r^{2}}{l^{2}}-\frac{\mathcal{I}(r)}{r},
\end{equation}
\begin{equation}\label{eq:I}
\mathcal{I}(r)=\frac{q^{2}\,r^{4\sigma-1}}{(4\sigma-1)\,(\beta q^{2})^{\sigma}}\;\hyp{\sigma,\sigma-\tfrac14}{\sigma+\tfrac34}{-\frac{r^{4}}{\beta q^{2}}},
\end{equation}
where $m_{0}$ is an integration constant and $\hyp{a,b}{c}{z}$ is the Gauss hypergeometric function. The function $\mathcal{I}$ is the integral of the magnetic energy density, and its derivative is elementary,
\begin{equation}\label{eq:Ip}
\mathcal{I}'(r)=\frac{q^{2}\,r^{4\sigma-2}}{\left(r^{4}+\beta q^{2}\right)^{\sigma}}.
\end{equation}
For $\sigma>1/4$ the magnetic energy $m_{M}$ is finite and the ADM mass is the sum $M=m_{0}+m_{M}$ \cite{ref089}, with
\begin{equation}\label{eq:mM}
m_{M}=\frac{q^{3/2}\,\Gamma\!\left(\sigma-\tfrac14\right)\Gamma\!\left(\tfrac54\right)}{2\,\beta^{1/4}\,\Gamma(\sigma)}.
\end{equation}
The metric function then takes the form
\begin{equation}\label{eq:fADM}
f(r)=1-\frac{2M}{r}+\frac{q^{2}}{r^{2}}\,\hyp{\sigma,\tfrac14}{\tfrac54}{-\frac{\beta q^{2}}{r^{4}}}+\frac{r^{2}}{l^{2}},
\end{equation}
in which $M$ is the total mass. For $\sigma\le 1/4$ the energy density is not integrable at the origin, so $m_{0}$ is not the constant of a convergent integral and $m_{M}$ in Eq.~\eqref{eq:mM} diverges at $\sigma=1/4$ through $\Gamma(\sigma-1/4)$. The closed form of $\mathcal{I}$ in Eq.~\eqref{eq:I} is singular at $\sigma=1/4$ for the same reason, and this value is omitted from the fixed-$m_0$ scan in that representation. For $\sigma\neq1/4$, Eq.~\eqref{eq:fADM} with $M=m_{0}+m_{M}$, where $m_{M}$ is given by Eq.~\eqref{eq:mM} (continued analytically for $\sigma<1/4$), is identical to Eq.~\eqref{eq:f}; this identity was verified numerically to $10^{-15}$ for $\sigma=0.12,\,0.3,\,0.9$. In the photon-sphere analysis below, as in Refs.~\cite{ref062,ref063,ref064,ref065,ref066,ref067,ref068}, the constant $m_{0}$ is held fixed, and the relation between $m_{0}$ and $M$ is discussed in Sec.~\ref{sec:ADM}. In the thermodynamic analysis of Sec.~\ref{sec:thermo} the total mass $M$ is fixed by the horizon condition and does not appear as a free parameter.

\subsection{Background circular-null-orbit functions}
With $g=f$ and $h=r^{2}$, Eqs.~\eqref{eq:H} and \eqref{eq:phi} give
\begin{align}\label{eq:Hmodel}
\mathbf{H}(r,\theta)&=\frac{\sqrt{f(r)}}{r\sin\theta},\nonumber\\
\phi^{r}&=\frac{r f'(r)-2f(r)}{2r^{2}\sin\theta},\\
\phi^{\theta}&=-\frac{\sqrt{f(r)}\cos\theta}{r^{2}\sin^{2}\theta},\nonumber
\end{align}
and, from Eqs.~\eqref{eq:f} and \eqref{eq:Ip},
\begin{equation}\label{eq:PScond}
r f'-2f=\frac{6m_{0}}{r}-2+\frac{3\,\mathcal{I}(r)}{r}-\mathcal{I}'(r).
\end{equation}
The radial component of $\phi$ coincides with the expression containing hypergeometric functions of $\sigma+1$ and $\sigma+3/4$ that is obtained by direct differentiation of $\mathbf{H}$; the two forms were compared numerically at several radii and values of $\sigma$ and agree to working precision. The background orbit condition $\phi^r(r,\pi/2)=0$ is explicitly independent of $l$, since $r^2/l^2$ cancels in Eq.~\eqref{eq:PScond}. The positive roots of $f(r)=0$ determine the horizon structure; the larger root, when two are present, is the outer event horizon.

All numerical values were obtained with arbitrary-precision evaluation of the hypergeometric functions. Roots were bracketed on a logarithmic grid and refined by Brent's method, and threshold values of $\sigma$ were determined by solving the double-root conditions, $f=f'=0$ for the horizons and $\phi^{r}=\partial_{r}\phi^{r}=0$ for the photon spheres, simultaneously for the radius and for $\sigma$. The winding numbers quoted in Secs.~\ref{sec:model} and \ref{sec:thermo} were also computed directly from Eq.~\eqref{eq:winding} by numerical integration along small closed counterclockwise curves around each zero point, and agree with the stated values. We set $m_{0}=1$ and $l=1$ and consider the two representative charges $q=0.5$ ($q<m_{0}$) and $q=1.2$ ($q>m_{0}$).

\subsection{Case \texorpdfstring{$q<m_{0}$}{q < m0}}
\begin{figure}[!tbp]
\centering\includegraphics[width=0.85\columnwidth]{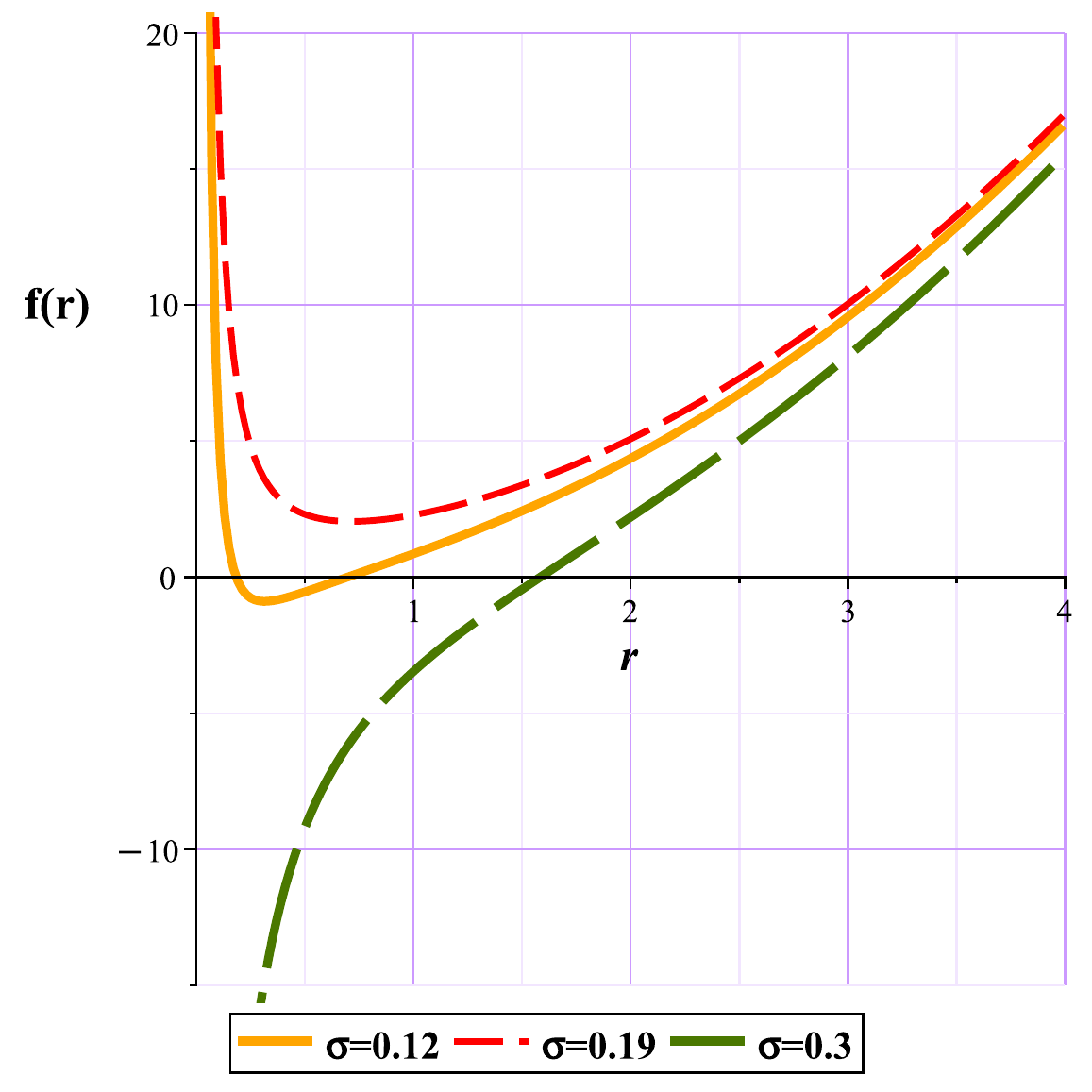}
\caption{Background metric function $f(r)$ for $q=0.5$, $m_{0}=1$, $l=1$, $\beta=0.1$ and $\sigma=0.12$ (two horizons), $0.19$ (no horizon) and $0.3$ (one horizon).}\label{fig:f1}
\end{figure}

For $q=0.5$ the horizon structure depends on $\sigma$ in the same way for all values of $\beta$ examined, with the thresholds listed in Table~\ref{tab:thr}. The geometry has two horizons for $\sigma<\sigma_{1}$, no horizon for $\sigma_{1}<\sigma<1/4$, and a single horizon for $\sigma>1/4$. The second threshold $\sigma_{2}$ in Table~\ref{tab:thr} is the value at which the pair of photon spheres of the horizonless geometry disappears.
\begin{table}[!t]
\caption{Background-metric thresholds in $\sigma$ for $q=0.5$, $m_0=l=1$. Two horizons exist for $\sigma<\sigma_{1}$ and none for $\sigma_{1}<\sigma<1/4$. In the horizonless geometry a stable--unstable photon-sphere pair exists for $\sigma_{1}<\sigma<\sigma_{2}$.}\label{tab:thr}
\begin{ruledtabular}
\begin{tabular}{ccc}
$\beta$ & $\sigma_{1}$ & $\sigma_{2}$\\
\hline
$0.001$ & $0.0748997$ & $0.0831101$\\
$0.1$ & $0.1465105$ & $0.1548466$\\
$1.2$ & $0.1865467$ & $0.1901076$\\
\end{tabular}
\end{ruledtabular}
\end{table}
Because the thresholds move with $\beta$, the discussion is carried out for $\beta=0.1$, for which the metric function is shown in Fig.~\ref{fig:f1}.

\begin{figure}[!tbp]
\centering\includegraphics[width=0.85\columnwidth]{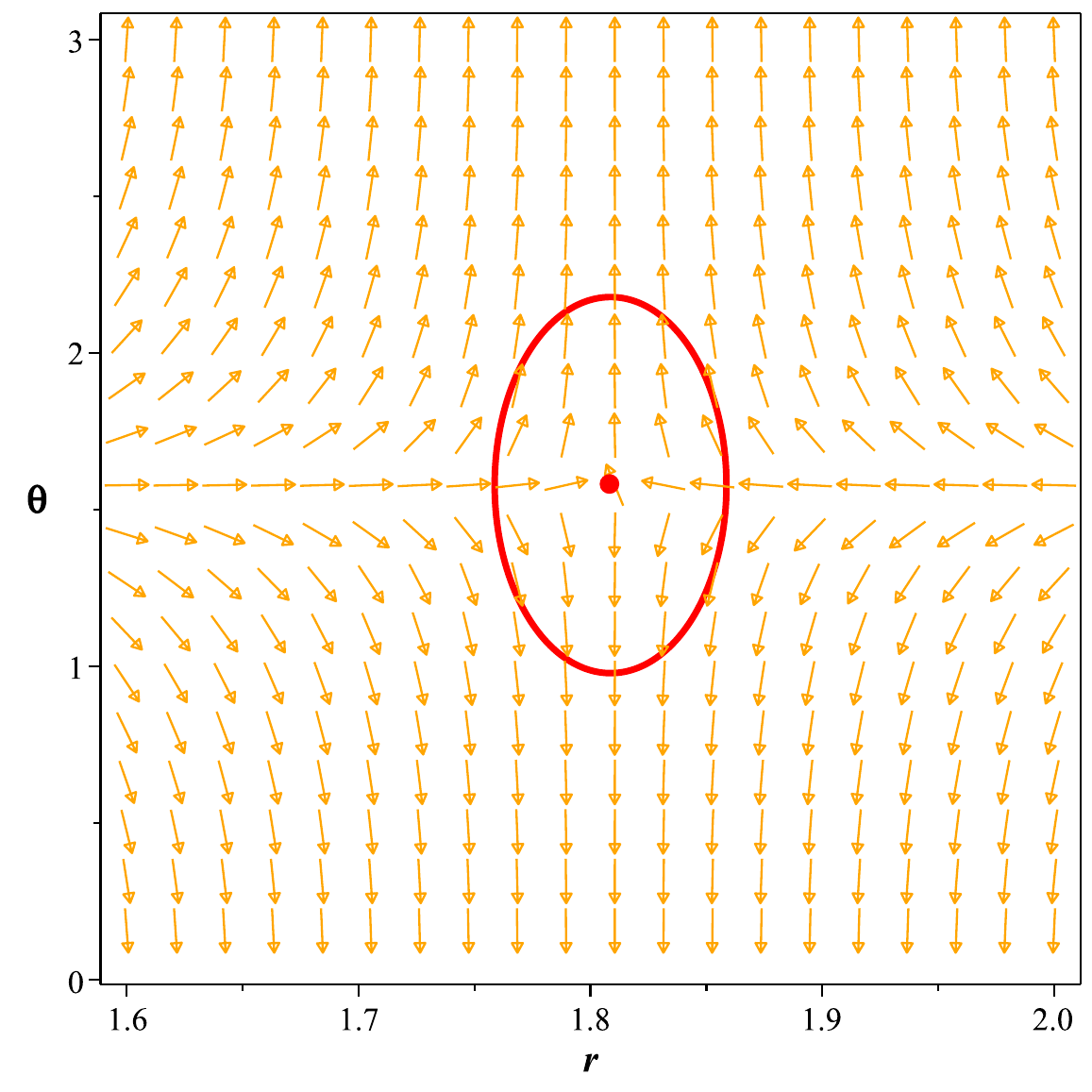}
\caption{Normalized topological field $n$ of the background metric in the $(r,\theta)$ plane for $\sigma=0.12$, $q=0.5$, $m_{0}=1$, $l=1$, $\beta=0.1$, in the region outside the event horizon $r_{+}=0.6953$. The unstable circular null orbit at $(r,\theta)=(1.8087,\pi/2)$ is a zero point of winding number $-1$ (red circle).}\label{fig:n1}
\end{figure}

For $\sigma=0.12$ the horizons are located at $r=0.1832$ and $r=0.6953$. The normalized field $n$ outside the event horizon is shown in Fig.~\ref{fig:n1}. It has a single zero point at $r=1.8087$, with winding number $-1$, which identifies an unstable photon sphere and $Q=-1$. The zero of $\phi^{r}$ at $r=0.2412$ lies between the two horizons and is not part of the exterior geometry.

\begin{figure}[!tbp]
\pan{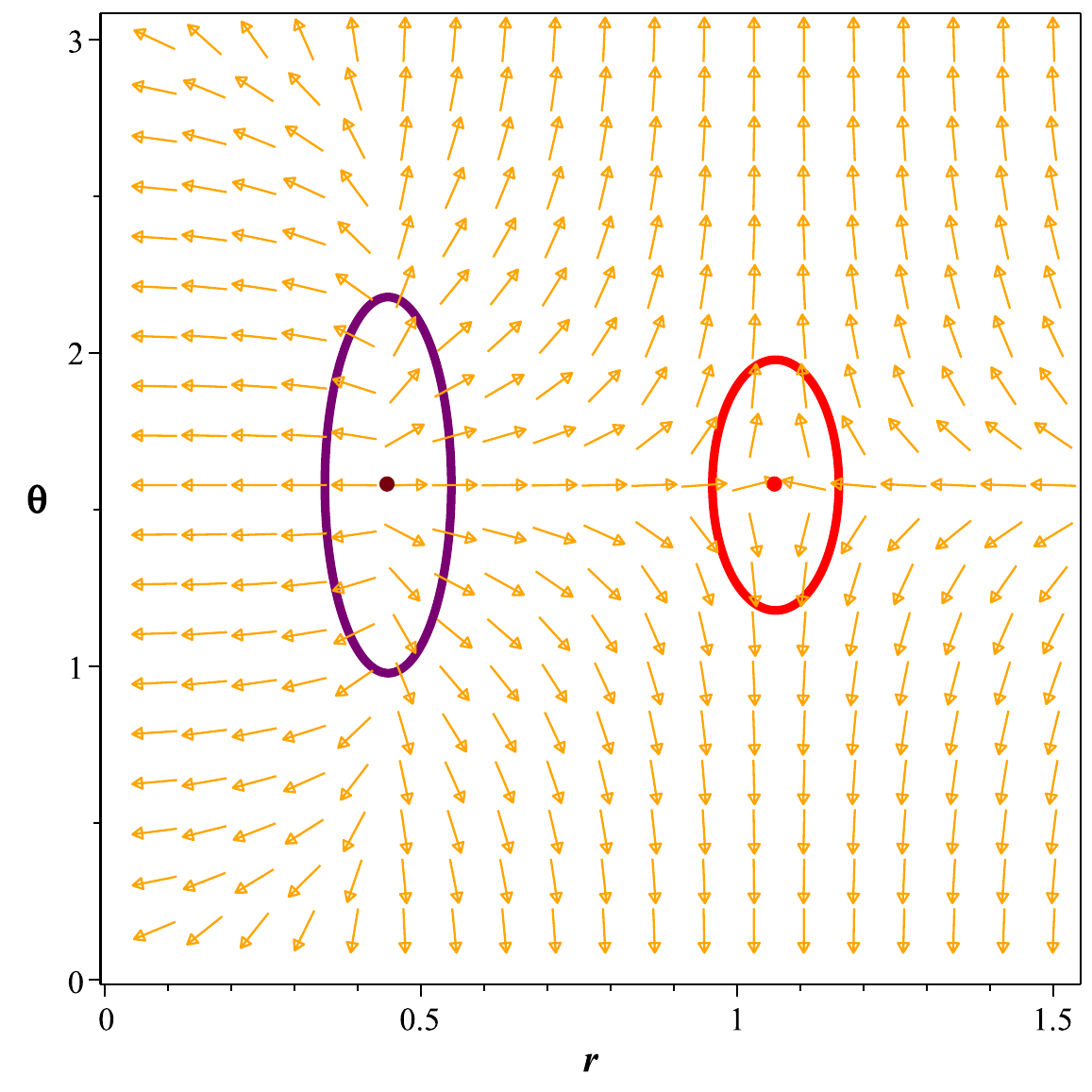}{(a)}\hfill\pan{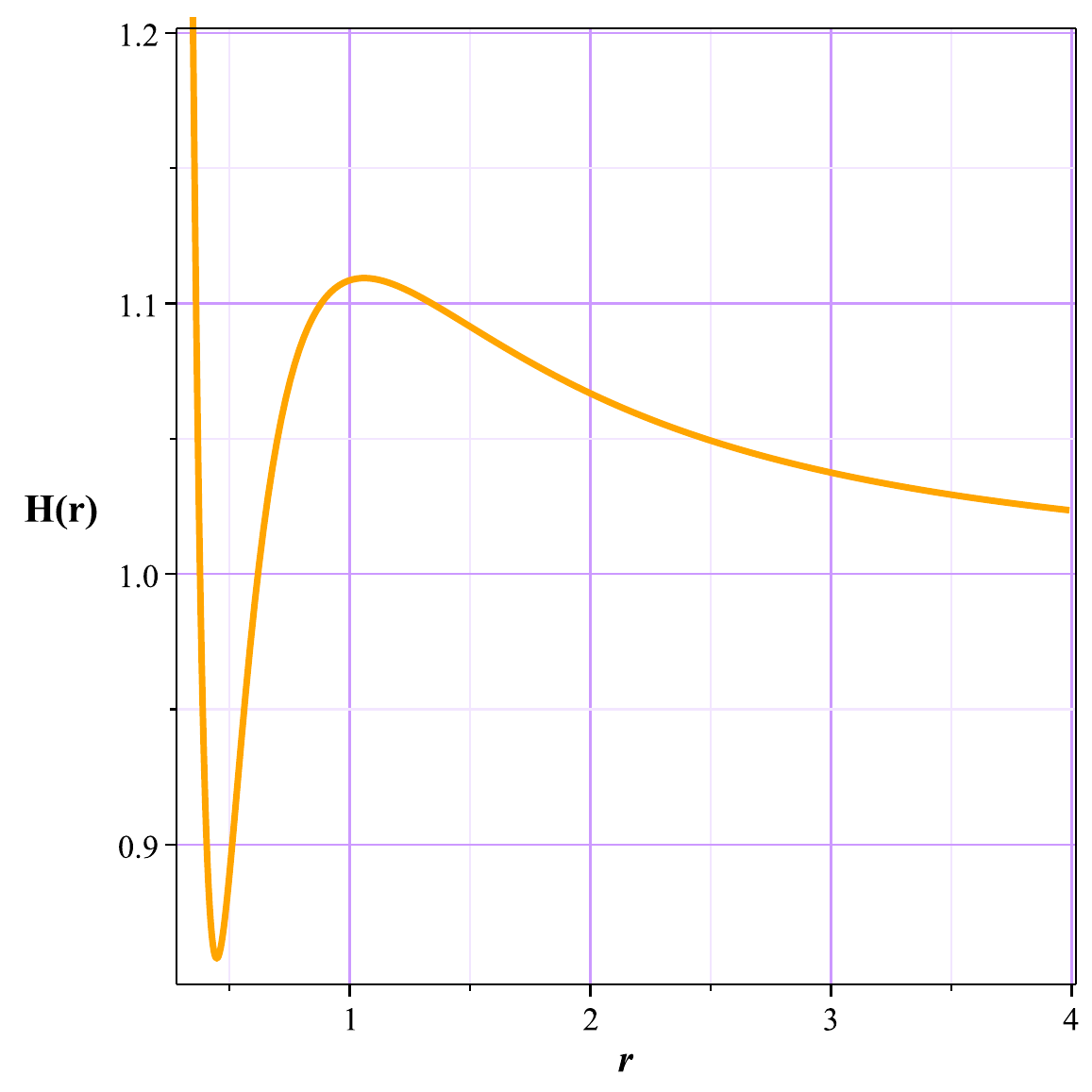}{(b)}
\caption{(a) Normalized topological field $n$ of the background metric for $\sigma=0.15$ with the stable circular null orbit at $r=0.4479$ ($\omega=+1$, purple) and the unstable circular null orbit at $r=1.0606$ ($\omega=-1$, red). (b) $\mathbf{H}(r,\pi/2)$ for the same parameters ($q=0.5$, $m_{0}=1$, $l=1$, $\beta=0.1$).}\label{fig:n2}
\end{figure}

For $\sigma=0.15$ the geometry has no horizon. Figure~\ref{fig:n2}(a) shows a stable photon sphere at $r=0.4479$ with $\omega=+1$ and an unstable one at $r=1.0606$ with $\omega=-1$, and Fig.~\ref{fig:n2}(b) shows the corresponding minimum and maximum of $\mathbf{H}$. The stable and unstable orbit indices sum to $Q=0$. This behavior holds in the whole interval $\sigma_{1}<\sigma<\sigma_{2}$, that is, $0.1465105<\sigma<0.1548466$ for $\beta=0.1$. For $\sigma_{2}<\sigma<1/4$ the field $\phi^{r}$ has no zero, and there is neither a horizon nor a background circular null orbit. This is consistent with the fixed-mass criterion of Sec.~\ref{sec:fixedM}: on this interval the total mass $M(\sigma)=m_{0}+m_{M}(\sigma)$ lies below $M_{\rm ps}^{\min}$, so no background photon-sphere pair can occur. The interval belongs to the NS0 class.

\begin{figure}[!tbp]
\pan{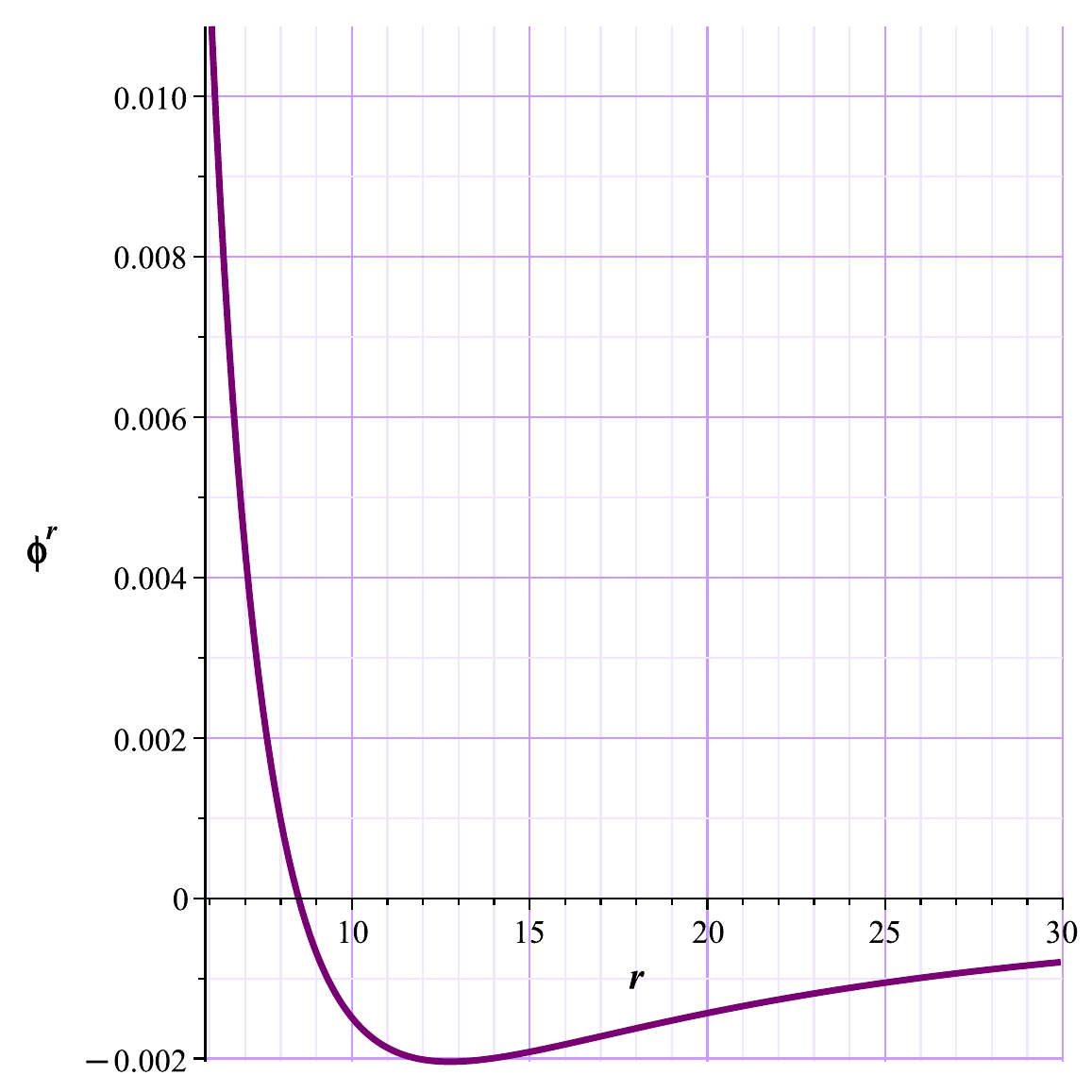}{(a)}\hfill\pan{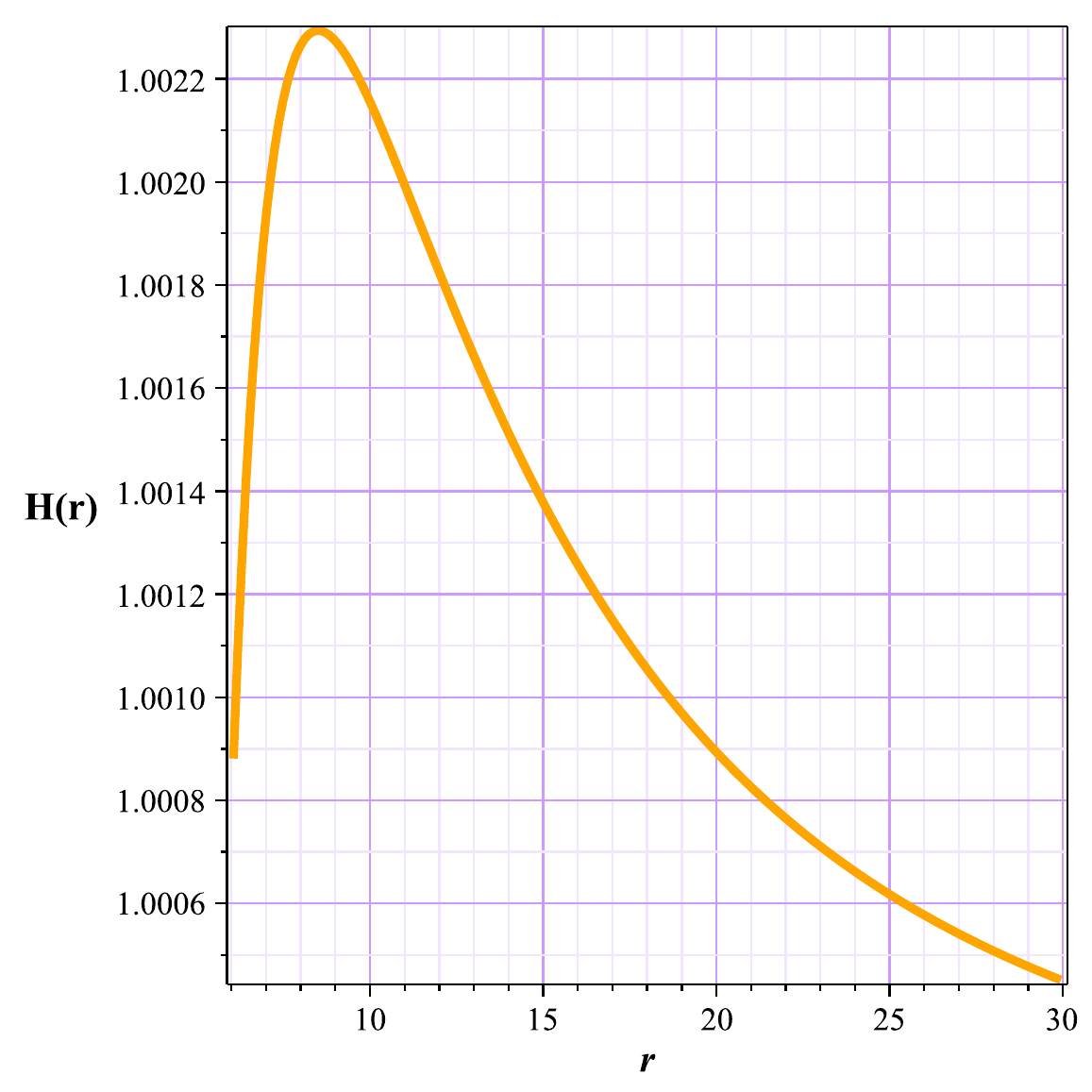}{(b)}
\caption{(a) Background radial component $\phi^{r}(r,\pi/2)$ for $\sigma=0.3$ ($q=0.5$, $m_{0}=1$, $l=1$, $\beta=0.1$), with its single zero at $r=8.5046$. (b) $\mathbf{H}(r,\pi/2)$ outside the horizon $r_{+}=1.5831$.}\label{fig:n3}
\end{figure}

For $\sigma>1/4$ a single horizon and a single unstable photon sphere were found at every value tested. For $\sigma=0.3$ they are located at $r_{+}=1.5831$ and $r=8.5046$, and Fig.~\ref{fig:n3} shows the single zero of $\phi^{r}$ and the maximum of $\mathbf{H}$. As $\sigma\to1/4^{+}$ the photon-sphere radius grows without bound, $r=27.42$ at $\sigma=0.26$ and $r=239.6$ at $\sigma=0.251$. The results are collected in Table~\ref{tab:cl1}.
\begin{table*}[!t]
\caption{Horizon and background circular-null-orbit classification for $m_0=l=1$ and $\beta=0.1$. BH denotes a black hole; NS and NS0 denote singular horizonless configurations with and without a stable--unstable background-orbit pair, respectively. U-PS and S-PS label unstable and stable background orbits. The listed radii correspond to the representative $\sigma$ values in the third column. An orbit-free domain has total index zero; for the NS0 rows, $\mathrm{TTC}=0$ denotes this empty sum and the topological theorem is not applied to it. The fixed-$m_0$ parametrization omits $\sigma=1/4$. The lower block gives the same classification for additional charges.}\label{tab:cl1}
\begin{ruledtabular}
\begin{tabular}{llcccccl}
$q$ & Interval of $\sigma$ & $\sigma$ & Horizons & Background orbits & TTC & Class & Radii\\
\hline
$0.5$ & $0<\sigma<0.1465105$ & $0.12$ & 2 & U-PS & $-1$ & BH & $r_{-}=0.1832$, $r_{+}=0.6953$; $r_{\mathrm{U}}=1.8087$\\
$0.5$ & $0.1465105<\sigma<0.1548466$ & $0.15$ & 0 & S-PS, U-PS & $0$ & NS & $r_{\mathrm{S}}=0.4479$; $r_{\mathrm{U}}=1.0606$\\
$0.5$ & $0.1548466<\sigma<1/4$ & $0.2$ & 0 & none & $0$ & NS0 & ---\\
$0.5$ & $\sigma>1/4$ & $0.3$ & 1 & U-PS & $-1$ & BH & $r_{+}=1.5831$; $r_{\mathrm{U}}=8.5046$\\
$1.2$ & $0<\sigma<1/4$ & $0.15$ & 0 & none & $0$ & NS0 & ---\\
$1.2$ & $\sigma>1/4$ & $0.9$ & 1 & U-PS & $-1$ & BH & $r_{+}=1.3280$; $r_{\mathrm{U}}=6.6879$\\
\end{tabular}
\end{ruledtabular}
\vspace{4pt}
\begin{ruledtabular}
\begin{tabular}{cccccc}
$q\backslash\sigma$ & $0.05$ & $0.12$ & $0.2$ & $0.3$ & $0.9$\\
\hline
$0.3$ & BH & BH & BH & BH & BH\\
$0.5$ & BH & BH & NS0 & BH & BH\\
$0.8$ & NS & NS0 & NS0 & BH & BH\\
$1.0$ & NS0 & NS0 & NS0 & BH & BH\\
$1.2$ & NS0 & NS0 & NS0 & BH & BH\\
\end{tabular}
\end{ruledtabular}
\end{table*}

\subsection{Case \texorpdfstring{$q>m_{0}$}{q > m0}}
\begin{figure}[!tbp]
\centering\includegraphics[width=0.85\columnwidth]{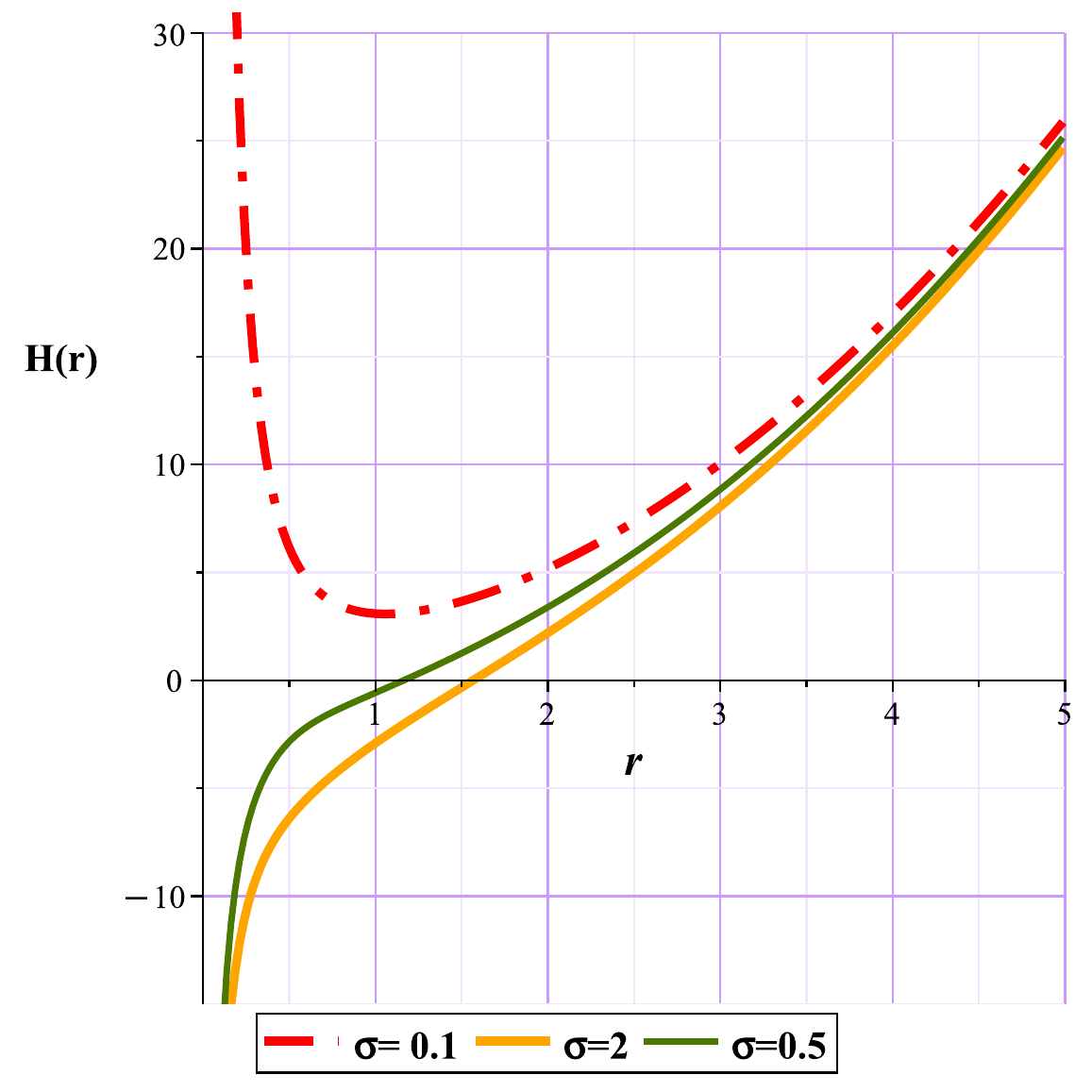}
\caption{Background metric function $f(r)$ for $q=1.2$, $m_{0}=1$, $l=1$, $\beta=0.1$ and $\sigma=0.1,\,0.5,\,2$. No horizon exists for $\sigma=0.1<1/4$; there is one horizon for $\sigma=0.5$ and $2$.}\label{fig:f2}
\end{figure}

\begin{figure}[!tbp]
\pan{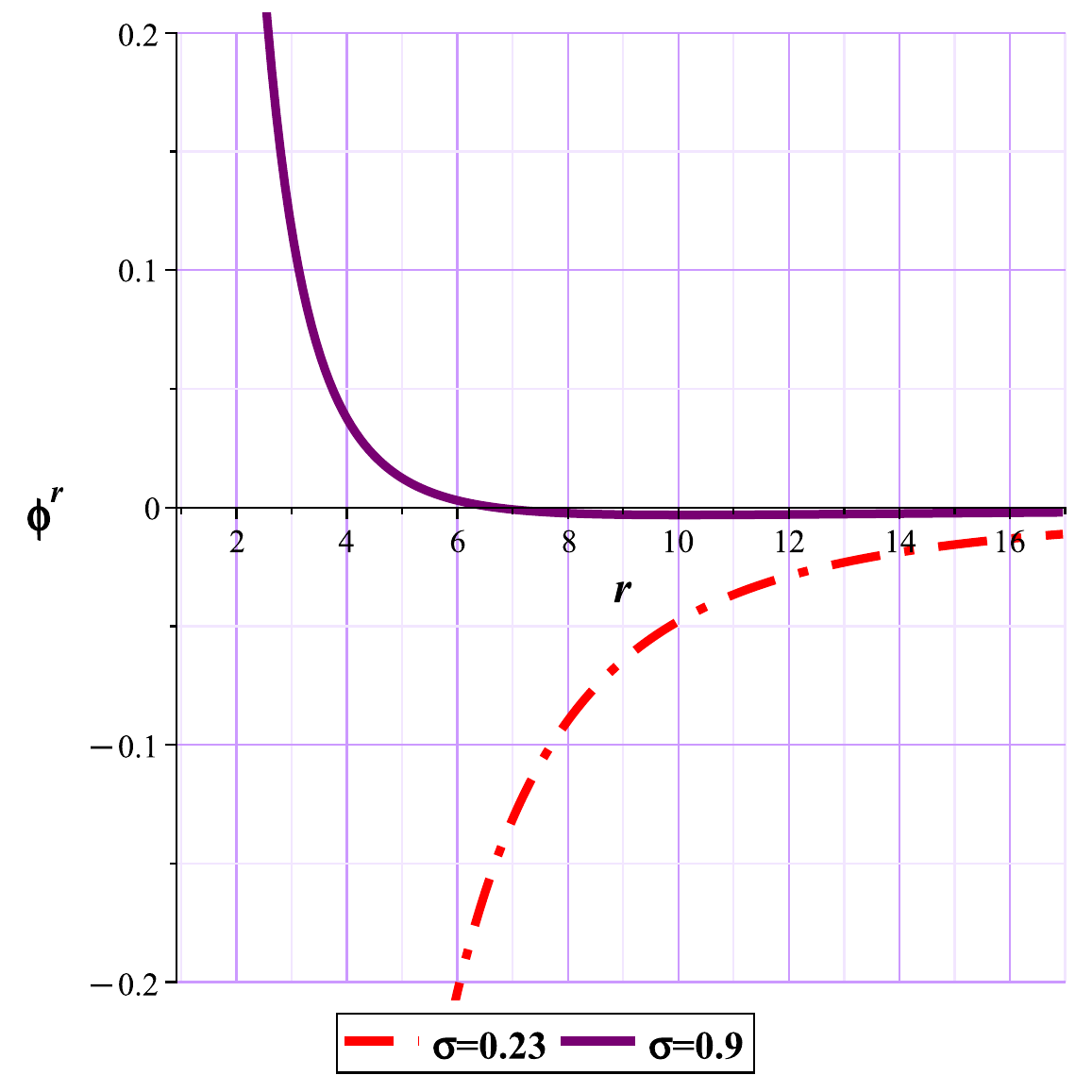}{(a)}\hfill\pan{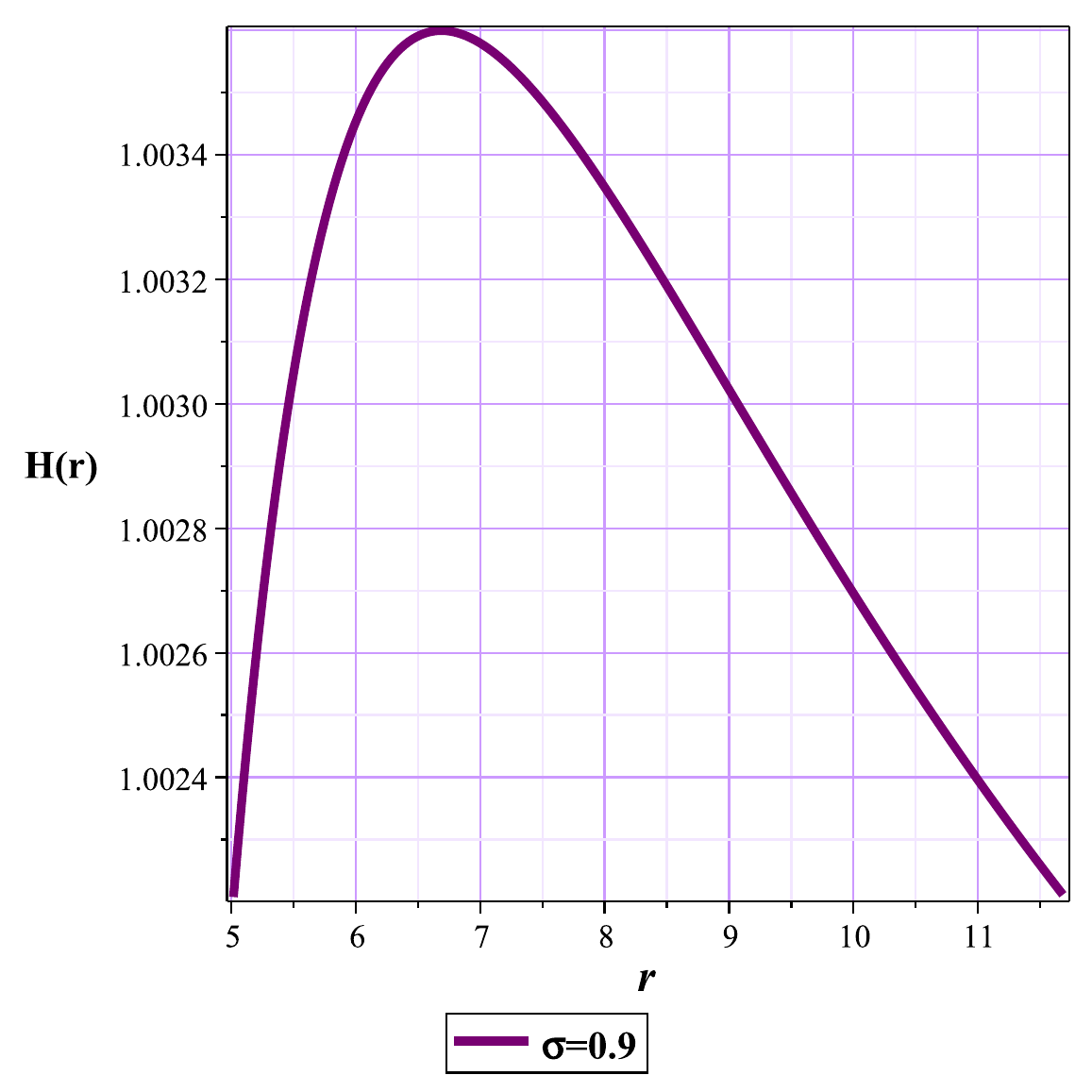}{(b)}
\caption{(a) Background radial component $\phi^{r}(r,\pi/2)$ for $\sigma=0.9$ (purple, zero at $r=6.6879$) and $\sigma=0.23$ (red, dash-dotted, no zero), with $q=1.2$, $m_{0}=1$, $l=1$, $\beta=0.1$. (b) $\mathbf{H}(r,\pi/2)$ outside the horizon $r_{+}=1.3280$.}\label{fig:n4}
\end{figure}

For $q=1.2$ no horizon is found for $0<\sigma<1/4$, for any of the values $\beta=0.001$, $0.1$ and $1.2$ (tested at $\sigma=0.02,\,0.08,\,0.15,\,0.2,\,0.24$, and for $\beta=0.1$ on a grid of step $0.01$ in $\sigma$), as shown by the metric function in Fig.~\ref{fig:f2}. The field $\phi^{r}$ has no zero in this range (Fig.~\ref{fig:n4}(a), curve for $\sigma=0.23$); this is consistent with the fixed-mass classification of Sec.~\ref{sec:fixedM}, and the range is classified as NS0 in the background-metric sector. For $\sigma>1/4$ there is one horizon and one unstable photon sphere. For $\sigma=0.9$ they are at $r_{+}=1.3280$ and $r=6.6879$ (Fig.~\ref{fig:n4}), and the photon-sphere radius grows without bound as $\sigma\to1/4^{+}$ ($r=93.82$ at $\sigma=0.26$ and $r=882.8$ at $\sigma=0.251$). The result is included in Table~\ref{tab:cl1}.

\subsection{Dependence on the magnetic charge}
The value $q=m_{0}$ is not the boundary between the two behaviors, which is a function of $\sigma$ and $\beta$. The lower block of Table~\ref{tab:cl1} gives the class for several charges at $\beta=0.1$. For $q=0.3$ the black hole persists at all tested values of $\sigma$. For $q=0.8<m_{0}$ the geometry is a naked singularity with a photon-sphere pair at $\sigma=0.05$ and belongs to NS0 at $\sigma=0.12$ and $0.2$. The values $q=1$ and $q=1.2$ give the same pattern.

\subsection{Relation to the ADM mass}\label{sec:ADM}
The parameter $m_{0}$ is not the total mass of the solution. For $\sigma\neq1/4$ the ADM mass is $M=m_{0}+m_{M}(\sigma)$, and at fixed $m_{0}$ it changes with $\sigma$ and diverges at $\sigma=1/4$. For $q=0.5$, $m_{0}=1$ and $\beta=0.1$ one finds $M=0.6950,\,0.5105,\,-0.2804,\,-6.571,\,9.145,\,2.854,\,1.369$ at $\sigma=0.12,\,0.15,\,0.20,\,0.24,\,0.26,\,0.30,\,0.90$, respectively. The mass decreases with $\sigma$ in $0<\sigma<1/4$, passes through zero at $\sigma\simeq0.189$ and tends to $-\infty$ as $\sigma\to1/4^{-}$, whereas it tends to $+\infty$ as $\sigma\to1/4^{+}$. The absence of horizons and photon spheres for $\sigma_{2}<\sigma<1/4$ therefore reflects an ADM mass that is small or negative in that interval, and the growth of the horizon and photon-sphere radii for $\sigma\to1/4^{+}$ reflects the growth of the mass. This accounts for the boundary at $\sigma=1/4$ in Table~\ref{tab:cl1}, which is a feature of the parametrization by $m_{0}$ and not of the metric written in terms of $M$, Eq.~\eqref{eq:fADM}, which is analytic across $\sigma=1/4$. The classification at fixed total mass is given in Sec.~\ref{sec:fixedM}. Central regularity requires $\sigma\geq1$ and $m_0=0$ \cite{ref089}. The configurations in the fixed-$m_0=1$ scan are therefore singular at the center. For $\sigma\leq1/4$, the magnetic energy integral from the origin diverges, although the fixed-$M$ exterior metric remains defined for $r>0$.

\subsection{Classification at fixed total mass}\label{sec:fixedM}
\begin{figure*}[!tbp]
\centering\includegraphics[width=0.90\textwidth]{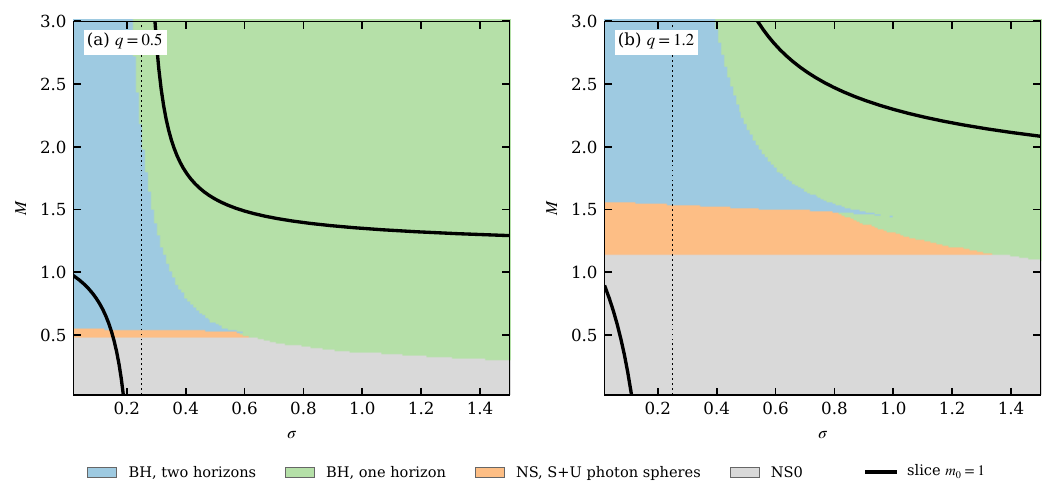}
\caption{Background classification in the $(\sigma,M)$ plane for $\beta=0.1$, $l=1$ and (a) $q=0.5$, (b) $q=1.2$. BH, NS and NS0 follow the definitions in Table~\ref{tab:cl1}; all orbit labels refer to the background metric. The black curve is the slice $M=m_{0}+m_{M}(\sigma)$ with $m_{0}=1$, which diverges at $\sigma=1/4$ (dotted line) and is shown where $0<M<3$.}\label{fig:mapM}
\end{figure*}

The thresholds of Table~\ref{tab:thr} refer to the slice $m_{0}=1$. To separate the structure of the solution from this choice, the classification was repeated with the total mass $M$ as the independent parameter, using Eq.~\eqref{eq:fADM}, which is analytic at $\sigma=1/4$. In this form the horizons satisfy $M=M_{h}(r)$ and the photon spheres satisfy $rf'-2f=0$, that is $M=M_{\rm ps}(r)$, with
\begin{align}\label{eq:Mcrit}
M_{h}(r)&=\frac{r}{2}\left[1+\frac{r^{2}}{l^{2}}+\frac{q^{2}}{r^{2}}\,\hyp{\sigma,\tfrac14}{\tfrac54}{-\frac{\beta q^{2}}{r^{4}}}\right],\nonumber\\
M_{\rm ps}(r)&=\frac{r}{6}\left[2+\frac{3q^{2}}{r^{2}}\,\hyp{\sigma,\tfrac14}{\tfrac54}{-\frac{\beta q^{2}}{r^{4}}}+\mathcal{I}'(r)\right].
\end{align}
The number of horizons and of photon spheres at a given $M$ is the number of intersections of the line $M=\mathrm{const}$ with these two curves, and the class of the geometry changes where $M$ equals a local extremum of $M_{h}$ or $M_{\rm ps}$. Figure~\ref{fig:mapM} shows the classes in the $(\sigma,M)$ plane for $\beta=0.1$ and $l=1$, together with the slice $M=m_{0}+m_{M}(\sigma)$ at $m_{0}=1$. Table~\ref{tab:Mcrit} gives the critical masses: for $\sigma\le1/4$ a pair of photon spheres exists for $M\ge M_{\rm ps}^{\min}$, and a horizon for $M\ge M_{\rm ext}$, the minimum of $M_{h}$.
\begin{table}[!t]
\caption{Background critical masses at fixed $M$ for $\beta=0.1$, $l=1$: the minimum $M_{\rm ps}^{\min}$ of $M_{\rm ps}(r)$, above which a photon-sphere pair exists, and the extremal mass $M_{\rm ext}$, the minimum of $M_{h}(r)$, above which a horizon exists. For $\sigma<1/4$, the classes are NS0 for $M<M_{\rm ps}^{\min}$ and NS for $M_{\rm ps}^{\min}<M<M_{\rm ext}$.}\label{tab:Mcrit}
\begin{ruledtabular}
\begin{tabular}{ccccc}
 & \multicolumn{2}{c}{$q=0.5$} & \multicolumn{2}{c}{$q=1.2$}\\
$\sigma$ & $M_{\rm ps}^{\min}$ & $M_{\rm ext}$ & $M_{\rm ps}^{\min}$ & $M_{\rm ext}$\\
\hline
$0.05$ & $0.4709$ & $0.5420$ & $1.1312$ & $1.5414$\\
$0.12$ & $0.4703$ & $0.5387$ & $1.1309$ & $1.5353$\\
$0.20$ & $0.4695$ & $0.5346$ & $1.1306$ & $1.5282$\\
$0.24$ & $0.4691$ & $0.5324$ & $1.1304$ & $1.5246$\\
$0.26$ & $0.4689$ & $0.5313$ & $1.1303$ & $1.5228$\\
$0.30$ & $0.4685$ & $0.5290$ & $1.1302$ & $1.5190$\\
$0.50$ & $0.4664$ & $0.5153$ & $1.1294$ & $1.4993$\\
\end{tabular}
\end{ruledtabular}
\end{table}
Two results follow. First, the critical masses vary smoothly with $\sigma$ across $\sigma=1/4$; no feature of the classification at fixed $M$ is associated with this value. Second, the thresholds of the fixed-$m_{0}$ scan are the values of $\sigma$ at which the curve $M=m_{0}+m_{M}(\sigma)$ crosses these critical masses. For $q=0.5$ the relations $m_{0}+m_{M}(\sigma_{1})=M_{\rm ext}(\sigma_{1})$ and $m_{0}+m_{M}(\sigma_{2})=M_{\rm ps}^{\min}(\sigma_{2})$ hold for each of the three values of $\beta$ in Table~\ref{tab:thr}, for example $0.53735$ and $0.46996$ at $\beta=0.1$, and the same masses are obtained from the two curves of Eq.~\eqref{eq:Mcrit}. The interval $\sigma_{2}<\sigma<1/4$ therefore corresponds to a total mass below $M_{\rm ps}^{\min}$, and for $q=1.2$ the fixed-$m_0$ slice lies in NS0 for $\sigma<1/4$, since $m_M<0$ there, so that $M<m_{0}=1<M_{\rm ps}^{\min}=1.13$. The single-horizon black hole for $\sigma>1/4$ at fixed $m_{0}$ reflects the large values of $M=m_{0}+m_{M}$, which exceed $M_{\rm ext}$ by a wide margin.

\subsection{Effective optical geometry}\label{sec:optical}
Electromagnetic disturbances in nonlinear electrodynamics propagate along characteristic cones determined by the field Lagrangian \cite{refOptNovello,refOptCosta}. We consider the geometric-optics limit on the fixed magnetic background and compare its extraordinary circular rays with the background null geodesics.

\subsubsection{Characteristic cones and propagation domain}
For $\mathcal F=F_{\mu\nu}F^{\mu\nu}/4$, the field equation and Bianchi identity are
\begin{equation}\label{eq:optfield}
\nabla_\mu(\mathcal L_{\mathcal F}F^{\mu\nu})=0,
\qquad \nabla_{[\lambda}F_{\mu\nu]}=0.
\end{equation}
Writing the leading disturbance as $b_{\mu\nu}=k_\mu a_\nu-k_\nu a_\mu$ gives the principal equation
\begin{equation}\label{eq:optprincipal}
\begin{split}
\mathcal L_{\mathcal F}[k^2a^\nu-k^\nu(k\cdot a)]
+\mathcal L_{\mathcal F\mathcal F}\zeta^\nu(\zeta\cdot a)&=0,\\
\zeta^\nu&=k_\mu F^{\mu\nu}.
\end{split}
\end{equation}
Away from coefficient degeneracies, the physical characteristic polynomial factors into an ordinary and an extraordinary cone \cite{refOptPolar}:
\begin{equation}\label{eq:optcones}
(g^{\mu\nu}k_\mu k_\nu)
(\gamma^{\alpha\beta}k_\alpha k_\beta)=0,
\end{equation}
The ordinary factor is the null-cone condition of the background metric $g_{\mu\nu}$. Consequently, the background circular null orbits studied in Sec.~\ref{sec:model} are the circular rays of the ordinary polarization mode; they are not, without further analysis, the circular rays of the extraordinary mode.
\begin{equation}\label{eq:optinverse}
\gamma^{\mu\nu}=g^{\mu\nu}
-\frac{\mathcal L_{\mathcal F\mathcal F}}{\mathcal L_{\mathcal F}}
F^\mu{}_{\alpha}F^{\alpha\nu}.
\end{equation}
The coefficient in Eq.~\eqref{eq:optinverse} follows the $1/4$ convention for $\mathcal F$; the two cones can coincide along special propagation directions.

For the magnetic field $F_{\theta\varphi}=q\sin\theta$, introduce
\begin{equation}\label{eq:optu}
u=\frac{\beta q^2}{r^4},\qquad A(u)=1+(1-\sigma)u,
\end{equation}
\begin{equation}\label{eq:optN}
N(u)=1+(2-5\sigma)u+(1-\sigma)(1-2\sigma)u^2.
\end{equation}
For $\sigma>1$, $A(u)$ vanishes at $u=1/(\sigma-1)$, i.e. at $r=[\beta q^2(\sigma-1)]^{1/4}$; on and beyond this radius the extraordinary optical geometry is not defined by Eq.~\eqref{eq:optinverse}. The extraordinary analysis is therefore restricted to the propagation domain in which $A(u)\neq0$, together with the conditions stated below. Differentiating Eq.~\eqref{eq:lag} yields
\begin{align}\label{eq:optderivatives}
\mathcal L_{\mathcal F}&=-\frac{A(u)}{4\pi(1+u)^{\sigma+1}},\\
\mathcal K\equiv\mathcal L_{\mathcal F}+2\mathcal F\mathcal L_{\mathcal F\mathcal F}
&=-\frac{N(u)}{4\pi(1+u)^{\sigma+2}}.\nonumber
\end{align}
The ratio
\begin{equation}\label{eq:opteta}
\eta(r)=\frac{\mathcal K}{\mathcal L_{\mathcal F}}
=\frac{N(u)}{(1+u)A(u)}
\end{equation}
defines the extraordinary propagation metric, up to a regular conformal factor:
\begin{equation}\label{eq:optmetric}
ds_{\rm e}^2=-f\,dt^2+\frac{dr^2}{f}+\frac{r^2}{\eta}d\Omega^2.
\end{equation}
Its static optical spatial metric is $dt^2=dr^2/f^2+r^2d\Omega^2/(f\eta)$, and $\eta\to1$ in the Maxwell limit.

We use the static component connected to the asymptotic weak-field region, with
\begin{equation}\label{eq:optdomain}
f>0,\qquad A(u)>0,\qquad N(u)>0.
\end{equation}
These conditions retain the Maxwell signs of $\mathcal L_{\mathcal F}$ and $\mathcal K$ and keep the principal part nondegenerate with the same static time direction. They specify a local propagation domain, rather than a full dynamical-stability criterion. For $\sigma=0.12$, $0.15$ and $0.30$, the coefficients of $A$ and $N$ are positive for $u\geq0$. For $\sigma=0.9$, the positive root of $N=1-2.5u-0.08u^2$ is $u_*\simeq0.3950070225$. At $q=1.2$ and $\beta=0.1$, it corresponds to $r_{\rm deg}\simeq0.777033$, inside the outer horizon $r_+\simeq1.328025$ of the example below.

\subsubsection{Circular rays and topological indices}
The functions entering the topological construction are
\begin{equation}\label{eq:optH}
H_{\rm e}=\frac{\sqrt{f\eta}}{r\sin\theta},
\qquad U_{\rm e}=H_{\rm e}^2(r,\pi/2)=\frac{f\eta}{r^2},
\end{equation}
\begin{align}\label{eq:optphi}
\phi_{\rm e}^r&=\frac{\sqrt\eta}{2r^2\sin\theta}
\left(rf'-2f+rf\frac{\eta'}{\eta}\right),\\
\phi_{\rm e}^{\theta}&=-\frac{\eta\sqrt f\cos\theta}{r^2\sin^2\theta}.\nonumber
\end{align}
The circular rays lie at $\theta=\pi/2$ and solve
\begin{equation}\label{eq:optcondition}
\mathcal C_{\rm e}(r)\equiv\eta(r)(rf'-2f)+rf\eta'=0,
\end{equation}
equivalently $U_{\rm e}'=0$ on the domain~\eqref{eq:optdomain}. A nondegenerate maximum of $U_{\rm e}$ is unstable and has winding number $-1$; a minimum is stable and has winding number $+1$, with the orientation of Eq.~\eqref{eq:winding}.

Writing $f=f_0+r^2/l^2$, with $f_0$ independent of $l$, gives
\begin{equation}\label{eq:optAdS}
\mathcal C_{\rm e}
=\eta(rf_0'-2f_0)+rf_0\eta'+\frac{r^3}{l^2}\eta'.
\end{equation}
The extraordinary orbit condition therefore depends explicitly on the AdS scale when $\eta'\neq0$.

\subsubsection{Comparison with the background orbits}
Table~\ref{tab:opticalcompare} compares selected background and extraordinary orbits at $m_0=l=1$ and $\beta=0.1$. Candidate roots were bracketed on a 6000-point logarithmic grid in the valid static component, up to $r=10^3$, and refined by differentiating $U_{\rm b}=f/r^2$ and $U_{\rm e}$ at 50 and 80 decimal digits. The displayed radii agree between the two precisions. The curvature signs and winding numbers agree; the latter were evaluated on two 4097-point elliptical contours with relative radial and angular amplitudes $10^{-3}$ and $10^{-4}$.

\begin{table*}[!t]
\caption{Background and extraordinary circular orbits at $m_0=l=1$, $\beta=0.1$. S and U denote stable and unstable orbits. Subscripts b and e denote background and extraordinary quantities, respectively. The relative shift is $100(r_{\rm e}-r_{\rm b})/r_{\rm b}$, and $\eta_{\rm e}=\eta(r_{\rm e})$. The black-hole rows refer to exterior orbits; both $\sigma=0.15$ rows belong to the same horizonless geometry.}
\label{tab:opticalcompare}
\begingroup\small\setlength{\tabcolsep}{5pt}
\begin{ruledtabular}
\begin{tabular}{ccccccccc}
$q$ & $\sigma$ & Type & $r_{\rm b}$ & $r_{\rm e}$ & Shift (\%) & $\omega_{\rm b}$ & $\omega_{\rm e}$ & $\eta_{\rm e}$\\
\hline
$0.5$ & $0.12$ & U & $1.808704839$ & $1.825562232$ & $+0.9320$ & $-1$ & $-1$ & $0.998923067$ \\
$0.5$ & $0.15$ & S & $0.447933720$ & $0.436704141$ & $-2.5070$ & $+1$ & $+1$ & $0.800652468$ \\
$0.5$ & $0.15$ & U & $1.060574675$ & $1.115589072$ & $+5.1872$ & $-1$ & $-1$ & $0.990533890$ \\
$0.5$ & $0.3$ & U & $8.504555429$ & $8.511686138$ & $+0.0838$ & $-1$ & $-1$ & $0.999994284$ \\
$1.2$ & $0.9$ & U & $6.687920296$ & $6.850593978$ & $+2.4324$ & $-1$ & $-1$ & $0.999764646$ \\
\end{tabular}
\end{ruledtabular}
\endgroup
\end{table*}

\begin{figure*}[!tbp]
\centering
\includegraphics[width=0.42\textwidth]{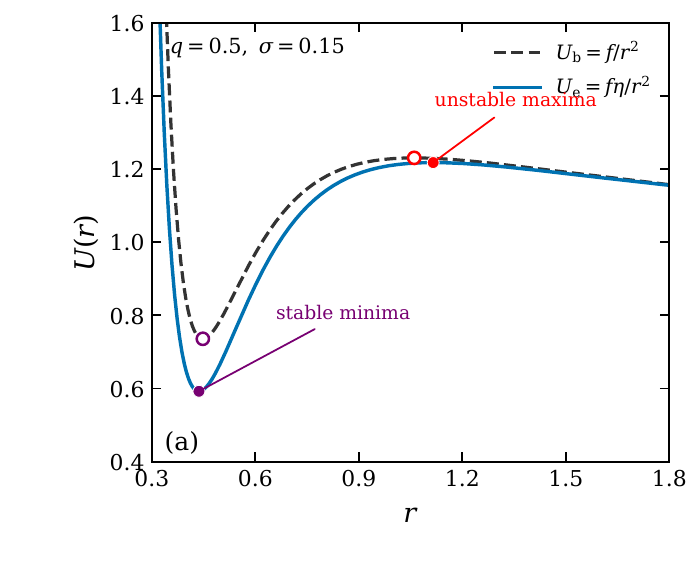}
\includegraphics[width=0.42\textwidth]{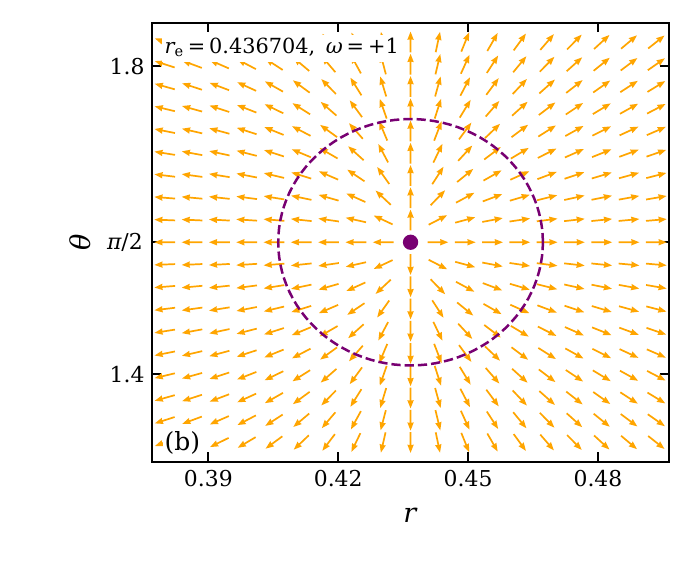}
\includegraphics[width=0.42\textwidth]{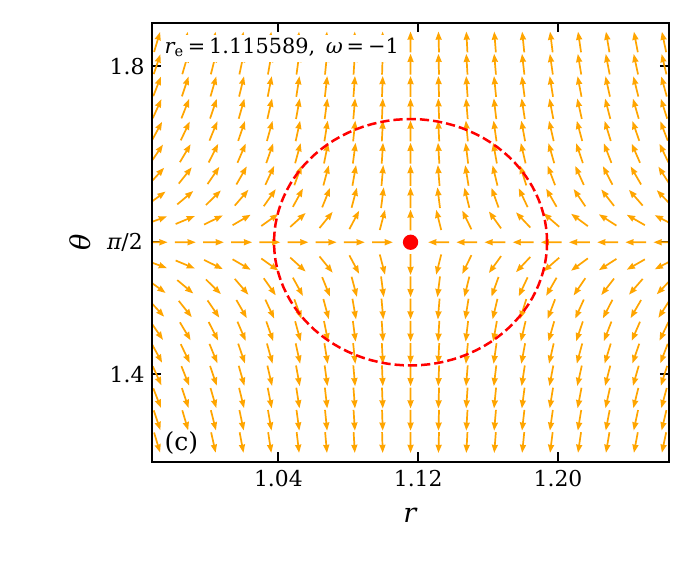}
\includegraphics[width=0.42\textwidth]{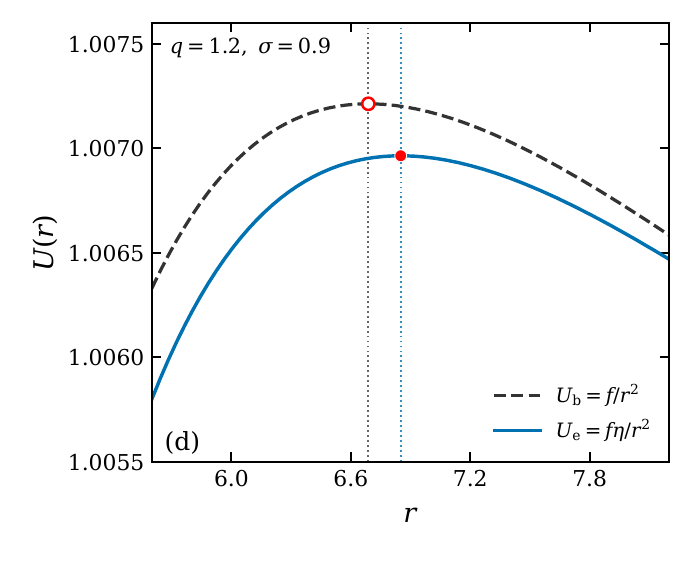}
\caption{Background and extraordinary optical orbit structure at $m_0=l=1$, $\beta=0.1$. (a) $U_{\rm b}=f/r^2$ (dark dashed curve) and $U_{\rm e}=f\eta/r^2$ (blue solid curve) for the horizonless geometry with $q=0.5$, $\sigma=0.15$. Open and filled circles mark background and extraordinary extrema, respectively; purple denotes stable minima and red unstable maxima. (b,c) Normalized extraordinary topological fields $\boldsymbol n_{\rm e}=\boldsymbol\phi_{\rm e}/\|\boldsymbol\phi_{\rm e}\|$ near the stable and unstable orbits. Orange arrows show the field direction, with the zero itself omitted. The purple and red dashed contours have counterclockwise winding numbers $+1$ and $-1$, verified independently of the displayed field grid. (d) The two potentials near the exterior unstable orbits for $q=1.2$, $\sigma=0.9$, using the same curve colours as in (a); dotted lines locate $r_{\rm b}$ and $r_{\rm e}$. All plotted regions satisfy Eq.~\eqref{eq:optdomain}.}
\label{fig:opticalcomparison}
\end{figure*}

Figure~\ref{fig:opticalcomparison} resolves the displacement of the two horizonless extrema and displays the effective vector fields near their zeros. The minimum moves inward and the maximum outward, with local indices $+1$ and $-1$. The exterior black-hole example also shifts outward. The largest relative displacement among the five listed orbits is $5.19\%$, while their stability assignments are unchanged.

This comparison concerns isolated orbits at the stated parameter values. The labels NS and NS0 in the preceding sections refer to the background-metric sector, equivalently to the ordinary polarization mode. The extraordinary merger boundaries require $\mathcal C_{\rm e}=\partial_r\mathcal C_{\rm e}=0$, together with the propagation-domain boundaries; they are not determined by the representative root search, so the background classification should not be extrapolated to the extraordinary sector without solving these conditions.

\section{Topological Phase Diagrams}\label{sec:thermo}
\subsection{Generalized free energy}
The generalized free energy is \cite{ref072}
\begin{equation}\label{eq:Fgen}
\mathfrak{F}=M-\frac{S}{\tau},
\end{equation}
where $\tau$ is the inverse temperature of the ensemble, and it is on shell only for $\tau=\tau_{h}=1/T_{h}$. In the extended phase space the pressure is $P=-\Lambda/8\pi=3/(8\pi l^{2})$, the entropy is $S=\pi r_{h}^{2}$, and the mass $M$ is an enthalpy \cite{ref089}. On shell, $\mathfrak{F}$ therefore equals the Gibbs free energy $G=M-TS$, and not the Helmholtz free energy. The condition $f(r_{h})=0$ in Eq.~\eqref{eq:fADM} gives
\begin{equation}\label{eq:Mrh}
M=\frac{r_{h}}{2}+\frac{4\pi P r_{h}^{3}}{3}+\frac{q^{2}}{2r_{h}}\,\hyp{\tfrac14,\sigma}{\tfrac54}{-\frac{\beta q^{2}}{r_{h}^{4}}},
\end{equation}
and the Hawking temperature $T=\partial M/\partial S=M'(r_{h})/(2\pi r_{h})$ is
\begin{equation}\label{eq:T}
T=\frac{1}{4\pi r_{h}}\left[1+8\pi P r_{h}^{2}-\frac{q^{2}}{r_{h}^{2}}\,\mathcal{F}_{1}+\frac{4\sigma\beta q^{4}}{5r_{h}^{6}}\,\mathcal{F}_{2}\right],
\end{equation}
with $\mathcal{F}_{1}=\hyp{\tfrac14,\sigma}{\tfrac54}{z}$, $\mathcal{F}_{2}=\hyp{\tfrac54,\sigma+1}{\tfrac94}{z}$ and $z=-\beta q^{2}/r_{h}^{4}$. This agrees with the temperature of Ref.~\cite{ref089}, and the derivative of Eq.~\eqref{eq:Mrh} was checked numerically against Eq.~\eqref{eq:T}. The vector field is
\begin{align}\label{eq:phith}
\phi&=\left(\frac{\partial\mathfrak{F}}{\partial r_{h}},\,-\cot\Theta\csc\Theta\right),\nonumber\\
\frac{\partial\mathfrak{F}}{\partial r_{h}}&=M'(r_{h})-\frac{2\pi r_{h}}{\tau},
\end{align}
with $0\le\Theta\le\pi$, so that $\phi^{\Theta}$ diverges at $\Theta=0,\pi$ and the field points outward there. The zero points lie at $\Theta=\pi/2$ and satisfy
\begin{equation}\label{eq:tau}
\begin{split}
\tau&=\tau_{h}(r_{h})=\frac{2\pi r_{h}}{M'(r_{h})}\\
&=\frac{20\pi r_{h}^{7}}{40\pi P r_{h}^{8}+4\sigma\beta q^{4}\mathcal{F}_{2}-5q^{2}r_{h}^{4}\mathcal{F}_{1}+5r_{h}^{6}}.
\end{split}
\end{equation}
A zero point with $d\tau_{h}/dr_{h}<0$ (a branch with positive heat capacity) has winding number $+1$, and one with $d\tau_{h}/dr_{h}>0$ has winding number $-1$. The functions in this section depend only on $(q,\beta,\sigma,P)$. The mass parameter $m_{0}$ and the AdS radius $l$ do not appear separately, since $M$ is determined by $r_{h}$ and $P$ fixes $l$. The partition of $\sigma$ by horizon number in Sec.~\ref{sec:model} therefore does not apply here, and the charges $q=0.5$ and $q=1.2$ are used only for continuity with that section. The analysis is restricted to non-extremal black holes, $T>0$; extremal configurations occur at $T=0$ and terminate the branches.

\subsection{Extended first law and Smarr relation}\label{sec:smarr}
In the extended phase space the mass of Eq.~\eqref{eq:Mrh} is a function of the four variables $S=\pi r_{h}^{2}$, $P$, $q$ and $\beta$, while the exponent $\sigma$ is a fixed parameter of the model. The conjugate quantities are defined by the partial derivatives
\begin{align}\label{eq:conj}
T&=\left(\frac{\partial M}{\partial S}\right)_{P,q,\beta}, & V&=\left(\frac{\partial M}{\partial P}\right)_{S,q,\beta}=\frac{4\pi r_{h}^{3}}{3},\nonumber\\
\Phi&=\left(\frac{\partial M}{\partial q}\right)_{S,P,\beta}, & \mathcal{B}&=\left(\frac{\partial M}{\partial \beta}\right)_{S,P,q}.
\end{align}
Differentiating Eq.~\eqref{eq:Mrh} with the rule $\frac{d}{dz}\,\hyp{a,b}{c}{z}=\frac{ab}{c}\,\hyp{a+1,b+1}{c+1}{z}$ gives the temperature of Eq.~\eqref{eq:T} together with
\begin{equation}\label{eq:PhiB}
\Phi=\frac{q}{r_{h}}\,\mathcal{F}_{1}-\frac{\sigma\beta q^{3}}{5r_{h}^{5}}\,\mathcal{F}_{2},\qquad
\mathcal{B}=-\frac{\sigma q^{4}}{10\,r_{h}^{5}}\,\mathcal{F}_{2},
\end{equation}
where $\mathcal{F}_{1}$ and $\mathcal{F}_{2}$ are defined below Eq.~\eqref{eq:T}. Here $\Phi$ is the magnetic potential and $\mathcal{B}$ is the quantity conjugate to $\beta$. Since $M$ is a state function, its differential is the extended first law,
\begin{equation}\label{eq:firstlaw}
dM=T\,dS+V\,dP+\Phi\,dq+\mathcal{B}\,d\beta .
\end{equation}
Equations~\eqref{eq:PhiB} and~\eqref{eq:firstlaw} coincide with the corresponding results of Ref.~\cite{ref089} (Eqs.~(25)--(27) of the arXiv version).

The conjugate quantities have a direct interpretation in terms of the field energy. With the energy density $\rho=-\mathcal{L}=q^{2}x^{4\sigma-4}/[8\pi(x^{4}+\beta q^{2})^{\sigma}]$ of the magnetic field, one has $\mathcal{I}'=8\pi x^{2}\rho$ by Eq.~\eqref{eq:Ip}, and the horizon condition together with $\mathcal{I}(\infty)=2m_{M}$ turns Eq.~\eqref{eq:Mrh} into
\begin{equation}\label{eq:Mint}
M=\frac{r_{h}}{2}+\frac{4\pi P r_{h}^{3}}{3}+4\pi\int_{r_{h}}^{\infty}\rho(x)\,x^{2}\,dx .
\end{equation}
The integral converges for every $\sigma>0$, because $\rho\,x^{2}\sim x^{-2}$ at large $x$, so that the relations of this subsection also hold for $\sigma<1/4$ without reference to the analytic continuation of $m_{M}$. In this form $\Phi=4\pi\int_{r_{h}}^{\infty}\partial_{q}\rho\,x^{2}dx$ and $\mathcal{B}=4\pi\int_{r_{h}}^{\infty}\partial_{\beta}\rho\,x^{2}dx$. Since $\partial_{\beta}\rho<0$, one has $\mathcal{B}<0$ for all parameter values. The temperature can be written as $T=f'(r_{h})/4\pi=[1+8\pi P r_{h}^{2}-8\pi r_{h}^{2}\rho(r_{h})]/(4\pi r_{h})$, which is equal to Eq.~\eqref{eq:T}.

In the units $G=c=1$ one has $[M]=[r_{h}]=[q]=L$, $[S]=L^{2}$, $[P]=L^{-2}$ and $[\beta]=L^{2}$, so that $M(\lambda^{2}S,\lambda^{-2}P,\lambda q,\lambda^{2}\beta)=\lambda M(S,P,q,\beta)$. Euler's theorem for homogeneous functions then gives the Smarr relation
\begin{equation}\label{eq:smarr}
M=2TS-2PV+\Phi q+2\beta\,\mathcal{B}.
\end{equation}
The term $2\beta\mathcal{B}$ cannot be dropped. Without it the right-hand side differs from $M$ by $2\beta\mathcal{B}\neq0$, since $\mathcal{B}<0$ for $\beta>0$ and $\sigma>0$. This term vanishes for $\beta\to0$, where $\mathcal{F}_{1}\to1$ and Eq.~\eqref{eq:smarr} reduces to the Smarr relation of the Reissner--Nordstr\"om--AdS black hole, $M=2TS-2PV+\Phi q$ with $\Phi=q/r_{h}$.

Both relations were verified numerically. For $\sigma=0.12$, $0.15$, $0.45$ and $0.9$, the closed forms of Eqs.~\eqref{eq:T} and~\eqref{eq:PhiB} agree with central-difference derivatives of Eq.~\eqref{eq:Mrh} to $4\times10^{-23}$ in absolute value at 30-digit working precision (step $10^{-10}$) and to $4\times10^{-36}$ at 50 digits (step $10^{-17}$); the same holds at $\sigma=0.25$ and $\sigma=1.2$. The Smarr relation, Eq.~\eqref{eq:smarr}, is satisfied by the closed forms to $10^{-51}$ at 50 digits (scripts \texttt{verify\_thermo\_precision.py} and \texttt{verify\_firstlaw\_path.py}). The first law, Eq.~\eqref{eq:firstlaw}, was integrated along the straight path from $(r_{h},P,q,\beta)=(1.5,\,0.001,\,0.5,\,0.1)$ to $(2.2,\,0.004,\,0.8,\,0.3)$; with 50-digit quadrature the residual is below $3\times10^{-51}$ for $\sigma=0.12$, $0.25$, $0.45$, $0.9$ and $1.2$.

\begin{figure*}[!tbp]
\centering
\includegraphics[width=0.98\textwidth]{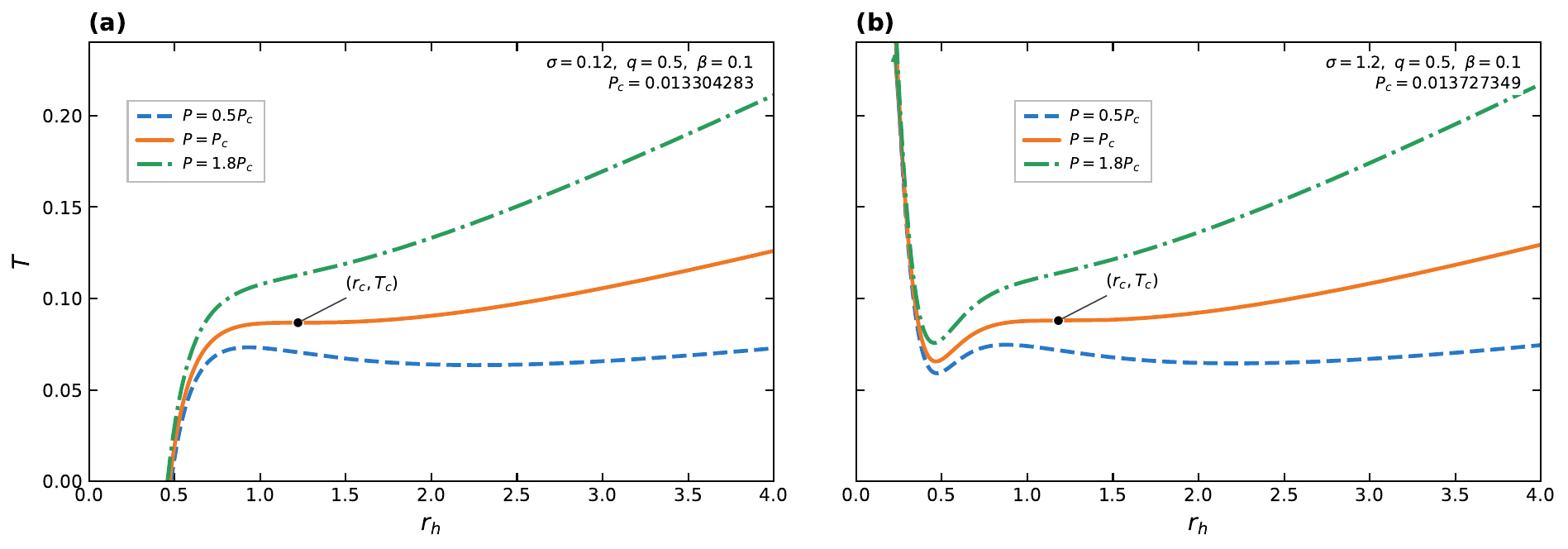}
\caption{Hawking temperature $T(r_h)$ at $q=0.5$ and $\beta=0.1$ for $P=0.5P_c$, $P=P_c$ and $P=1.8P_c$: (a) $\sigma=0.12$, with $P_c=0.013304283174$; (b) $\sigma=1.2$, with $P_c=0.013727349243$. The critical pressures are calculated separately at fixed $q$ by imposing $\partial T/\partial r_h=\partial^2T/\partial r_h^2=0$. The color-coded legends identify the three pressure levels, and black circles mark the critical points $(r_c,T_c)$. Only the physical segments with $T>0$ are shown. For $\sigma=1.2$, the small-$r_h$ branches leave the upper edge because $T\to+\infty$ as $r_h\to0^+$; the arrows indicate continuation toward smaller $r_h$. Each point on an isobar belongs to a solution whose mass is fixed by the horizon condition, rather than to one geometry with fixed mass. On the plotted branches $T>0$ and $M_h$ is monotonic in $r_h$.}\label{fig:TisoCompare}
\end{figure*}

The comparison of the temperature isobars across the nonlinear exponent is shown in Fig.~\ref{fig:TisoCompare}. In both panels we fix $q=0.5$ and $\beta=0.1$ and use the pressure fractions $P/P_c=0.5,1,1.8$, with the critical pressure evaluated separately for each value of $\sigma$. The horizon radius $r_h$ is varied along each isobar, and the mass $M_h(r_h)$ is reconstructed from the horizon condition, Eq.~\eqref{eq:Mrh}; the curves therefore describe a family of black-hole solutions, not the radial profile of a single geometry with fixed mass. Only the branches with $T>0$ are displayed. For $\sigma=0.12$ the temperature changes sign once, at the radii quoted below, and is positive beyond it; for $\sigma=1.2$ it is positive for all $r_h$ at the three pressures considered. Hence $M_h(r_h)$ is monotonic on each displayed branch and the mass $M_h(r_h)$ is not attained at any larger radius, so that each displayed horizon is the outermost root of $f$. The relation $dM_h/dr_h=2\pi r_hT>0$ alone would not establish this, because $M_h$ may repeat a value at a larger radius when $T$ changes sign. For $\sigma=0.12$, the physical branches begin at their respective $T=0$ crossings, at $r_h\simeq0.47939$, $0.47068$ and $0.45858$ for $P/P_c=0.5$, $1$ and $1.8$. For $\sigma=1.2$, the temperature is positive at small radii and behaves as $T\sim1/(4\pi r_h)$ as $r_h\to0^+$; hence the small-radius branches rise beyond the upper temperature limit of the plot. The horizontal axis is shown from zero for comparison, but $r_h=0$ is not included as a physical point.

\subsection{Critical points}
Writing $T=A(r_{h})+2Pr_{h}$, the conditions $\partial_{r_h}T=\partial_{r_h}^{2}T=0$ give $A''(r_{c})=0$ and $P_{c}=-A'(r_{c})/2$. Table~\ref{tab:crit} lists physical solutions with $P_{c}>0$. The final two entries, $(q,\sigma)=(3.249107,1.2)$ and $(3.920545,1.5)$, additionally satisfy the regularity condition $m_{0}=M-m_{M}=0$: for each exponent, $q$ and $r_c$ were solved together with the criticality equations. Thus these are regular critical black holes, rather than generic critical points at a fixed charge. For both solutions, $T_c>0$ and the larger positive root of $f(r)=0$ is $r_+=r_c$; the corresponding inner horizons are $r_-=0.752053$ and $0.982012$, respectively. As a test of the procedure, the parameters of Ref.~\cite{ref089}, $q=1$, $\beta=0.1$ and $\sigma=0.75$, give $r_{c}=2.4374$ and $P_{c}=0.003332$, in agreement with the value $P_{c}\approx0.003$ reported there.
\begin{table*}[!t]
\caption{Critical point $(r_{c},P_{c},T_{c})$ for $\beta=0.1$ and its ratio to the Reissner--Nordstr\"om--AdS value at the same $q$ ($r_c^{\rm RN}=\sqrt6\,q$, $P_c^{\rm RN}=1/(96\pi q^2)$, $T_c^{\rm RN}=\sqrt6/(18\pi q)$, with $P_cr_c/T_c=3/16=0.1875$). The rows with $q=0.5$ and $\sigma=1.2,1.5$ are generic fixed-charge critical solutions and are not assumed regular. The final two rows are regular critical solutions with $m_0=0$, obtained by solving for $q$ together with the criticality conditions.}\label{tab:crit}
\begin{ruledtabular}
\begin{tabular}{ccccccccc}
$q$ & $\sigma$ & $r_{c}$ & $P_{c}$ & $T_{c}$ & $r_{c}/r_{c}^{\rm RN}$ & $P_{c}/P_{c}^{\rm RN}$ & $T_{c}/T_{c}^{\rm RN}$ & $P_{c}r_{c}/T_{c}$\\
\hline
$0.5$ & $0.12$ & $1.220937360$ & $0.013304283174$ & $0.086748664761$ & $0.9969$ & $1.0031$ & $1.0013$ & $0.1872$ \\
$0.5$ & $0.9$ & $1.1935$ & $0.013599$ & $0.087563$ & $0.9745$ & $1.0253$ & $1.0107$ & $0.1854$ \\
$1.2$ & $0.45$ & $2.9334$ & $0.002307$ & $0.036129$ & $0.9980$ & $1.0019$ & $1.0009$ & $0.1873$ \\
$1$ & $0.75$ & $2.4374$ & $0.003332$ & $0.043408$ & $0.9951$ & $1.0049$ & $1.0021$ & $0.1871$ \\
$0.5$ & $1.2$ & $1.181395415$ & $0.013727349243$ & $0.087911491712$ & $0.9646$ & $1.0350$ & $1.0148$ & $0.1845$ \\
$0.5$ & $1.5$ & $1.168084$ & $0.01386592$ & $0.08828421$ & $0.9537363$ & $1.0454654$ & $1.0190601$ & $0.1834592$ \\
$3.249107$ & $1.2$ & $7.952774$ & $0.000314320$ & $0.01333603$ & $0.9992613$ & $1.0007386$ & $1.0003164$ & $0.1874406$ \\
$3.920545$ & $1.5$ & $9.597246$ & $0.000215855$ & $0.01105159$ & $0.9993660$ & $1.0006339$ & $1.0002716$ & $0.1874490$ \\
\end{tabular}
\end{ruledtabular}
\end{table*}
For $\beta\to0$ the model reduces to the Reissner--Nordstr\"om--AdS black hole with magnetic charge $q$, for which $r_{c}=\sqrt{6}\,q$, $P_{c}=1/(96\pi q^{2})$, $T_{c}=\sqrt{6}/(18\pi q)$ and $P_{c}r_{c}/T_{c}=3/16$ in the present units. Table~\ref{tab:crit} also lists the ratios of the critical quantities to these values at the same charge. The fixed-charge examples show that the nonlinear electrodynamics shifts $r_c$, $P_c$ and $T_c$ relative to RN. The last two rows are regular critical solutions for which the charge was selected by $m_0=0$; they are not part of a fixed-$q$ scan in $\sigma$. The condition $m_0=0$ is imposed on the critical state itself, in the full state space with the usual independent variables. Since $m_0=M-m_M$ and $M=M_h(r_h,P)$ varies along any path through this state at fixed $q$, $\beta$ and $\sigma$, neighbouring states, including the two coexisting phases at pressures slightly below $P_c$, have $m_0\neq0$ and a singular center. The regularity statement therefore applies to the critical point and not to the neighbouring states. Their ratios remain close to the RN values, as expected from the small values of $\beta q^2/r_c^4$ at the selected critical points.
For $P>P_{c}$ the temperature of the cases $\sigma=0.12$ and $\sigma=0.45$ has no extrema in the region $T>0$ and there is no phase transition. At $P=P_{c}$ the temperature has an inflection point, which is the critical point of the second-order transition. For $P<P_{c}$ the temperature has a local maximum and a local minimum, so that a first-order transition is possible. Whether it occurs is decided by the free energy (Sec.~\ref{sec:G}). For $\sigma=0.9$ and $1.2$ the temperature also develops an additional small-radius minimum, because $T\to\infty$ as $r_h\to0$. For $\sigma=1.2$ and $q=0.5$, at $P=10^{-3}$ the extrema occur at $r_h=0.4790$ (minimum), $0.8044$ (maximum) and $6.2469$ (minimum); the extra small-black-hole branch changes the topological charge pattern as detailed below. This small-radius minimum does not, by itself, imply phase coexistence.

\subsection{Topological charges at \texorpdfstring{$P=10^{-3}$}{P = 10 to the minus 3}}
The benchmark cases below have $P=10^{-3}<P_{c}$. The zero points and the extrema of $T(r_{h})$ are given in Table~\ref{tab:zp}. The extrema are the radii at which a pair of zero points of opposite charge is created or annihilated as $\tau$ is varied. They bound the unstable intermediate branch.
\begin{table*}[!t]
\caption{Zero points of $\phi$ for $\beta=0.1$ and $P=10^{-3}$, with the winding number in parentheses. The extrema of $T(r_{h})$ are given as $r_{h}$ ($T$). Zero points exist at the stated $\tau$; the intervals of $\tau$ with the maximal number of zero points are given in the text.}\label{tab:zp}
\begin{ruledtabular}
\begin{tabular}{ccclcl}
$q$ & $\sigma$ & $\tau$ & Zero points $r_{h}$ & TTC & Extrema of $T$\\
\hline
$0.5$ & $0.12$ & $20$ & $0.6337\,(+1)$, $1.5075\,(-1)$, $23.2926\,(+1)$ & $+1$ & $0.8692$ $(0.06315)$, $6.2469$ $(0.02515)$\\
$0.5$ & $0.9$ & $35$ & $3.6839\,(-1)$, $10.5076\,(+1)$ & $0$ & $0.3948$ $(0.03134)$, $0.8269$ $(0.06432)$, $6.2469$ $(0.02515)$\\
$0.5$ & $0.9$ & $25$ & $0.3363\,(-1)$, $0.4825\,(+1)$, $2.0960\,(-1)$, $17.7616\,(+1)$ & $0$ & same as for $\tau=35$\\
$1.2$ & $0.45$ & $35$ & $1.8321\,(+1)$, $2.8904\,(-1)$, $10.5699\,(+1)$ & $+1$ & $2.2125$ $(0.02984)$, $5.9041$ $(0.02473)$\\
$0.5$ & $1.2$ & $17$ & $0.4155\,(-1)$, $0.5919\,(+1)$, $1.1427\,(-1)$, $27.9907\,(+1)$ & $0$ & $0.4790$ $(0.05353)$, $0.8044$ $(0.06488)$, $6.2469$ $(0.02515)$ \\
$0.5$ & $1.2$ & $25$ & $2.0961\,(-1)$, $17.7616\,(+1)$ & $0$ & same as for $\tau=17$ \\
\end{tabular}
\end{ruledtabular}
\end{table*}

\begin{figure}[!tbp]
\pan{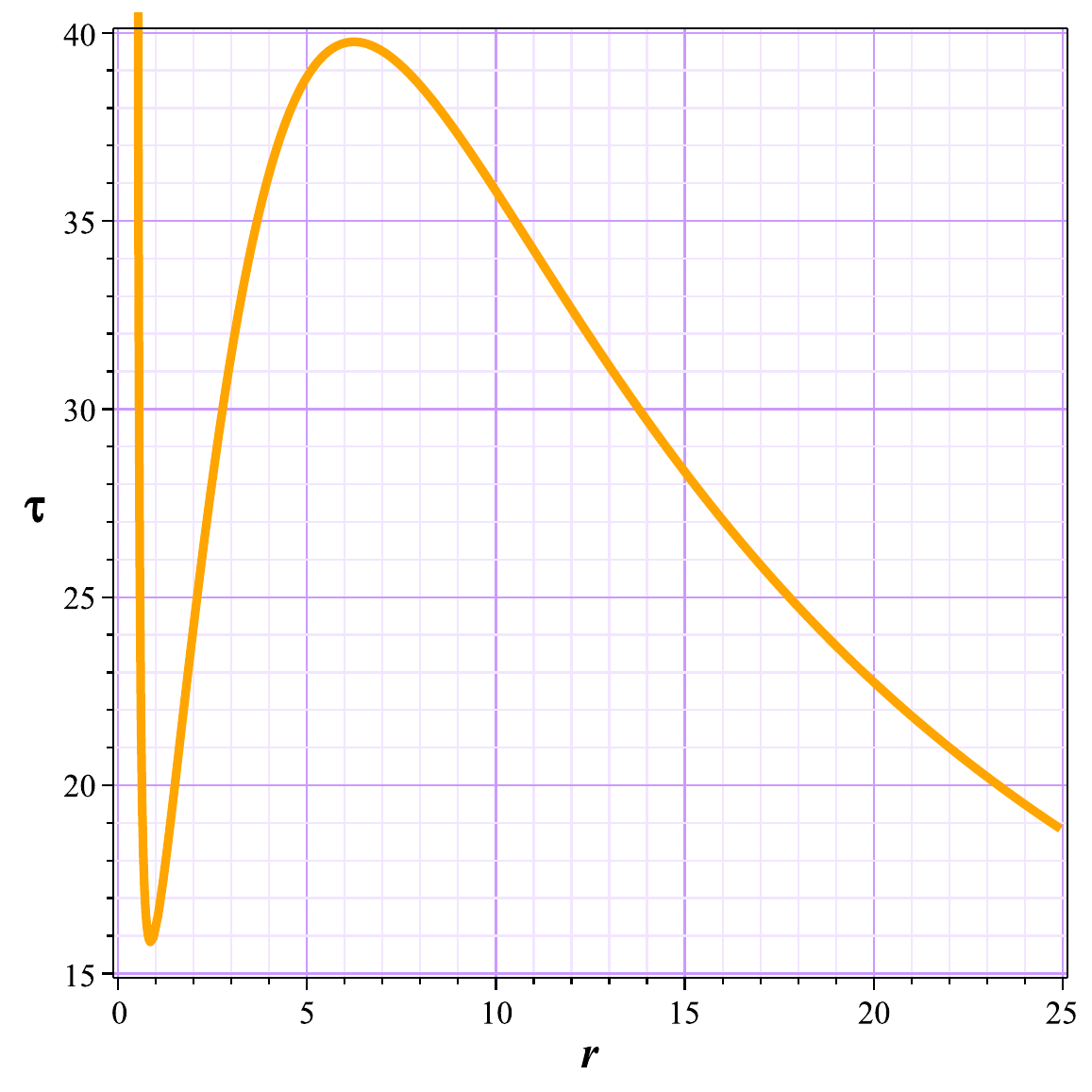}{(a)}\hfill\pan{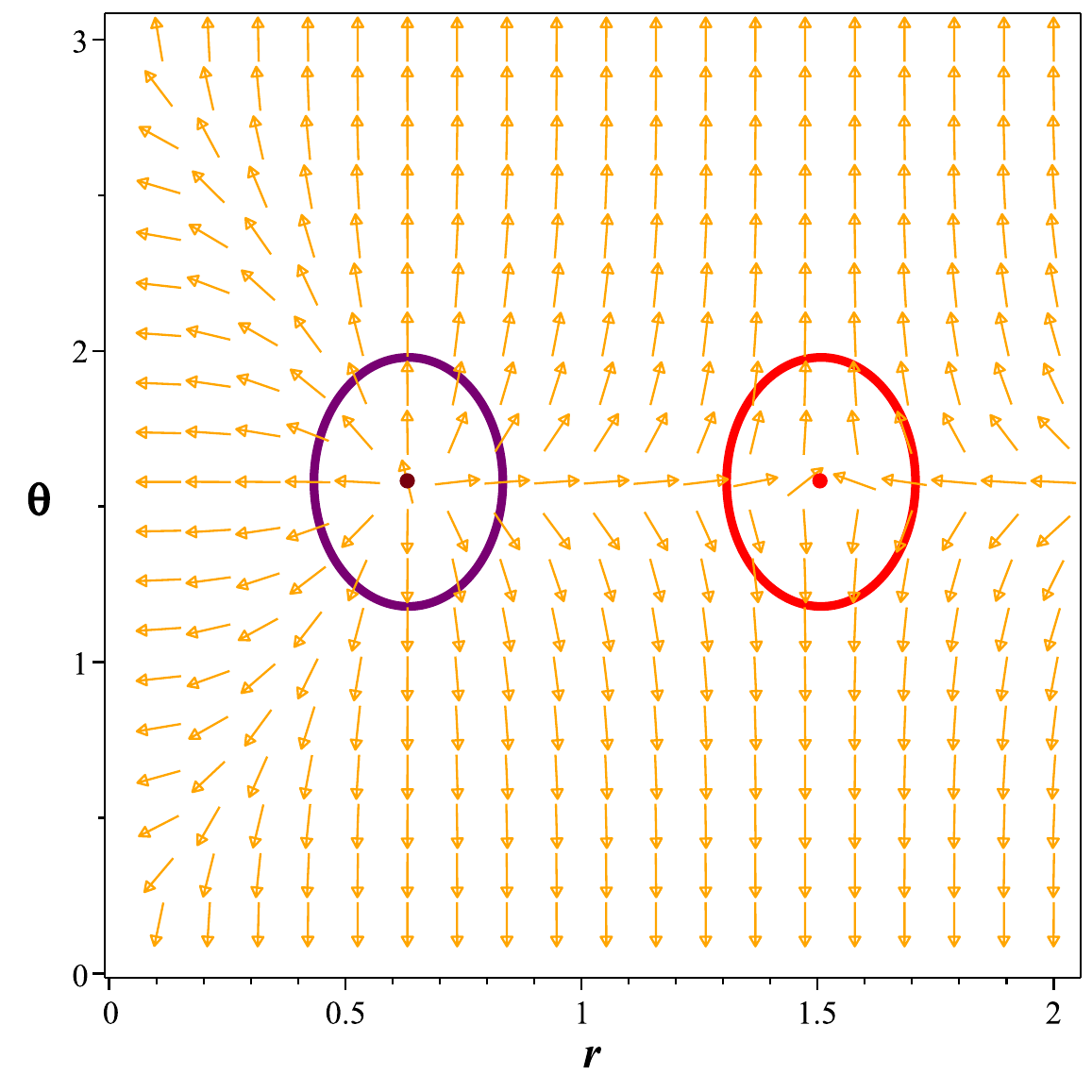}{(b)}\\[4pt]
\pan{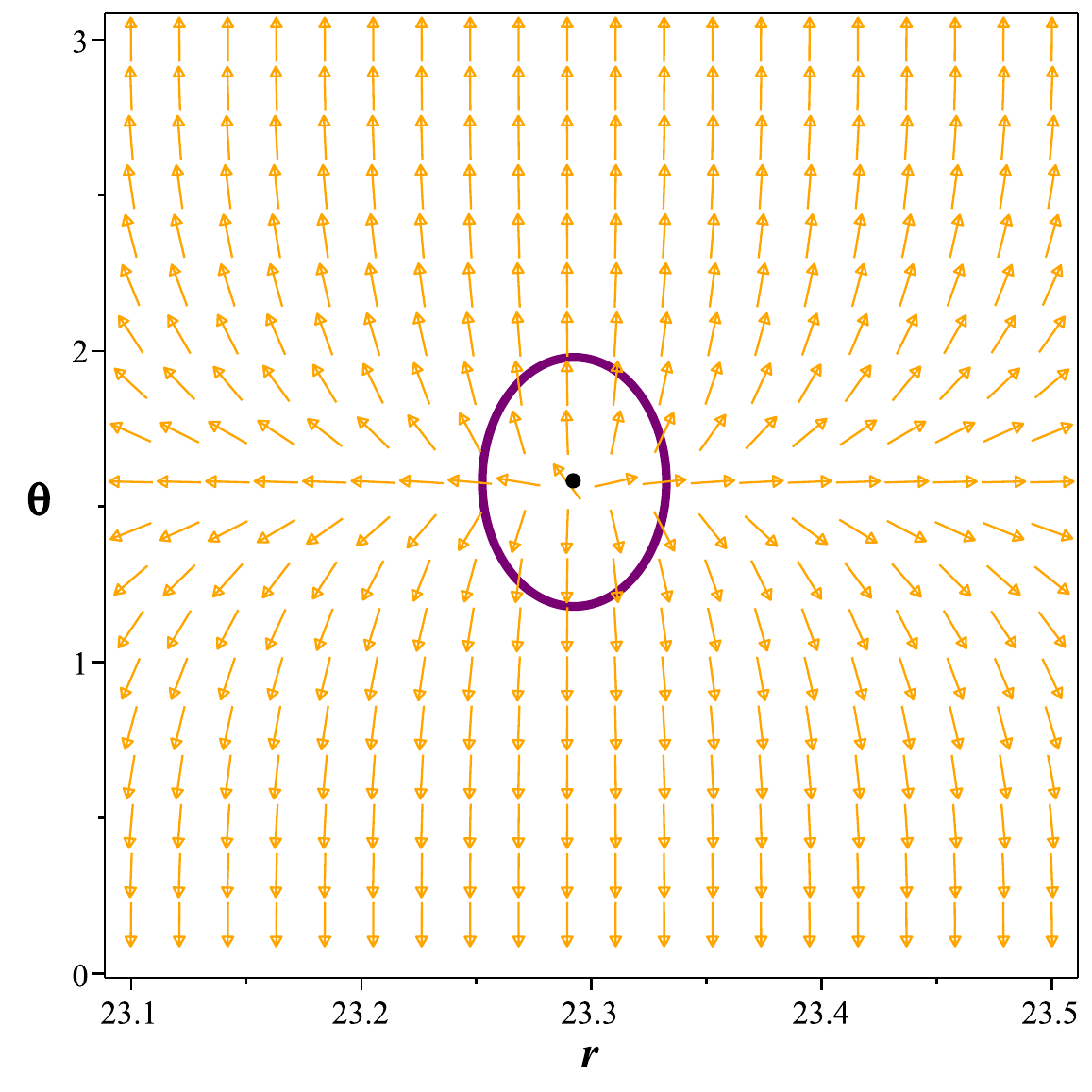}{(c)}
\caption{$\sigma=0.12$, $q=0.5$, $\beta=0.1$, $P=10^{-3}$. (a) $\tau_{h}(r_{h})$ and the line $\tau=20$. (b), (c) Normalized field $n$ in the $(r_{h},\Theta)$ plane for $\tau=20$, with zero points at $r_{h}=0.6337$ ($+1$, purple), $1.5075$ ($-1$, red) and $23.2926$ ($+1$).}\label{fig:th1}
\end{figure}

\paragraph*{$q=0.5$, $\sigma=0.12$.} Figure~\ref{fig:th1} shows $\tau_{h}(r_{h})$ and the field for $\tau=20$. There are three zero points, a small stable black hole at $r_{h}=0.6337$ ($+1$), an unstable intermediate black hole at $1.5075$ ($-1$) and a large stable black hole at $23.2926$ ($+1$), so $\mathrm{TTC}=+1$. The unstable branch extends between the local maximum of $T$ at $r_{h}=0.8692$ and its local minimum at $r_{h}=6.2469$, and three zero points exist for $15.84<\tau<39.76$. For $\tau>39.76$ only the small black hole remains, so that the total charge is $+1$ for every $\tau$.

\begin{figure*}[!tbp]
\includegraphics[width=0.90\textwidth]{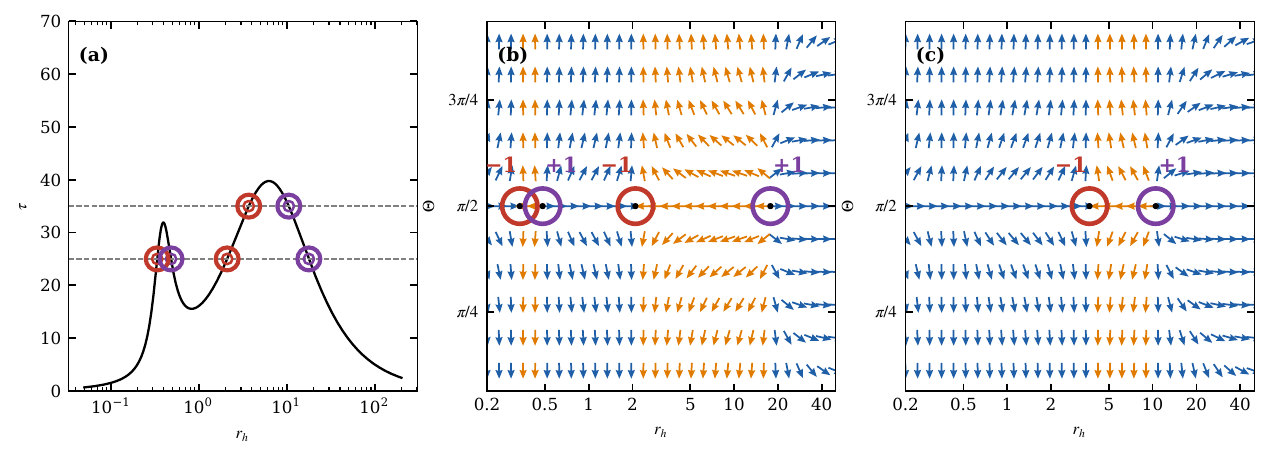}
\caption{Thermodynamic topology for $\sigma=0.9$, $q=0.5$, $\beta=0.1$, and $P=10^{-3}$. (a) $\tau_h(r_h)$ for $\tau=25,35$, with intersections marked by circles. (b,c) Normalized field $n(r_h,\Theta)$ for $\tau=25,35$, respectively. Red and purple circles denote zero points with winding numbers $\omega=-1$ and $+1$; blue and orange arrows indicate $\phi^{r_h}>0$ and $\phi^{r_h}<0$. The $r_h$ axes are logarithmic.}\label{fig:th2b}
\end{figure*}

\paragraph*{$q=0.5$, $\sigma=0.9$.} Here $T(r_{h})\to\infty$ for $r_{h}\to0$, in contrast with $\sigma=0.12$ where $T\to-\infty$, and the temperature has four branches separated by the extrema at $r_{h}=0.3948$ (minimum), $0.8269$ (maximum) and $6.2469$ (minimum). The branches alternate in sign, $-1$, $+1$, $-1$, $+1$. For $\tau=35$ (Fig.~\ref{fig:th2b}(c)) only the last two are intersected, giving the zero points $3.6839$ ($-1$) and $10.5076$ ($+1$) and $\mathrm{TTC}=0$. For $15.55<\tau<31.91$, for example $\tau=25$, four zero points appear (Fig.~\ref{fig:th2b}) and the sum is again zero. The total charge $0$ is therefore independent of $\tau$ and differs from the class $+1$ of the case $\sigma=0.12$ because of the small black hole branch with negative heat capacity. Although $P=10^{-3}<P_{c}=0.013599$, the stable branches $r_{h}\in(0.3948,0.8269)$ and $r_{h}>6.2469$ are not related by a first-order transition at this pressure. Where both exist, for $0.03134<T<0.06432$, the large black hole has the lower Gibbs free energy (Sec.~\ref{sec:G}), so that the small black hole is metastable. A first-order small--large transition exists for $\sigma=0.9$ only for $P_{*}<P<P_{c}$ with $P_{*}=0.001307$. Here $\mathrm{TTC}=0$ expresses the balance of positive and negative local indices; the two- and four-zero patterns illustrate that this sum does not fix the branch count. Equilibrium coexistence is determined by the free energy. For $P>P_{c}$ only the minimum at small $r_{h}$ remains, and the zero points are two, with charges $-1$ and $+1$.

\begin{figure*}[!tbp]
\centering
\includegraphics[width=0.99\textwidth]{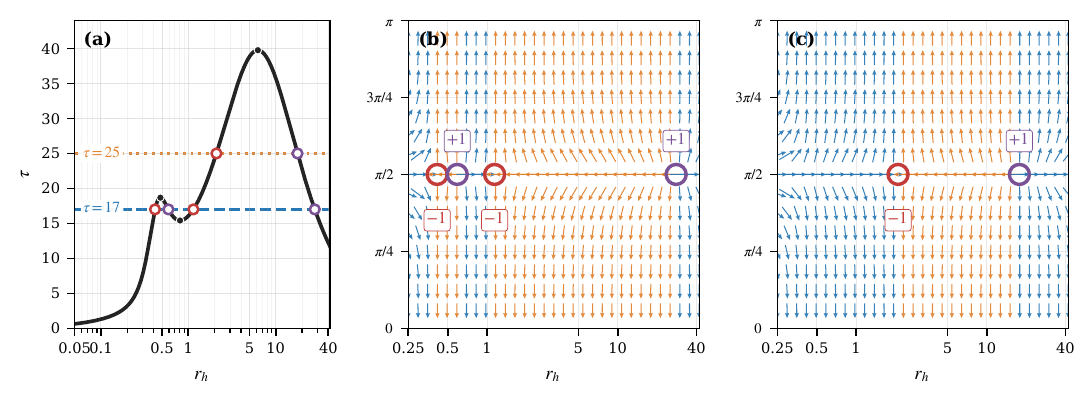}
\caption{Thermodynamic topology for $q=0.5$, $\sigma=1.2$, $\beta=0.1$, and $P=10^{-3}$. (a) $\tau_h(r_h)$ for $\tau=17,25$, with intersections marked by circles. (b,c) Normalized field $n(r_h,\Theta)$ for $\tau=17,25$, respectively. Red and purple circles denote zero points with winding numbers $\omega=-1$ and $+1$; blue and orange arrows indicate $\phi^{r_h}>0$ and $\phi^{r_h}<0$. The $r_h$ axes are logarithmic.}
\label{fig:tau12topo}
\end{figure*}

\paragraph*{$q=0.5$, $\sigma=1.2$.} At $P=10^{-3}$ the inverse-temperature curve $\tau_h(r_h)=1/T(r_h)$ has extrema at $r_h=0.479013$, $0.804373$ and $6.246925$, where $\tau_h$ is approximately $18.6795$, $15.4121$ and $39.7600$, respectively. For $\tau=17$, the equation $T(r_h)=1/\tau$ gives four zero points at $r_h=0.4155011$, $0.5918585$, $1.1426777$ and $27.9907209$, with winding numbers $-1,+1,-1,+1$. At $\tau=25$, only two zero points remain, at $r_h=2.0960916$ and $17.7616232$, with winding numbers $-1$ and $+1$. Both cases therefore have total topological charge $Q=0$. The three-panel plot in Fig.~\ref{fig:tau12topo} compares both inverse-temperature levels and the corresponding normalized vector fields. 

\begin{figure}[!tbp]
\pan{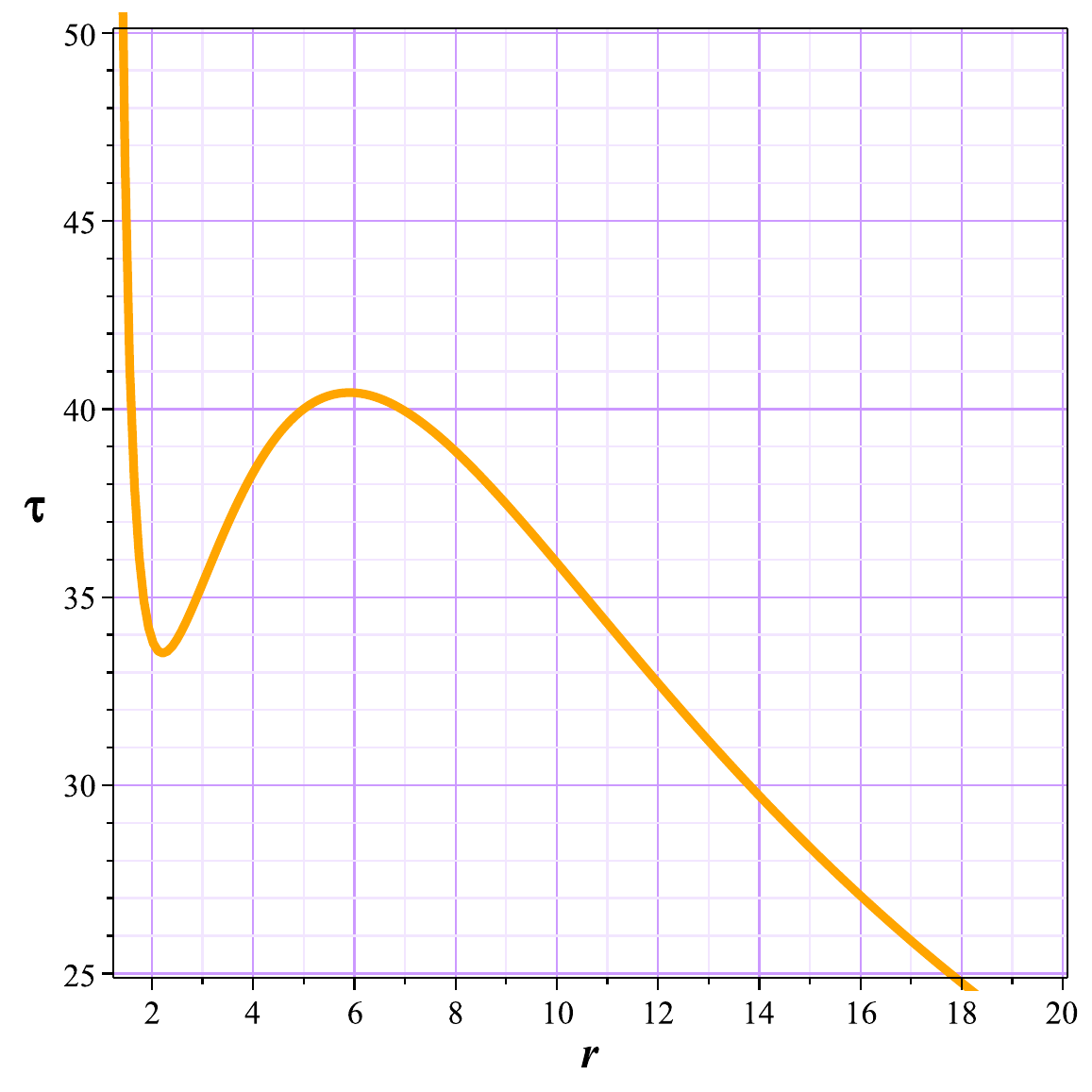}{(a)}\hfill\pan{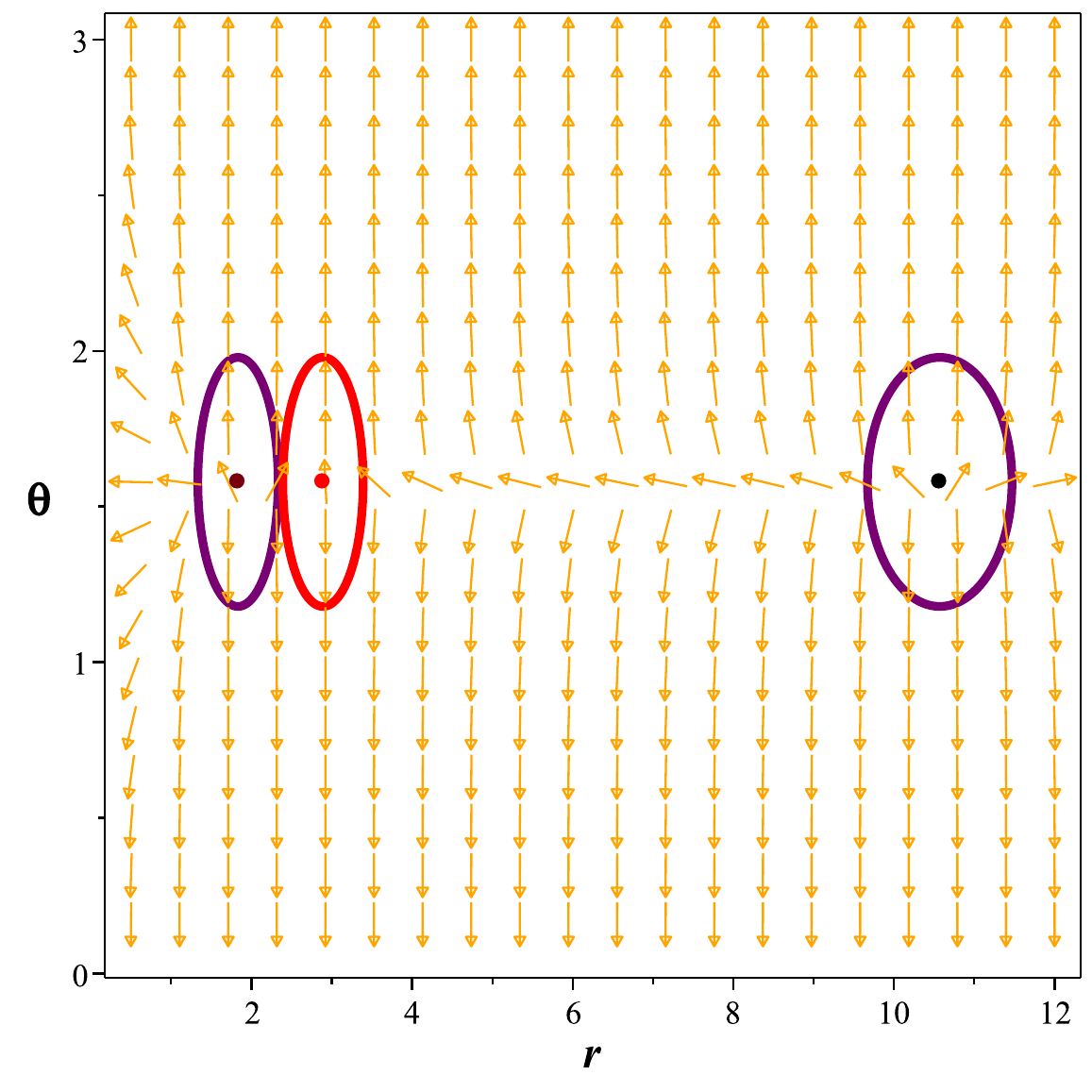}{(b)}
\caption{$\sigma=0.45$, $q=1.2$, $\beta=0.1$, $P=10^{-3}$. (a) $\tau_{h}(r_{h})$ and the line $\tau=35$. (b) Normalized field $n$ for $\tau=35$, with zero points at $r_{h}=1.8321$ ($+1$), $2.8904$ ($-1$) and $10.5699$ ($+1$).}\label{fig:th3}
\end{figure}

\paragraph*{$q=1.2$, $\sigma=0.45$.} For $\tau=35$ (Fig.~\ref{fig:th3}) there are three zero points, at $r_{h}=1.8321$ ($+1$), $2.8904$ ($-1$) and $10.5699$ ($+1$), so $\mathrm{TTC}=+1$. The unstable branch lies between the extrema at $r_{h}=2.2125$ and $5.9041$, and three zero points exist for $33.51<\tau<40.44$.

\subsection{Free energy}\label{sec:G}
\begin{figure}[!tbp]
\includegraphics[width=\columnwidth]{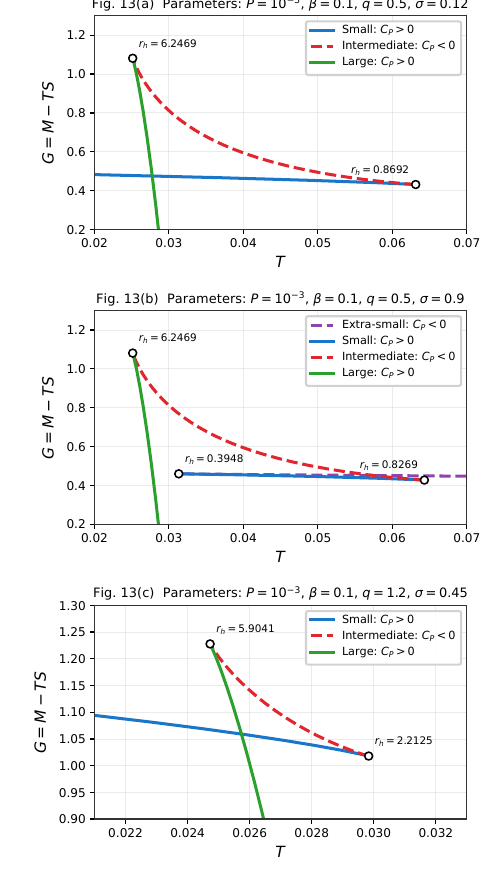}
\caption{Gibbs free energy $G=M-TS$ at $P=10^{-3}$, $\beta=0.1$ as a function of $T$: (a) $\sigma=0.12$, $q=0.5$; (b) $\sigma=0.9$, $q=0.5$; (c) $\sigma=0.45$, $q=1.2$. Circles mark the extrema of $T(r_{h})$.}\label{fig:G}
\end{figure}

The on-shell free energy $G(T)=M-TS$ is shown parametrically in Fig.~\ref{fig:G}. For $\sigma=0.12$ and $\sigma=0.45$ it has a single swallowtail whose cusps are at the extrema of $T$ in Table~\ref{tab:zp}. The crossing of the two stable branches fixes the first-order transition temperature $T_{t}$, at which $T(r_{s})=T(r_{l})$ and $G(r_{s})=G(r_{l})$ for the small and the large black hole (Table~\ref{tab:coex}). For $\sigma=0.9$ and $1.2$ the curve has an additional small-radius cusp associated with a negative-heat-capacity branch. At $P=10^{-3}$ the stable branches do not cross for either of these two values of $\sigma$. For $\sigma=0.9$, the crossing appears for $P>P_{*}=0.001307$; for $\sigma=1.2$, a first-order crossing is found at $P=0.005$, with the coexistence values listed in Table~\ref{tab:coex}. The free energy therefore complements the topological assignment: the local indices label the thermal stability of the computed zeros, whereas the free energy selects the equilibrium branch.
\begin{table*}[!t]
\caption{First-order coexistence of the small ($r_{s}$) and large ($r_{l}$) black holes, defined by $T(r_{s})=T(r_{l})$ and $G(r_{s})=G(r_{l})=G_{t}$, for $\beta=0.1$. The quantity $\Delta S=\pi(r_{l}^{2}-r_{s}^{2})$ is the entropy jump. For $\sigma=0.9$ and $1.2$ at $P=10^{-3}$ the stable branches do not cross.}\label{tab:coex}
\begin{ruledtabular}
\begin{tabular}{cccccccc}
$q$ & $\sigma$ & $P$ & $r_{s}$ & $r_{l}$ & $T_{t}$ & $G_{t}$ & $\Delta S$\\
\hline
$0.5$ & $0.12$ & $0.001$ & $0.54484$ & $9.8632$ & $0.027774$ & $0.47518$ & $304.69$\\
$1.2$ & $0.45$ & $0.001$ & $1.6240$ & $8.0204$ & $0.025741$ & $1.0592$ & $193.80$\\
$0.5$ & $0.9$ & $0.001$ & \multicolumn{5}{c}{no coexistence}\\
$0.5$ & $0.9$ & $0.005$ & $0.58360$ & $3.7248$ & $0.058227$ & $0.44037$ & $42.516$\\
$0.5$ & $1.2$ & $0.001$ & \multicolumn{5}{c}{no coexistence} \\
$0.5$ & $1.2$ & $0.005$ & $0.50987$ & $3.7396$ & $0.058295$ & $0.43738$ & $43.118$ \\
\end{tabular}
\end{ruledtabular}
\end{table*}

\begin{figure*}[!tbp]
\centering
\includegraphics[width=0.98\textwidth]{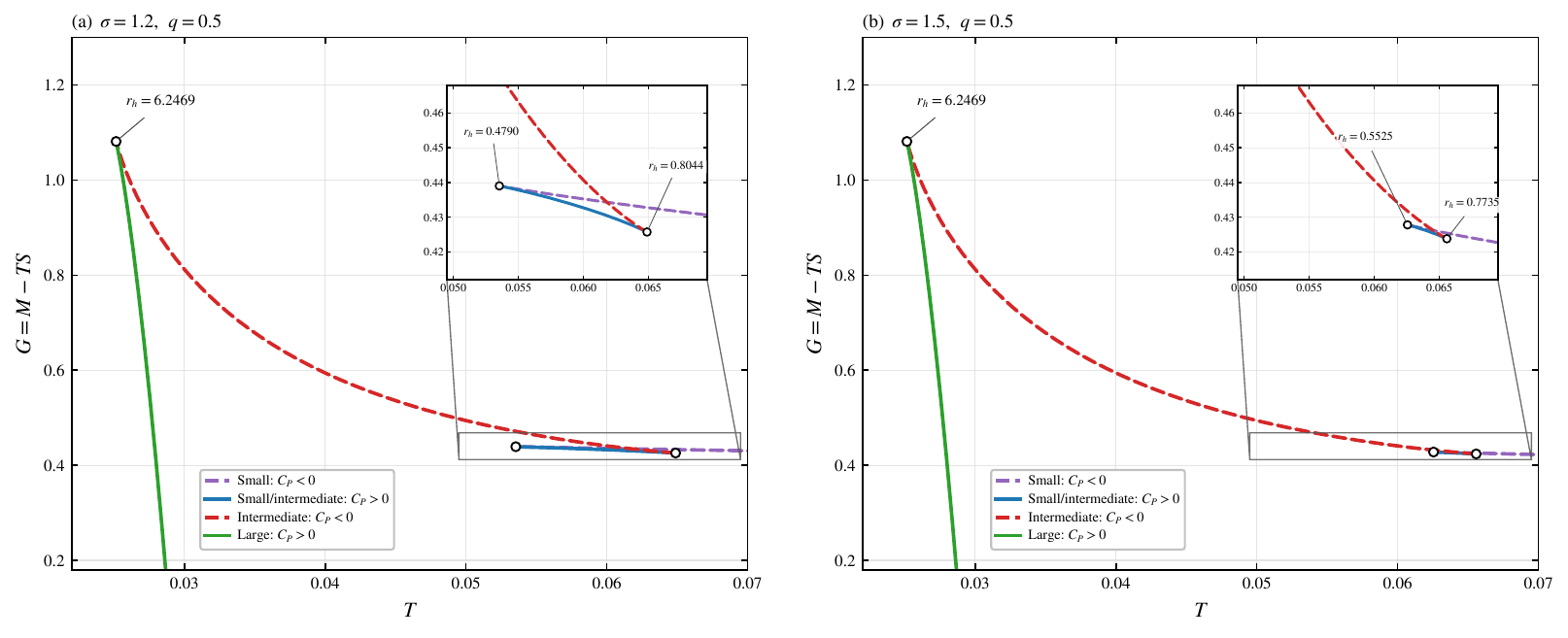}
\caption{Bulk Gibbs free energy $G=M-TS$ versus $T$ for $P=10^{-3}$, $q=0.5$, and $\beta=0.1$: (a) $\sigma=1.2$ and (b) $\sigma=1.5$. Solid and dashed curves denote $C_P>0$ and $C_P<0$, respectively; open circles mark extrema of $T(r_h)$. Insets enlarge the small-black-hole branches. Only $T>0$ solutions are shown.}
\label{fig:Ghighsigma}
\end{figure*}

Figure~\ref{fig:Ghighsigma} extends the Gibbs-free-energy analysis to the high-$\sigma$ regime at fixed $q=0.5$. For both $\sigma=1.2$ and $1.5$, $T(r_h)$ has three extrema, yielding four branches in the positive-temperature domain. The heat capacity, $C_P=2\pi r_h T/(dT/dr_h)$, distinguishes the locally stable segments ($C_P>0$, solid curves) from the unstable ones ($C_P<0$, dashed curves). The insets resolve the small-black-hole branches, whose folds are compressed in the full view. Over the temperature range shared by the stable intermediate- and large-radius branches, the large-radius branch has lower $G$; the two stable branches therefore do not cross, and no first-order small--large coexistence occurs at $P=10^{-3}$.

\section{Holographic Thermodynamics}\label{sec:holo}
Assuming a holographic interpretation, the black hole is described by a thermodynamic state of a three-dimensional conformal field theory (CFT) on a sphere of radius $R$. The boundary theory is characterized here only through its thermodynamic variables, defined by an extension of the bulk--boundary dictionary to the present solution; it is not identified microscopically. In the holographic extension of the thermodynamics of Refs.~\cite{refH1,refH2,refH3}, a variation of the cosmological constant is a variation of the central charge $C$ of the CFT, while the radius $R$ is an independent parameter that fixes the CFT volume $\mathcal{V}=4\pi R^{2}$. This section constructs the dictionary for the present black hole, which carries a magnetic charge and the coupling $\beta$, verifies the first law and the Euler relation, and then repeats for this black hole the analysis of the charged AdS black hole in Ref.~\cite{refH2}: the phase structure in the canonical, grand-canonical and mixed ensembles, the critical point and the critical exponents.

\subsection{Dictionary}
In the units $G=1$ the CFT variables are
\begin{align}\label{eq:dict}
E&=\frac{l\,M}{R}, & \tilde{T}&=\frac{l\,T}{R}, & S&=\pi r_{h}^{2}, & \mathcal{V}&=4\pi R^{2},\nonumber\\
C&=\frac{l^{2}}{4}, & \tilde{Q}&=q\,l, & \tilde{\Phi}&=\frac{\Phi}{R}, & \tilde{\beta}&=\beta,
\end{align}
where $M$, $T$ and $\Phi$ are given by Eqs.~\eqref{eq:Mrh}, \eqref{eq:T} and \eqref{eq:PhiB}, and $l^{2}=3/(8\pi P)$. The first seven entries are the dictionary of Refs.~\cite{refH1,refH2,refH3} for a conformal factor $\omega=R/l$, with the electric charge replaced by the magnetic charge $q$, and with the normalization of $C$ of Ref.~\cite{refH2}. The coupling is carried over unchanged as a formal thermodynamic variable, $\tilde{\beta}=\beta$, with dimension $[L^{2}]$, and is held fixed in the ensembles below. We use $\tilde{\beta}=0.1$ in units $R=1$. Its conjugate quantity is $\tilde{\Psi}=l\,\mathcal{B}/R$. $\tilde{Q}=ql$ is the magnetic flux of the bulk field through the sphere, held fixed in the canonical ensemble, and $\tilde{\Phi}=\Phi/R$ is defined as its thermodynamic conjugate through the first law, Eq.~\eqref{eq:PhiB}. For Maxwell theory this replacement is the electric--magnetic dual of the electric dictionary; for the present nonlinear Lagrangian it is a definition by the first law, and the ensembles at fixed $\tilde{\Phi}$ are Legendre transforms of $E$. No field-theory interpretation of $\tilde{\beta}$ is given. The pressure and the chemical potential of the CFT are
\begin{equation}\label{eq:pmu}
p=\frac{E}{2\mathcal{V}},\qquad
\mu=\frac{E-\tilde{T}S-\tilde{\Phi}\tilde{Q}-\tilde{\Psi}\tilde{\beta}}{C}.
\end{equation}
The expression for $p$ is the equation of state of a conformal fluid in two spatial dimensions. The term $\tilde{\Psi}\tilde{\beta}$ in $\mu$ is the contribution of the coupling to the Euler relation, Eq.~\eqref{eq:euler}. In terms of the dimensionless variables $x=r_{h}/l$ and $y=q/l=\tilde{Q}/(4C)$, and of the ratio $\hat{\beta}=\beta/l^{2}=\tilde{\beta}/(4C)$, with $z=-\hat{\beta}y^{2}/x^{4}$, one finds in units $R=1$
\begin{align}\label{eq:xy}
\frac{E R}{C}&=2x(1+x^{2})+\frac{2y^{2}F_{1}}{x}, & S&=4\pi C x^{2},\nonumber\\
\tilde{T}R&=\frac{1}{4\pi x}\Big[1+3x^{2}-\frac{y^{2}F_{1}}{x^{2}}\nonumber\\
&\qquad\qquad+\frac{4\sigma\hat{\beta}y^{4}F_{2}}{5x^{6}}\Big],\nonumber\\
\tilde{\Phi}R&=\frac{y}{x}F_{1}-\frac{\sigma\hat{\beta}y^{3}F_{2}}{5x^{5}},\nonumber\\
\mu R&=x-x^{3}-\frac{y^{2}F_{1}}{x}+\frac{2\sigma\hat{\beta}y^{4}F_{2}}{5x^{5}},
\end{align}
and $p=E/(8\pi R^{2})$,
with $F_{1}={}_{2}F_{1}(\tfrac14,\sigma;\tfrac54;z)$ and $F_{2}={}_{2}F_{1}(\tfrac54,\sigma+1;\tfrac94;z)$. For $\hat{\beta}=0$ these are the expressions of Ref.~\cite{refH2}. The last term of $\mu R$ is $-\tilde{\Psi}\tilde{\beta}/(CR^{-1})$. The dimensionless quantities $\tilde{T}R$, $\mu R$, $\tilde{\Phi}R$, $E/(CR^{-1})$ and $S/C$ depend only on $(x,y)$ and $(\sigma,\hat{\beta})$. Thus the phase structure depends on $(\sigma,\hat{\beta},y)$ when $C$ is fixed. In our fixed-$C$ canonical and grand-canonical examples, $R=1$, $\tilde{\beta}=0.1$ and $C=5/6$, so $\hat{\beta}=0.03$. If instead $C$ varies at fixed $\tilde{Q}$ and $\tilde{\beta}$, then $\hat{\beta}$ must vary as $\hat{\beta}=\kappa y$, where $\kappa=\tilde{\beta}/\tilde{Q}$; the fixed-$\hat{\beta}$ curves cannot be reused for that variation.

\subsection{First law and Euler relation}\label{sec:euler}
The first law of the CFT is
\begin{equation}\label{eq:cftfirst}
dE=\tilde{T}\,dS-p\,d\mathcal{V}+\mu\,dC+\tilde{\Phi}\,d\tilde{Q}+\tilde{\Psi}\,d\tilde{\beta},\qquad \tilde{\Psi}=\frac{l\,\mathcal{B}}{R}.
\end{equation}
It follows from the bulk first law, Eq.~\eqref{eq:firstlaw}, with $dP=-3\,dl/(4\pi l^{3})$, $dq=d\tilde{Q}/l-\tilde{Q}\,dl/l^{2}$, $d\beta=d\tilde{\beta}$, and from $E\propto1/R$ at fixed $S$, $C$, $\tilde{Q}$ and $\tilde{\beta}$, which gives $p=-\partial E/\partial\mathcal{V}=E/(2\mathcal{V})$. The last term is the response to a variation of the coupling and vanishes for all processes and ensembles considered below, in which $\tilde{\beta}$ is held fixed ($d\tilde{\beta}=0$). It is retained so that the first law is stated for general variations of the parameters, in parallel with the term $\mathcal{B}\,d\beta$ of Eq.~\eqref{eq:firstlaw}. The Euler relation is
\begin{equation}\label{eq:euler}
E=\tilde{T}S+\mu\,C+\tilde{\Phi}\,\tilde{Q}+\tilde{\Psi}\,\tilde{\beta},
\end{equation}
without a term $p\mathcal{V}$, in agreement with Refs.~\cite{refH1,refH2,refH3}. The term $\tilde{\Psi}\tilde{\beta}$ is present because $\tilde{\beta}=\beta$ scales as $\lambda^{2}$ under the transformation $(r_{h},l,q,\beta)\to(\lambda r_{h},\lambda l,\lambda q,\lambda^{2}\beta)$ of Sec.~\ref{sec:smarr}, in the same way as $S$, $C$, $\tilde{Q}$ and $ER=lM$. The function $ER$ is thus homogeneous of degree one in $(S,C,\tilde{Q},\tilde{\beta})$ at fixed $R$, and Eq.~\eqref{eq:euler} is Euler's theorem for it. The relation is the counterpart of the Smarr relation, Eq.~\eqref{eq:smarr}, and the two expressions for $\mu$, from Eq.~\eqref{eq:euler} and as $\partial E/\partial C$ at fixed $(S,\mathcal{V},\tilde{Q},\tilde{\beta})$, coincide only if the Smarr relation holds including the term $2\beta\mathcal{B}$.

One could instead hold fixed the dimensionless ratio $\hat{\beta}=\beta/l^{2}$, which is invariant under the scaling. That alternative convention removes the coupling term from the Euler relation and gives $\mu_{\hat{\beta}}=\mu+\tilde{\Psi}\tilde{\beta}/C=x-x^{3}-y^{2}F_{1}/x$ in units $R=1$, with conjugate quantity $\tilde{\Psi}_{\hat{\beta}}=l^{3}\mathcal{B}/R$. Although both conventions can be formulated consistently, we do not use fixed $\hat{\beta}$ when $C$ varies: throughout this work the physical coupling $\tilde{\beta}=\beta$ is fixed. The conventions coincide for the fixed-$C$ phase diagrams, where $l$ and $\hat{\beta}$ are automatically constant, but differ for variations of $C$ at fixed $\tilde Q$. The dictionary was verified by comparing Eqs.~\eqref{eq:dict}--\eqref{eq:cftfirst} with partial derivatives of $E(S,\mathcal{V},C,\tilde{Q},\tilde{\beta})$. The original benchmark cases agree to better than $10^{-30}$ in 30-digit arithmetic; the added $\sigma=1.2$ benchmark agrees to numerical precision. The last column of Table~\ref{tab:dict} reports the difference between $\mu$ and $\partial E/\partial C$.

\subsection{Ensembles and free energies}
Following Ref.~\cite{refH2}, the thermodynamic variables $(\tilde{Q},\tilde{\Phi})$, $(p,\mathcal{V})$ and $(\mu,C)$ form three conjugate pairs, and the choice of one variable from each pair defines one of eight ensembles. The corresponding potentials, all of which follow from Eq.~\eqref{eq:euler} and $E=2p\mathcal{V}$, are listed in Table~\ref{tab:ens}. The potential of the ensemble $(\tilde{\Phi},\mathcal{V},\mu)$ vanishes identically in Ref.~\cite{refH2}; here it equals $\tilde{\Psi}\tilde{\beta}$ by Eq.~\eqref{eq:euler}, which is zero for $\beta=0$. The coupling $\tilde{\beta}$ is held fixed in all ensembles.
The phase structure below is studied in the ensembles at fixed volume $\mathcal{V}$ with controls $(\tilde{Q},\mathcal{V},C)$, $(\tilde{\Phi},\mathcal{V},C)$ and $(\tilde{Q},\mathcal{V},\mu)$. In the four ensembles in which $p$ is fixed the volume of the CFT varies, and for $\hat{\beta}=0$ Ref.~\cite{refH2} finds no transitions in them. For the present black hole the dependence on $p$ and $\mathcal{V}$ is examined through the pressure--volume isotherms of Sec.~\ref{sec:pV}.

\subsection{Canonical ensemble \texorpdfstring{$(\tilde{Q},\mathcal{V},C)$}{(Q, V, C)}}\label{sec:canon}
\begin{figure*}[!tbp]
\centering\includegraphics[width=0.78\textwidth]{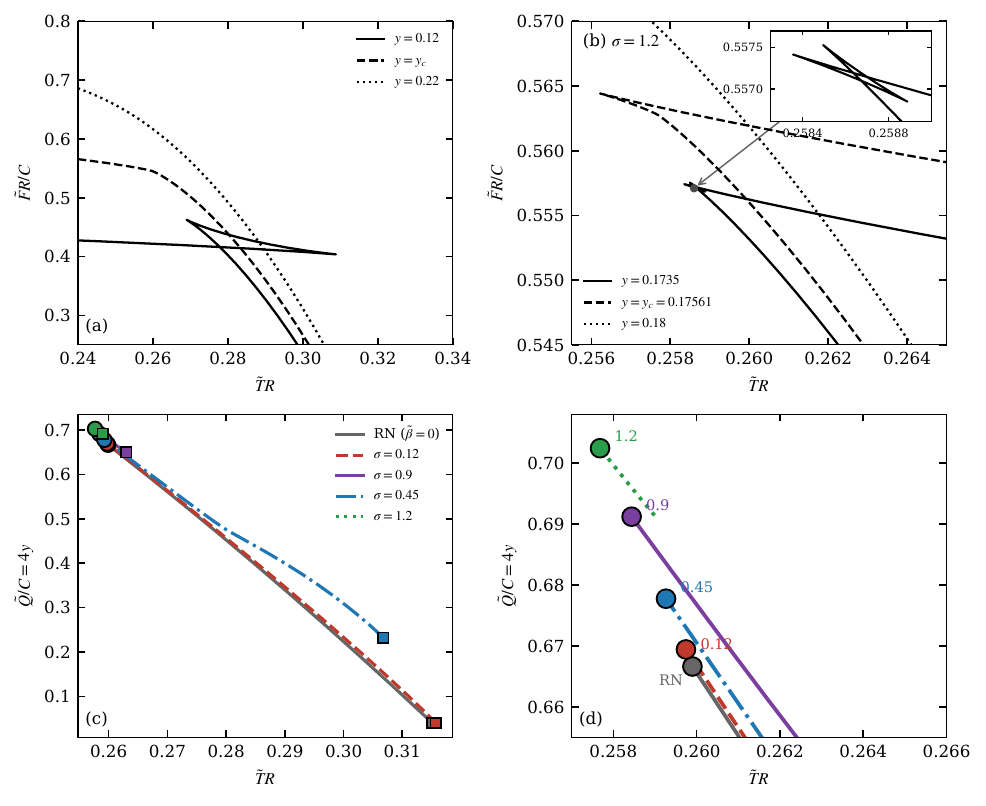}
\caption{Free energy and coexistence lines of the fixed-$C$ canonical ensemble for fixed $\tilde{\beta}=0.1$ and $C=5/6$ (thus $\hat{\beta}=0.03$). (a) $\tilde{F}R/C$ against $\tilde{T}R$ for $\sigma=0.12$ at $y=0.12$, $y_{c}=0.16736$ and $0.22$. (b) The same quantity for $\sigma=1.2$ at $y=0.1735$, $y_{c}=0.17561$ and $0.18$; the swallowtail at $y=0.1735$ is very small and is enlarged in the inset. (c) Coexistence lines in the $(\tilde{T}R,\tilde{Q}/C)$ plane for RN and for $\sigma=0.12$, $0.9$, $0.45$ and $1.2$. The circles mark the critical points, at which each line ends, and the squares mark the opposite end of each line, at $y=0.01$ for RN and $\sigma=0.12$, at $y_{*}=0.1618$ for $\sigma=0.9$, at $y=0.0574$ for $\sigma=0.45$ and at $y_{*}=0.172858$ for $\sigma=1.2$. (d) Enlargement of the region around the critical points.}\label{fig:h1}
\end{figure*}

At fixed $(C,\tilde{Q},\tilde{\beta})$ the bulk parameters $l$, $q$ and $\beta$ are fixed, and the dimensionless temperature $\tilde{T}R=l\,T(r_{h})$ does not depend on $R$. The phase structure is that of Sec.~\ref{sec:thermo} at the pressure $P=3/(8\pi l^{2})=3/(32\pi C)$ and is controlled by $y=\tilde{Q}/(4C)$. The critical point is the solution of $\partial_{x}\tilde{T}=\partial_{x}^{2}\tilde{T}=0$ at fixed $y$, and it is given in Table~\ref{tab:cftcrit} for $\hat{\beta}=0.03$ and for $\hat{\beta}=0$. The latter reproduces the values of Ref.~\cite{refH2}: $x_{c}=1/\sqrt{6}$, $y_{c}=1/6$, $C/\tilde{Q}_{c}=3/2$ and $\tilde{T}_{c}R=\sqrt{6}/(3\pi)$, which confirms the implementation. For $\hat{\beta}>0$ the fixed-$C$ critical values $C/\tilde{Q}_{c}$, $S_{c}/C$, $\mu_{c}R$ and $p_{c}$ decrease, while $\tilde{\Phi}_{c}R$ increases; the shifts grow with $\sigma$. The critical temperature changes by less than $0.6\%$.
\begin{table*}[!t]
\caption{Canonical critical point in units $R=1$. For the nonlinear models, $\tilde{\beta}=0.1$ and $C=5/6$ are held fixed, so $\hat{\beta}=0.03$; RN is the $\tilde{\beta}=0$ limit. The listed ratio is $C/\tilde{Q}_{c}=1/(4y_{c})$ at fixed $C$, not a critical locus obtained by varying $C$ at fixed $\tilde Q$. $S_{c}$ is in units of $C$, and $p_{c}$ is in units of $C/R^{3}$.}\label{tab:cftcrit}
\begin{ruledtabular}
\begin{tabular}{ccccccccccc}
Case & $\sigma$ & $\hat{\beta}$ & $x_{c}$ & $y_{c}$ & $C/\tilde{Q}_{c}$ & $S_{c}/C$ & $\tilde{T}_{c}R$ & $\mu_{c}R$ & $\tilde{\Phi}_{c}R$ & $p_{c}R^{3}/C$\\
\hline
RN & --- & $0$ & $0.40825$ & $0.16667$ & $1.5000$ & $2.0944$ & $0.25990$ & $0.27217$ & $0.40825$ & $0.04332$\\
A & $0.12$ & $0.03$ & $0.40655$ & $0.16736$ & $1.4937$ & $2.0770$ & $0.25974$ & $0.27061$ & $0.41107$ & $0.04318$\\
B & $0.9$ & $0.03$ & $0.39055$ & $0.17280$ & $1.4467$ & $1.9167$ & $0.25844$ & $0.25606$ & $0.43651$ & $0.04186$\\
C & $0.45$ & $0.03$ & $0.40110$ & $0.16944$ & $1.4754$ & $2.0217$ & $0.25927$ & $0.26562$ & $0.41997$ & $0.04273$\\
D & $1.2$ & $0.03$ & $0.37870$ & $0.17561$ & $1.4236$ & $1.8022$ & $0.25768$ & $0.24548$ & $0.45409$ & $0.04087$ \\
\end{tabular}
\end{ruledtabular}
\end{table*}
For $y<y_{c}$ the temperature $\tilde T(x)$ has a local maximum and minimum and the free energy $F=E-\tilde{T}S$ has a swallowtail (Fig.~\ref{fig:h1}a). A first-order transition between a small-entropy and a large-entropy phase at a temperature $\tilde{T}_{t}(y)$ occurs on the model-dependent interval of $y$ in which the two stable branches of the swallowtail cross. Table~\ref{tab:cftcoex} gives the transition for several values of $y$, that is for several values of $\tilde{Q}/C$, and Fig.~\ref{fig:h1}c shows the coexistence line in the $(\tilde{T},\tilde{Q}/C)$ plane, which ends at the critical point. For $\sigma=0.12$ and for $\hat{\beta}=0$ the line extends to small $y$. For $\sigma=0.9$ the transition exists only in the window $y_{*}\leq y\leq y_{c}$ with $y_{*}=0.1618$. For $\sigma=1.2$ the window is much narrower and lies close to the critical point, $0.172858\lesssim y\leq y_c=0.175606$, so that its swallowtail is small; it is shown for $y=0.1735$ in Fig.~\ref{fig:h1}b, where the cusps are resolved in the inset, and the corresponding coexistence line is the short segment ending at $\tilde{Q}/C\simeq0.70$ in Figs.~\ref{fig:h1}c and~\ref{fig:h1}d. Below the model-dependent lower threshold there is no small--large free-energy crossing between the stable branches. This $y$-window belongs to the fixed-$C$ canonical diagram and is not identified with the bulk pressure window $P_{*}<P<P_{c}$, which would require a separate fixed-$\tilde Q$ calculation at fixed $\tilde{\beta}$. For $\sigma=0.45$ the transition exists for $y\geq0.0574$, but for $y\lesssim0.12$ the small phase is a very small black hole, $x_{s}\lesssim0.1$, and for $0.105\lesssim y\lesssim0.115$ the function $\tilde{T}(x)$ has in addition a narrow stable island at $\tilde{T}R\simeq0.31$--$0.34$, which lies above the lower envelope of the free energy and does not take part in the transition. The coexistence lines were obtained from crossings of stable branches of $\tilde{T}(x)$ on the lower envelope of $F$; an independently written root-finder (code package) reproduces Table~\ref{tab:cftcoex}. Since $\tilde{Q}=4Cy$, these are coexistence lines in the $(\tilde{Q},\tilde{T})$ plane at fixed $C=5/6$ and fixed $\tilde{\beta}=0.1$. They are not the coexistence lines in the $(1/C,\tilde{T})$ plane at fixed $\tilde{Q}$, which are given in Sec.~\ref{sec:varyC}.
\begin{table*}[!t]
\caption{First-order transition in the canonical ensemble for fixed $\tilde{\beta}=0.1$ and $C=5/6$ (hence $\hat{\beta}=0.03$) in units $R=1$. $x_{s}$ and $x_{l}$ are the values of $r_{h}/l$ of the small and large phase, $\Delta S/C=4\pi(x_{l}^{2}-x_{s}^{2})$, $\tilde{T}_{t}\Delta S/C$ is the latent heat and $\tilde{F}_{t}=F/C$ is the common free energy. The cases are RN ($\hat{\beta}=0$), A ($\sigma=0.12$), B ($\sigma=0.9$), C ($\sigma=0.45$) and D ($\sigma=1.2$).}\label{tab:cftcoex}
\begin{ruledtabular}
\begin{tabular}{cccccccc}
$y=\tilde{Q}/(4C)$ & Case & $\tilde{T}_{t}R$ & $x_{s}$ & $x_{l}$ & $\Delta S/C$ & $\tilde{T}_{t}\Delta S/C$ & $\tilde{F}_{t}R/C$\\
\hline
$0.1$ & RN & $0.28471$ & $0.1310$ & $0.7634$ & $7.1086$ & $2.0239$ & $0.3578$\\
$0.1$ & A & $0.28518$ & $0.1218$ & $0.7677$ & $7.2203$ & $2.0591$ & $0.3543$\\
$0.1$ & B & \multicolumn{6}{c}{no transition}\\
$0.1$ & C & $0.28998$ & $0.0102$ & $0.8088$ & $8.2198$ & $2.3836$ & $0.3168$\\
$0.14$ & RN & $0.27009$ & $0.2243$ & $0.6243$ & $4.2652$ & $1.1520$ & $0.4752$\\
$0.14$ & A & $0.27032$ & $0.2174$ & $0.6276$ & $4.3566$ & $1.1777$ & $0.4741$\\
$0.14$ & B & \multicolumn{6}{c}{no transition}\\
$0.14$ & C & $0.27115$ & $0.1886$ & $0.6394$ & $4.6913$ & $1.2720$ & $0.4699$\\
$0.16$ & RN & $0.26249$ & $0.3123$ & $0.5123$ & $2.0725$ & $0.5440$ & $0.5278$\\
$0.16$ & A & $0.26263$ & $0.3047$ & $0.5172$ & $2.1943$ & $0.5763$ & $0.5273$\\
$0.16$ & B & \multicolumn{6}{c}{no transition}\\
$0.16$ & C & $0.26309$ & $0.2805$ & $0.5311$ & $2.5560$ & $0.6725$ & $0.5256$\\
$0.165$ & RN & $0.26055$ & $0.3593$ & $0.4593$ & $1.0286$ & $0.2680$ & $0.5402$\\
$0.165$ & A & $0.26067$ & $0.3478$ & $0.4681$ & $1.2339$ & $0.3216$ & $0.5398$\\
$0.165$ & B & $0.26185$ & $0.2603$ & $0.5177$ & $2.5165$ & $0.6589$ & $0.5360$\\
$0.165$ & C & $0.26107$ & $0.3177$ & $0.4890$ & $1.7372$ & $0.4535$ & $0.5386$\\
$0.17$ & B & $0.25965$ & $0.3159$ & $0.4654$ & $1.4671$ & $0.3809$ & $0.5495$\\
$0.17$ & D & \multicolumn{6}{c}{no transition} \\
$0.173$ & D & $0.258918$ & $0.27860$ & $0.45992$ & $1.68276$ & $0.43570$ & $0.555705$ \\
\end{tabular}
\end{ruledtabular}
\end{table*}
The coexistence line is, as in a van der Waals fluid, a line of first-order transitions ending at a critical point. For the fixed-$C$ canonical ensemble considered here, the van der Waals correspondence is $\tilde{Q}\leftrightarrow P$ and $\tilde{\Phi}$, $\mu\leftrightarrow V$, as in Ref.~\cite{refH2}. The alternative identification $1/C\leftrightarrow P$ is treated in Sec.~\ref{sec:varyC}, where $\hat{\beta}$ changes with $C$ at fixed $\tilde{\beta}$ and $\tilde Q$; it is not represented by Fig.~\ref{fig:h1}. At fixed $C$ the temperature at the critical point is almost independent of $\sigma$ and of the chosen nonlinear coupling (Table~\ref{tab:cftcrit}).

\subsection{Critical point at fixed $\tilde{Q}$ with varying $C$}\label{sec:varyC}
\begin{figure*}[!tbp]
\centering\includegraphics[width=1\textwidth]{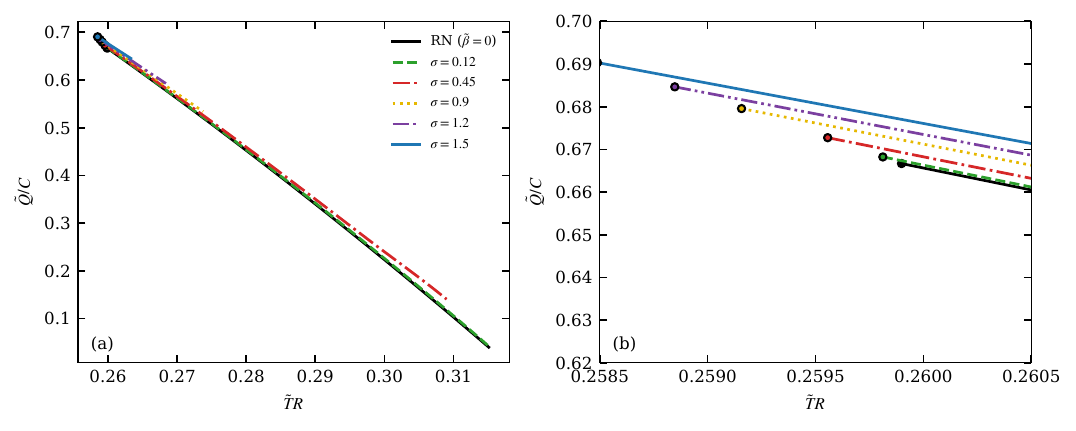}
\caption{Canonical coexistence lines in the $(\tilde T R,\tilde Q/C)$ plane at fixed $\tilde Q$ and $\tilde\beta=0.1$ ($\kappa=0.1$), so that $\hat\beta=\kappa y$ varies along each line. Curves denote RN (black), $\sigma=0.12$ (green), $\sigma=0.45$ (red), $\sigma=0.9$ (yellow), $\sigma=1.2$ (purple) and $\sigma=1.5$ (blue). Circles mark the critical points. (b) Enlargement of the region around the critical points.}\label{fig:varyC}
\end{figure*}

In the identification $1/C\leftrightarrow P$ the charge $\tilde Q$ and the coupling $\tilde\beta$ are fixed while $C$, and with it $l$, varies. With $\kappa=\tilde\beta/\tilde Q=0.1$ one has $y=\tilde Q/(4C)$ and $\hat\beta=\kappa y$, so that $\hat\beta$ changes along the family. The temperature $\tilde T R(x,y)$ is then Eq.~\eqref{eq:xy} with $\hat\beta=\kappa y$. Since $y$ is held fixed in the derivatives with respect to $x$, the critical point is again defined by $\partial_{x}\tilde T=\partial_{x}^{2}\tilde T=0$, now evaluated with $\hat\beta=\kappa y$. The coexistence line is obtained for each $y$ from the conditions $\tilde T(x_{s})=\tilde T(x_{l})$ and $\tilde F(x_{s})=\tilde F(x_{l})$ at that value of $\hat\beta$. The critical points and the lower end of the coexistence lines are given in Table~\ref{tab:varyC} and the lines are shown in Fig.~\ref{fig:varyC}. The critical temperature changes by about $0.5\%$ from RN to $\sigma=1.5$, and the critical ratio $C/\tilde Q_{c}$ decreases from $3/2$ with $\sigma$ less strongly than at fixed $C$ (Table~\ref{tab:cftcrit}), because $\hat\beta_{c}=\kappa y_{c}$ is then smaller than $0.03$ and grows with $\tilde Q/C$. The coexistence lines of the cases with a small black hole branch of negative heat capacity ($\sigma=0.9$, $1.2$ and $1.5$) end at larger $y$ than those of RN, $\sigma=0.12$ and $\sigma=0.45$, as in the fixed-$C$ ensemble.
\begin{table}[!t]
\caption{Critical point at fixed $\tilde Q$ with $\tilde\beta=0.1$ held fixed ($\kappa=\tilde\beta/\tilde Q=0.1$, $R=1$, varying $C$). The last column is the value of $\tilde Q/C$ at which the coexistence line ends; for RN and $\sigma=0.12$ the line was followed down to $\tilde Q/C=0.04$.}\label{tab:varyC}
\begin{ruledtabular}
\begin{tabular}{lccccc}
Case & $y_{c}$ & $C/\tilde Q_{c}$ & $\tilde T_{c}R$ & $\hat\beta_{c}$ & $\tilde Q/C$ (end)\\
\hline
RN & $0.16667$ & $1.5000$ & $0.25990$ & $0$ & $\le0.04$\\
$\sigma=0.12$ & $0.16706$ & $1.4965$ & $0.25981$ & $0.01671$ & $\le0.04$\\
$\sigma=0.45$ & $0.16818$ & $1.4865$ & $0.25956$ & $0.01682$ & $0.134$\\
$\sigma=0.9$ & $0.16989$ & $1.4716$ & $0.25916$ & $0.01699$ & $0.521$\\
$\sigma=1.2$ & $0.17116$ & $1.4606$ & $0.25885$ & $0.01712$ & $0.593$\\
$\sigma=1.5$ & $0.17258$ & $1.4486$ & $0.25849$ & $0.01726$ & $0.640$\\
\end{tabular}
\end{ruledtabular}
\end{table}
\subsection{Critical exponents}
Near the critical point the exponents are those of mean-field theory, $\alpha=0$, $\beta_{\rm c}=1/2$, $\gamma=1$ and $\delta=3$, for $\hat{\beta}=0$, in agreement with Ref.~\cite{refH2}. The same exponents hold for the present black hole because $\tilde{T}(x,y)$ is analytic at the critical point and its expansion there is nondegenerate: with $\partial_x\tilde T=\partial_x^2\tilde T=0$, the cubic coefficient $\partial_x^3\tilde T R$ takes the values $11.46$, $11.28$, $10.65$, $9.30$ and $7.49$, and the control-parameter coefficient $\partial_y\tilde T R$ the values $-0.390$, $-0.394$, $-0.406$, $-0.431$ and $-0.463$, for RN and for $\sigma=0.12$, $0.45$, $0.9$ and $1.2$, respectively, all at $\hat\beta=0.03$ (script \texttt{verify\_critical\_coeffs.py}); both are nonzero, as required for $\delta=3$ and $\beta_{\rm c}=1/2$; the nonlinear numerical checks below are performed in the fixed-$C$ ensemble, where fixed $\tilde{\beta}$ and fixed $\hat{\beta}$ are equivalent. Three of the exponents were verified numerically by local fits for the cases RN, A, B and C; for case D ($\sigma=1.2$) the exponents follow from the nondegeneracy of the expansion given above. The order parameter $x_{l}-x_{s}$ along the coexistence line behaves as $(1-y/y_{c})^{\beta_{\rm c}}$ with local exponents $0.5000$--$0.5011$ for $1-y/y_{c}=3\times10^{-3},10^{-3},3\times10^{-4},10^{-4}$. The critical isotherm $\tilde{T}=\tilde{T}_{c}$ gives $\tilde{Q}-\tilde{Q}_{c}\propto(\tilde{\Phi}-\tilde{\Phi}_{c})^{\delta}$ with local exponents $3.027$, $3.012$ and $3.006$ for $r_{h}$ at $1\%$, $0.5\%$ and $0.2\%$ from the critical value. The susceptibility $(\partial\tilde{\Phi}/\partial\tilde{Q})_{\tilde{T}}$ at $\tilde{\Phi}=\tilde{\Phi}_{c}$ behaves as $(\tilde{T}/\tilde{T}_{c}-1)^{-\gamma}$ with local exponents $0.989$--$0.991$ and $0.999$ for $\tilde{T}/\tilde{T}_{c}-1=10^{-3}$--$10^{-4}$ and $10^{-4}$--$10^{-5}$, so that $\gamma=1$. The results are collected in Table~\ref{tab:exp}. The heat capacity $C_{\tilde{\Phi}\mathcal{V}C}/C$ at the critical point is finite, with the values $-16.76$, $-16.43$, $-13.66$ and $-15.42$ for $\hat{\beta}=0$ and for $\sigma=0.12$, $0.9$ and $0.45$ at $\hat{\beta}=0.03$, in agreement with $\alpha=0$ for the variable pair $(\tilde{Q},\tilde{\Phi})$ in the van der Waals correspondence. The finiteness of $C_{\tilde\Phi\mathcal{V}C}$ concerns only the absence of a divergence; its negative sign is not a statement about stability.

\begin{table}[!t]
\caption{Critical exponents of the canonical ensemble at fixed $C$: mean-field values and numerical local exponents quoted in the text. The exponent $\beta_{\rm c}$ is the order-parameter exponent and is unrelated to the coupling $\beta$; $t=\tilde T/\tilde T_c-1$, and $C_{\tilde\Phi}\equiv C_{\tilde\Phi\mathcal{V}C}/C$.}
\label{tab:exp}
\begingroup
\footnotesize
\setlength{\tabcolsep}{3pt}
\begin{ruledtabular}
\begin{tabular}{cccc}
\shortstack{Exponent} & \shortstack{Definition} & \shortstack{Mean\\field} & \shortstack{Numerical} \\
\hline
$\alpha$ & \shortstack{$C_{\tilde\Phi}\sim$\\$|t|^{-\alpha}$} & $0$ & \shortstack{finite\\$(-16.76$ to $-13.66)$} \\
$\beta_{\rm c}$ & \shortstack{$x_l-x_s\sim$\\$(1-y/y_c)^{\beta_{\rm c}}$} & $1/2$ & \shortstack{$0.5000$--\\$0.5011$} \\
$\gamma$ & \shortstack{$(\partial\tilde\Phi/\partial\tilde Q)_{\tilde T}$\\$\sim t^{-\gamma}$} & $1$ & \shortstack{$0.989$--$0.991$,\\$0.999$} \\
$\delta$ & \shortstack{$\tilde Q-\tilde Q_c\sim$\\$(\tilde\Phi-\tilde\Phi_c)^{\delta}$} & $3$ & \shortstack{$3.027$, $3.012$,\\$3.006$} \\
\end{tabular}
\end{ruledtabular}
\endgroup
\end{table}

\subsection{Grand-canonical ensemble \texorpdfstring{$(\tilde{\Phi},\mathcal{V},C)$}{(Phi, V, C)}}\label{sec:W}
In the grand-canonical calculations we again hold $C=5/6$ and $\tilde{\beta}=0.1$ fixed, so $\hat{\beta}=0.03$. At fixed $\tilde{\Phi}$ the charge $y(x)$ follows from $\tilde{\Phi}R=\phi$, and the potential is $W/C=\mu R+\tilde{\Psi}\tilde{\beta}/(CR^{-1})=x-x^{3}-y^{2}F_{1}/x$. In the RN limit $y=\phi x$, and $\tilde{T}R=(1+3x^{2}-\phi^{2})/(4\pi x)$ has a single minimum. The temperature has therefore a single stable branch, the black hole with positive heat capacity, and the transition is the confinement--deconfinement (Hawking--Page) transition at $W=0$ between thermal AdS and the black hole (Fig.~\ref{fig:h2}a). Table~\ref{tab:HP} gives the transition temperature. At $\tilde{\Phi}R=0.3$, Fig.~\ref{fig:h2}(a) shows the curves for RN and for $\sigma=0.12$, $0.45$, $0.9$ and $1.2$; the corresponding transition temperatures are listed in Table~\ref{tab:HP}. The plotted curves nearly coincide on the main scale, consistent with the very small shifts in the Hawking--Page temperature; the inset resolves the separation of their $W=0$ crossings. For $\sigma=1.2$ the Hawking--Page point lies at $\tilde{T}_{\rm HP}R=0.303638$, which is $1.1\times10^{-5}$ below the RN value $0.303648$; this is the largest shift among the cases shown, and in the inset the $\sigma=1.2$ curve is the lowest of the five, in the same order as the exponent. For $\tilde{\Phi}R\le0.4$, the shifts with $\sigma$ are confined to the last reported decimal or are smaller. As for $\hat{\beta}=0$, it decreases with $\tilde{\Phi}$. No swallowtail appears in $W(\tilde{T})$ for $\tilde{\Phi}R\leq0.4$.
\begin{table}[!t]
\caption{Hawking--Page temperature $\tilde{T}_{\rm HP}R$ of the ensemble $(\tilde{\Phi},\mathcal{V},C)$ for fixed $\tilde{\beta}=0.1$, $C=5/6$ and hence $\hat{\beta}=0.03$ (RN denotes $\tilde{\beta}=0$).}\label{tab:HP}
\begin{ruledtabular}
\begin{tabular}{cccccc}
$\tilde{\Phi}R$ & RN & $\sigma=0.12$ & $\sigma=0.9$ & $\sigma=0.45$ & $\sigma=1.2$\\
\hline
$0.1$ & $0.31671$ & $0.31671$ & $0.31671$ & $0.31671$ & $0.31671$\\
$0.2$ & $0.31188$ & $0.31188$ & $0.31188$ & $0.31188$ & $0.31188$\\
$0.3$ & $0.30365$ & $0.30365$ & $0.30364$ & $0.30364$ & $0.30364$\\
$0.4$ & $0.29174$ & $0.29173$ & $0.29171$ & $0.29172$ & $0.29170$\\
\end{tabular}
\end{ruledtabular}
\end{table}

\subsection{Mixed ensemble \texorpdfstring{$(\tilde{Q},\mathcal{V},\mu)$}{(Q, V, mu)}}\label{sec:Gens}
\begin{figure*}[!tbp]
\centering\includegraphics[width=1\textwidth]{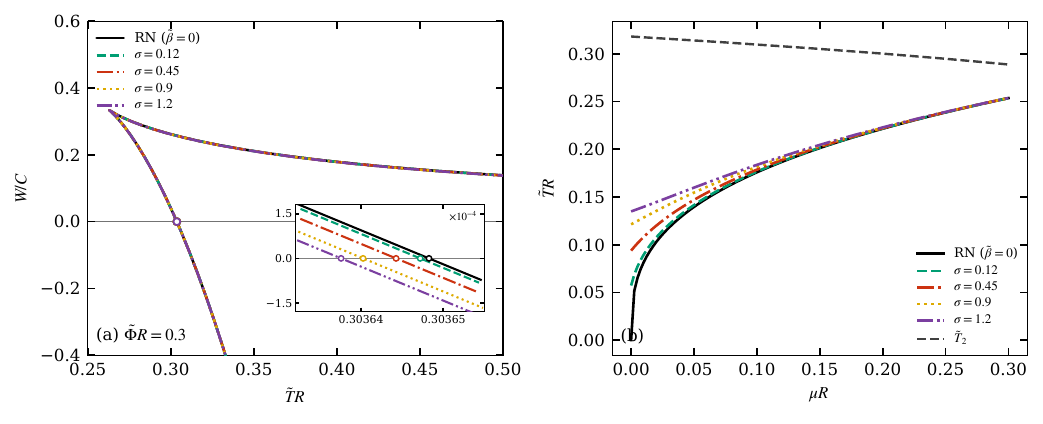}
\caption{(a) Potential $W/C$ of the ensemble $(\tilde{\Phi},\mathcal{V},C)$ against $\tilde{T}R$ for $\tilde{\Phi}R=0.3$; the Hawking--Page temperature is the zero. The inset enlarges the closely spaced zero crossings. Curves denote RN (black), $\sigma=0.12$ (green), $\sigma=0.45$ (red), $\sigma=0.9$ (yellow) and $\sigma=1.2$ (purple). (b) Cusp temperature $\tilde{T}_{\min}R$ and larger-$x$ endpoint temperature $\tilde{T}_{1}R$ for the mixed ensemble, plotted against $\mu R$; the endpoint curve is dashed gray. The colors identify the same five cases, for $\tilde\beta/\tilde Q=0.1$.}\label{fig:h2}
\end{figure*}

In the mixed ensemble we hold $\tilde Q$, $\mathcal V$, $\mu$ and the physical coupling $\tilde\beta=0.1$ fixed while the central charge $C$, and hence $l$, varies. Taking $\tilde Q=1$ and $R=1$ gives $\kappa=\tilde\beta/\tilde Q=0.1$ and $\hat\beta=\kappa y$, so $\hat\beta$ varies along the curve rather than remaining fixed at $0.03$. The condition $\mu R=\mathrm{const}$ reads $y^{2}F_{1}/x-2\sigma\kappa y^{5}F_{2}/(5x^{5})=x-x^{3}-\mu R$, which determines $y(x)$. The entropy is $S=\pi\tilde Q x^{2}/y$, and the heat capacity is $C_{\mu}/\tilde Q=\tilde T\,d(\pi x^{2}/y)/dx\,/\,(d\tilde T/dx)$. For $\mu R>0$, the allowed interval in $x$ is bounded by the two roots of $x-x^{3}=\mu R$. We denote the temperature at the larger-$x$ endpoint by $\tilde T_{1}$ and the temperature at the smaller-$x$ endpoint by $\tilde T_{2}$; both endpoints occur at $y\to0$, with $\tilde T_{i}R=(1+3x_{i}^{2})/(4\pi x_{i})$. The two branches meet at a cusp with $\tilde T_{0}=\tilde T_{\min}(\mu)$. The finite branch discontinuity in the free energy occurs at the separate endpoint $\tilde T_{1}$, not at the cusp. Whether this discontinuity represents a transition between locally stable equilibrium phases depends on the stability of the lower-entropy branch at that temperature and is examined below. Table~\ref{tab:G} lists $\tilde T_{\min}$ and the larger-$x$ endpoint temperature $\tilde T_{1}$ for $\kappa=0.1$, while Fig.~\ref{fig:h2}b shows $\tilde T_{\min}(\mu)$. The difference between $\tilde T_{\min}$ and its Reissner--Nordstr\"om value increases with $\sigma$ and decreases with $\mu$. Among the cases in Fig.~\ref{fig:h2}b, $\sigma=1.2$ lies highest: at $\mu R=0.1$ it exceeds the RN value by about $4\%$ ($0.18368$ against $0.17590$), and at $\mu R=0.25$ by only $0.2\%$. For $\mu=0$, the RN branch has $\tilde T_{\min}\to0$, whereas the nonlinear model has a finite lower endpoint for $\tilde\beta>0$.
\begin{table}[!t]
\caption{Lower temperature $\tilde{T}_{\min}R$ at the cusp and larger-$x$ endpoint temperature $\tilde{T}_{1}R=(1+3x_{+}^{2})/(4\pi x_{+})$ of the mixed ensemble for $\tilde\beta/\tilde Q=0.1$, where $x_{+}$ is the larger root of $x-x^{3}=\mu R$ and $y=0$. For $\hat{\beta}=0$ and $\mu=0$, the RN temperature $\tilde{T}R=x/\pi$ has no positive-temperature minimum and $\tilde{T}_{\min}=0$.}\label{tab:G}
\begin{ruledtabular}
\begin{tabular}{ccccccc}
$\mu R$ & RN & $\sigma=0.12$ & $\sigma=0.9$ & $\sigma=0.45$ & $\sigma=1.2$ & $\tilde{T}_{1}R$\\
\hline
$0.0$ & $0.00000$ & $0.05718$ & $0.12133$ & $0.09376$ & $0.13485$ & $0.31831$\\
$0.1$ & $0.17590$ & $0.17662$ & $0.18161$ & $0.17866$ & $0.18368$ & $0.30991$\\
$0.2$ & $0.22162$ & $0.22175$ & $0.22262$ & $0.22211$ & $0.22296$ & $0.30036$\\
$0.25$ & $0.23873$ & $0.23878$ & $0.23909$ & $0.23891$ & $0.23922$ & $0.29496$\\
\end{tabular}
\end{ruledtabular}
\end{table}
\begin{figure*}[!tbp]
\centering
\includegraphics[width=1\textwidth]{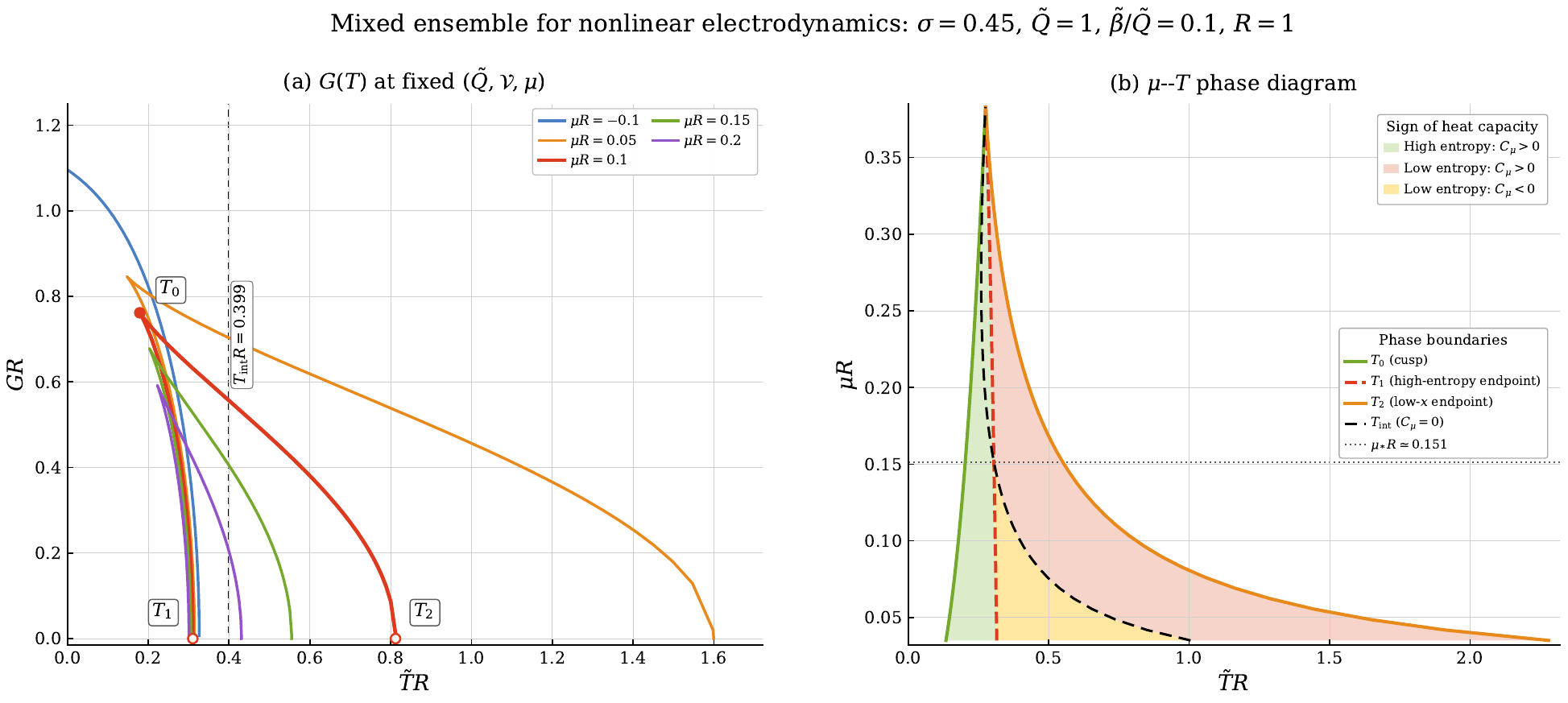}
\caption{Mixed-ensemble free-energy branches and phase boundaries for nonlinear electrodynamics with $\sigma=0.45$, $\tilde Q=1$, $\tilde\beta/\tilde Q=0.1$ and $R=1$. (a) The potential $GR$ as a function of $\tilde T R$ for several values of $\mu R$. For the highlighted curve $\mu R=0.1$, $T_{0}$ denotes the cusp, $T_{1}$ the endpoint of the high-entropy (larger-$x$) branch, and $T_{2}$ the endpoint of the lower-$x$ branch. The vertical dashed line marks $T_{\mathrm{int}}$, where $C_{\mu}=0$ on the lower-entropy branch. The distinct values of $GR$ at $T_{1}$ are shown by separate markers and are not connected by an extra segment. (b) Numerical $\mu$--$\tilde T$ phase diagram showing the same characteristic boundaries. The external legend identifies the sign of $C_{\mu}$ in each shaded region; this sign alone does not establish full local thermodynamic stability.}\label{fig:mixedKong}
\end{figure*}

Figure~\ref{fig:mixedKong} presents the corresponding branch structure and numerical $\mu$--$\tilde T$ phase diagram for the representative value $\sigma=0.45$. In panel (a), the branches meet continuously at the cusp $T_{0}=T_{\min}$. At the distinct endpoint $T_{1}$, the high-entropy branch reaches $GR=0$, whereas the lower-entropy solution at the same temperature has a finite value of $GR$; the two endpoint values are marked separately, without an added segment between them. The low-entropy branch changes the sign of $C_{\mu}$ at $T_{\mathrm{int}}$. In panel (b), the boundaries $T_{0}$, $T_{1}$, $T_{2}$ and $T_{\mathrm{int}}$ delimit the branch structure over the positive-$\mu R$ range sampled numerically. The external legend indicates the sign of $C_{\mu}$ in the shaded regions, but this sign alone does not establish full local thermodynamic stability. A zeroth-order equilibrium transition can be claimed only where the destination branch also satisfies the full local-stability conditions discussed in Appendix~\ref{app:hessian}.

\begin{figure*}[!tbp]
\centering
\includegraphics[width=1\textwidth]{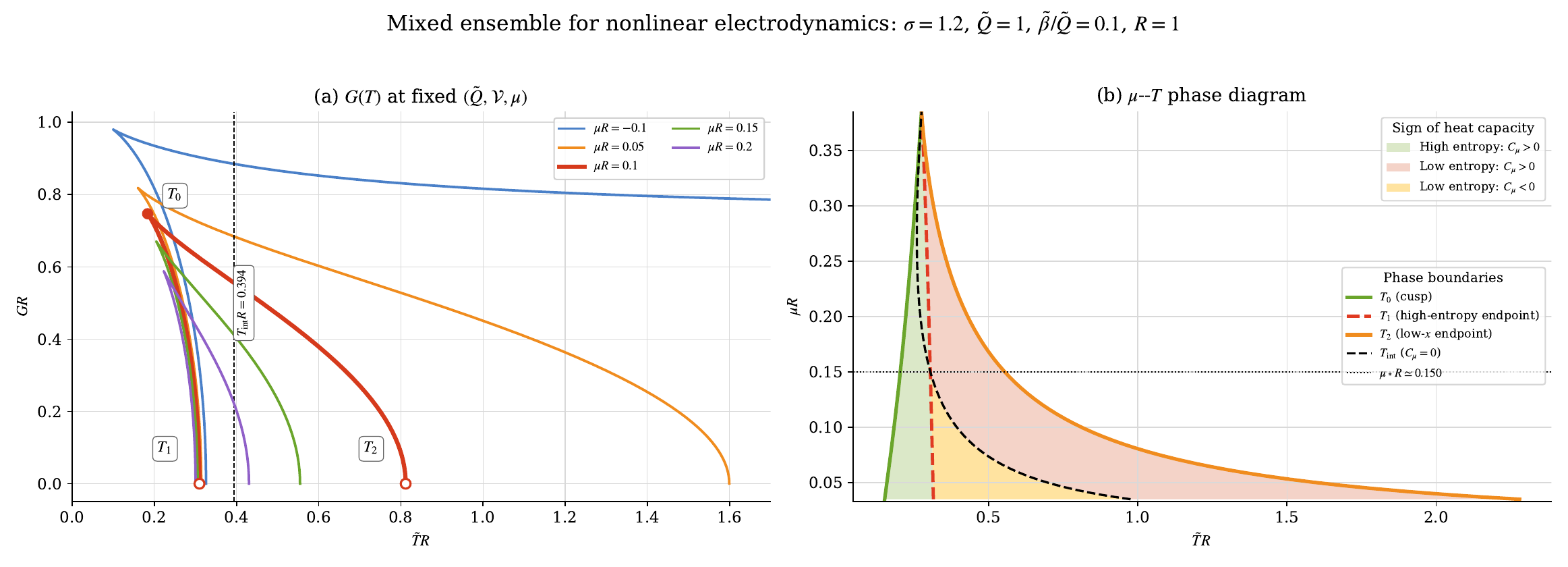}
\caption{Mixed-ensemble free-energy branches and phase boundaries for nonlinear electrodynamics with $\sigma=1.2$, $\tilde Q=1$, $\tilde\beta/\tilde Q=0.1$ and $R=1$. The quantities and notation are those of Fig.~\ref{fig:mixedKong}. (a) The potential $GR$ as a function of $\tilde TR$ for several values of $\mu R$; for the highlighted curve $\mu R=0.1$, $T_{0}$ denotes the cusp, and $T_{1}$ and $T_{2}$ the endpoints of the larger-$x$ and lower-$x$ branches. The dashed vertical line marks $T_{\mathrm{int}}R=0.394$. The curve for $\mu R=-0.1$ is shown for $\tilde TR\le1.7$. (b) The corresponding $\mu$--$\tilde T$ phase diagram; the dotted line marks $\mu_{*}R\simeq0.150$.}\label{fig:mixedS12}
\end{figure*}

For $\sigma=1.2$ the same construction is shown in Fig.~\ref{fig:mixedS12}. The boundary curves $T_{0}$, $T_{1}$, $T_{2}$ and $T_{\mathrm{int}}$ keep the ordering found for $\sigma=0.45$, and the only visible changes are quantitative. For $\mu R=0.1$ the cusp temperature rises from $T_{0}R=0.17866$ to $0.18368$ and $T_{\mathrm{int}}R$ decreases from $0.3988$ to $0.3940$, while $T_{1}R=0.30991$ and $T_{2}R=0.81177$ are fixed by $\mu$ alone and therefore do not depend on $\sigma$. The value of $\mu$ at which $T_{\mathrm{int}}$ crosses $T_{1}$ moves from $\mu_{*}R\simeq0.1514$ to $0.1500$. A qualitative difference appears for negative $\mu$. At $\mu R=-0.1$ the temperature does not start from zero as it does for $\sigma=0.45$: because $T\to\infty$ as $x\to0$ for $\sigma>1/4$, the curve has a minimum at $\tilde T R\simeq0.0995$, where $GR\simeq0.98$, and continues to larger temperatures on a second branch, whose temperature is cut at $\tilde TR=1.7$ in the figure.

\subsection{Pressure--volume plane}\label{sec:pV}
Ref.~\cite{refH2} finds no criticality in the $(p,\mathcal{V})$ plane for the charged AdS black hole. For the present black hole, at fixed $(\tilde{T},C,\tilde{Q})$ one has $R=\tau(x)/\tilde{T}$ with $\tau=\tilde T R$, $\mathcal{V}=4\pi R^{2}$ and $p=C\tilde T^{3}\,h(x)/(2\pi)$, where $h(x)=\varepsilon(x)/\tau(x)^{3}$ and $\varepsilon=M/l$. A critical point requires $\partial_{x}h=\partial_{x}^{2}h=0$ with $\partial_{x}\tau\neq0$, and corresponds to the merging of two extrema of the isotherm $p(\mathcal{V})$. The factor $\tilde T^{3}$ shows that the shape of the isotherm does not depend on $\tilde T$: for a conformal theory on a sphere the combination $p\,\mathcal{V}^{3/2}$ is a function of $x$ and $y$ only, and a critical point is a line in the $(p,\mathcal{V})$ plane parametrized by $\tilde T$, with $p\mathcal{V}^{3/2}$ constant along it. In this subsection a critical point of the isotherm $p(\mathcal{V})$ means the merging of two of its extrema, a property of the equation of state at fixed $(\tilde T,C,\tilde Q)$. We counted the extrema of $h(x)$ on the branch $\tau>0$ for $y\in[0.005,0.40]$ and $\hat\beta=0$ and $\hat\beta=0.03$, and located every change in their number by solving $\partial_{x}h=\partial_{x}^{2}h=0$ (Table~\ref{tab:pV}). For RN and $\sigma=0.12$ the number of extrema does not change with $y$, so that there is no critical point in the range scanned, in agreement with Ref.~\cite{refH2}. For $\sigma=0.9$, $\sigma=1.2$ and $\sigma=1.5$ a pair of extrema is created at one critical point and persists for larger $y$. For $\sigma=0.45$ the number of extrema increases from one to three at $y=0.11168$ and returns to one at $y=0.11774$, where two of them merge, so that there are two critical points in a narrow interval of $\tilde Q/C$. In all cases with critical points the temperature $\tau(x)$ is monotonic between the extrema, so that $\mathcal{V}$ is a single-valued function of $x$ and the isotherm $p(\mathcal{V})$ has a van der Waals-type loop. At each critical point of Table~\ref{tab:pV}, $d\tau/dx$ lies between $0.11$ and $0.26$, so that the heat capacity at fixed $(\mathcal{V},C,\tilde Q)$, $8\pi C x\tau/\tau'$, is positive there. Extrema that leave the domain through its boundary at $x\to\infty$ for $y\to1/\sqrt6$ do not correspond to critical points.
\begin{table}[!t]
\caption{Critical points of the isotherms $p(\mathcal{V})$ at fixed $(\tilde T,C,\tilde Q)$ for $\hat\beta=0.03$ ($\tilde\beta=0.1$, $C=5/6$ in units $R=1$). The last column is $p_{c}\mathcal{V}_{c}^{3/2}/C=4\sqrt{\pi}\,\varepsilon_{c}$.}\label{tab:pV}
\begin{ruledtabular}
\begin{tabular}{lccccc}
Case & $x_{c}$ & $y_{c}$ & $\tilde Q/C$ & $\tilde T_{c}R$ & $p_{c}\mathcal{V}_{c}^{3/2}/C$\\
\hline
RN, $\sigma=0.12$ & \multicolumn{5}{c}{no critical point} \\
$\sigma=0.45$ & $0.12052$ & $0.11168$ & $0.4467$ & $0.33080$ & $0.7636$\\
$\sigma=0.45$ & $0.07044$ & $0.11774$ & $0.4710$ & $0.28461$ & $0.7352$\\
$\sigma=0.9$ & $0.23583$ & $0.17470$ & $0.6988$ & $0.24710$ & $1.320$\\
$\sigma=1.2$ & $0.28090$ & $0.19938$ & $0.7975$ & $0.23470$ & $1.555$\\
$\sigma=1.5$ & $0.31768$ & $0.21827$ & $0.8731$ & $0.22923$ & $1.751$\\
\end{tabular}
\end{ruledtabular}
\end{table}

\subsection{Heat capacities and stability}\label{sec:cap}
At fixed $C$ one has $S=4\pi C x^{2}$, and the heat capacity at fixed values of the variables $X$ of such an ensemble is
\begin{equation}\label{eq:cap}
C_{X}=\tilde{T}\Big(\frac{\partial S}{\partial\tilde{T}}\Big)_{X}=\frac{8\pi C\,x\,\tilde{T}}{(d\tilde{T}/dx)_{X}},
\end{equation}
where $\tilde{T}(x)$ is evaluated along the curve defined by $X$: $y$ constant for $(\tilde{Q},\mathcal{V},C)$, $y(x)$ from $\tilde{\Phi}R=\mathrm{const}$ for $(\tilde{\Phi},\mathcal{V},C)$ and $y(x)$ from $\mu R=\mathrm{const}$ for $(\tilde{Q},\mathcal{V},\mu)$, while Eq.~\eqref{eq:cap} does not apply to the mixed ensemble, where $C$ varies, $S=\pi\tilde Q x^{2}/y(x)$ and $C_{\mu}=\tilde T\,(dS/dx)/(d\tilde T/dx)$ as in Sec.~\ref{sec:Gens}. A branch is thermally stable if $C_{X}>0$; the relation of this criterion to full local stability is examined in Appendix~\ref{app:hessian}. Fig.~\ref{fig:h3} shows $C_{X}/C$ for the three ensembles and Table~\ref{tab:capdiv} gives the points at which $C_{X}$ diverges, that is the extrema of $\tilde{T}(x)$.
\begin{figure*}[!tbp]
\centering\includegraphics[width=0.90\textwidth]{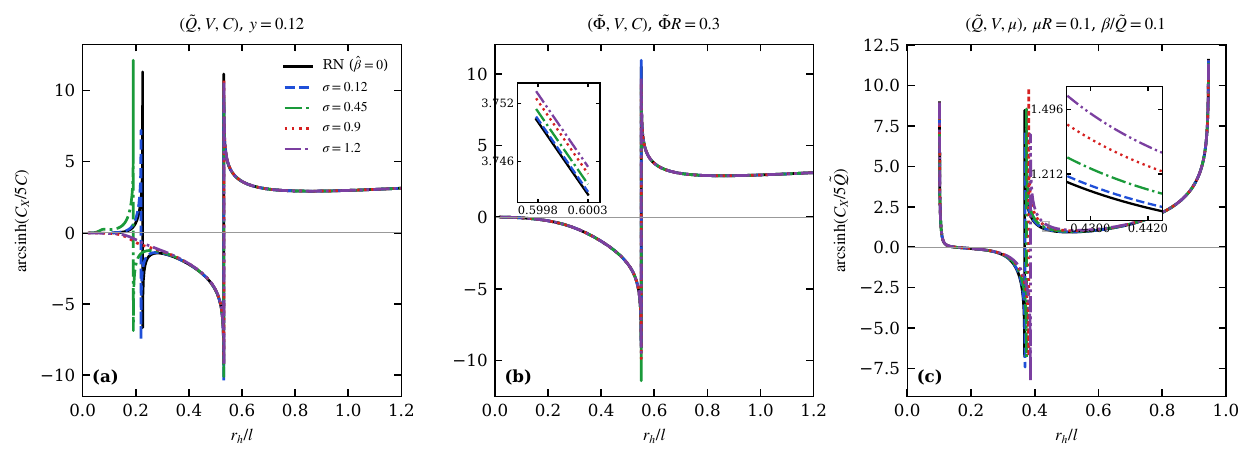}
\caption{Heat capacity $C_{X}/C$, plotted as $\mathrm{arcsinh}(C_{X}/5C)$ against $x=r_{h}/l$. In (a) and (b), the colored curves use fixed $\tilde{\beta}=0.1$ and $C=5/6$, hence $\hat{\beta}=0.03$; in (c), $\tilde{\beta}=0.1$ and $\tilde Q=1$ are fixed, so $\kappa=0.1$ and $\hat{\beta}=\kappa y$ varies with $C$. Colors denote $\sigma=0.12$ (blue dashed), $0.45$ (green dash-dotted), $0.9$ (red dotted) and $1.2$ (purple dash-dot-dotted); RN is black. Panel (a) is the canonical ensemble at $y=0.12$, (b) the grand-canonical ensemble at $\tilde{\Phi}R=0.3$, and (c) the mixed ensemble at $\mu R=0.1$. The inverse-hyperbolic-sine scaling displays the divergences. The insets in (b) and (c) enlarge the ranges $0.5996\le x\le0.6004$ and $0.425\le x\le0.445$, respectively, where the curves of the different cases are separated only slightly on the main scale.}\label{fig:h3}
\end{figure*}
In the canonical ensemble at $y=0.12$ the heat capacity has two divergences for $\sigma=0.12$, $0.45$ and for the RN case: a small black hole branch with $C_{X}>0$ for $x<x_{1}$, an unstable branch with $C_{X}<0$ between $x_{1}$ and $x_{2}$, and a large black hole branch with $C_{X}>0$ for $x>x_{2}$. This is the van der Waals structure whose free energy has the swallowtail of Fig.~\ref{fig:h1}a, and the first-order transition of Table~\ref{tab:cftcoex} connects the two stable branches. The large-black-hole divergence $x_{2}\simeq0.532$ is almost independent of $\sigma$. The small-black-hole divergence $x_{1}$ is lowered by the nonlinear electrodynamics and moves to larger $\tilde{T}$ for $\sigma=0.45$. For $\sigma=0.9$ and $\sigma=1.2$ it is absent for $x\gtrsim0.02$, so that $y=0.12$ lies outside the window of the first-order transition found above. In the grand-canonical ensemble there is a single divergence at $x\simeq0.5508$ for $\tilde{\Phi}R=0.3$, at the minimum of $\tilde{T}$, which separates the unstable small and the stable large black hole, in agreement with the absence of a swallowtail and with the Hawking--Page transition of Sec.~\ref{sec:W}. In the mixed ensemble, the heat capacity diverges at the cusp temperature $\tilde{T}_{\min}(\mu)$, where the branches meet continuously. The finite discontinuity in the branch-valued potential instead occurs at the distinct endpoint $\tilde{T}_{1}$ of the larger-$x$ branch (Fig.~\ref{fig:mixedKong}). A zeroth-order equilibrium transition can be inferred only if the competing branch is locally stable there; the sampled endpoint states fail the Hessian test (Table~\ref{tab:endpointH}). The cusp and endpoint temperatures shift with $\sigma$. At large $x$ the heat capacity of the stable branch grows as $C_{X}\simeq8\pi C x^{2}=2S$ in the canonical and grand-canonical ensembles, as for a neutral AdS black hole. In the mixed ensemble with $\mu R>0$ the branch is confined between the two roots of $x-x^{3}=\mu R$ and has no large-$x$ limit.

A positive heat capacity, $C_X>0$, tests thermal stability against entropy fluctuations at fixed ensemble controls. For the grand-canonical and mixed ensembles, full local stability also requires the corresponding energy Hessian to be positive semidefinite. In the mixed ensemble, the lower-$x$ interval includes states with $C_\mu>0$ but $E_{CC}<0$, so this condition is not sufficient for full stability. The criteria, numerical checks, and their implications for the branch-end discontinuity are given in Appendix~\ref{app:hessian}. The sampled endpoint states are locally unstable.

\section{Conclusions}\label{sec:concl}

We have examined the background circular null orbits, effective electromagnetic propagation, thermodynamic topology, and boundary thermodynamic description of a magnetically charged AdS solution in nonlinear electrodynamics.

The background classification is controlled by horizon formation and the creation or annihilation of circular-null-orbit pairs. For $m_0=l=1$, $q=0.5$ and $\beta=0.1$, the background thresholds are $\sigma=0.1465105$ for horizon merger and $\sigma=0.1548466$ for merger of the background orbit pair. At fixed $m_0$, the apparent boundary near $\sigma=1/4$ follows the divergent relation between $m_0$ and the total mass. The background mass thresholds are smooth across this value when expressed in terms of $M$. Central regularity remains a separate condition, requiring $\sigma\geq1$ and $m_0=0$.

The extraordinary optical geometry introduces the factor $\eta$ into the radial potential and an explicit AdS-scale dependence into the orbit condition. In the four configurations examined, the five isolated orbit indices are preserved while their radii shift by up to $5.19\%$. The horizonless example contains a stable--unstable pair whose two extrema move apart, as shown in Fig.~\ref{fig:opticalcomparison}.

The thermodynamic phase structure depends on the small-radius temperature and the crossings of stable Gibbs-free-energy branches. At $P=10^{-3}$, the cases $(q,\sigma)=(0.5,0.12)$ and $(1.2,0.45)$ have $\mathrm{TTC}=+1$ and exhibit first-order small--large black-hole transitions. For $q=0.5$ and $\sigma=0.9$ or $1.2$, the temperature instead diverges to $+\infty$ as $r_h\to0$, and an additional small-radius branch with negative heat capacity appears. Their stable Gibbs-free-energy branches do not cross at $P=10^{-3}$. For $\sigma=0.9$, a first-order crossing occurs in the interval $0.001307<P<P_c$, with $P_c=0.013599$, whereas for $\sigma=1.2$ a crossing is found at $P=0.005$, where $T_t=0.0582955$. The change in thermodynamic behavior is therefore associated with the small-radius temperature and the resulting branch structure, rather than with a transition at $\sigma=1$. Regular critical configurations are obtained by imposing $m_0=0$ together with the criticality conditions. The resulting solutions have positive critical pressure and temperature and an outer event horizon, at $(q,\sigma)=(3.249,1.2)$ and $(3.921,1.5)$ (Table~\ref{tab:crit}). At these points, the ratio $P_c r_c/T_c$ differs from the Reissner--Nordstr\"om value $3/16$ by approximately $0.03\%$. The extended first law and the Smarr relation hold for every $\sigma>0$, including $\sigma=1/4$, since they follow from the horizon condition and from integrals over $r>r_h$ that converge for all $\sigma>0$ and do not involve $m_0$; they require the quantity $\mathcal{B}=\partial M/\partial\beta$ conjugate to $\beta$ to be included; omitting the term $2\beta\mathcal{B}$ violates the Smarr relation by exactly that amount.

In the holographic description, the bulk relations map to a first law and a fixed-$R$ Euler relation for the dual CFT,
\[
E=\tilde{T}S+\mu C+\tilde{\Phi}\tilde{Q}
+\tilde{\Psi}\tilde{\beta},
\]
with no independent $p\mathcal{V}$ term in the Euler relation. At fixed $C$, the canonical ensemble exhibits van der Waals-type criticality, with critical temperatures that depend only weakly on $\sigma$ and remain close to the Reissner--Nordstr\"om value (Table~\ref{tab:cftcrit}). The ratio $C/\tilde{Q}_c$ decreases with increasing $\sigma$ from the Reissner--Nordstr\"om value $3/2$ at $\tilde{\beta}=0$. In the benchmark cases examined, the exponents $\beta_{\rm c}$, $\gamma$, and $\delta$ are numerically consistent with their mean-field values, while the finite heat capacity at the critical point is consistent with $\alpha=0$. The range of $\tilde{Q}/C$ supporting the first-order transition narrows as $\sigma$ increases and becomes particularly small for $\sigma=1.2$, where it is $0.1729\lesssim y\leq0.1756$. When $1/C$ is treated as an analogue of pressure at fixed $\tilde{Q}$, the critical locus follows the same qualitative trend. In this construction, $\tilde{\beta}$ is held fixed while $\hat{\beta}=\tilde{\beta}/(4C)$ varies with $C$. The corresponding isotherms in the $(p,\mathcal{V})$ plane exhibit critical points for $\sigma\geq0.45$ among the cases studied, but not for the Reissner--Nordstr\"om model or for $\sigma=0.12$.

The grand-canonical ensemble displays a Hawking--Page transition, with its temperature changing only in the fifth decimal place across the models considered. In the mixed ensemble, the free-energy branches meet continuously at the cusp $\tilde{T}_{\min}(\mu)$, while a distinct finite discontinuity occurs at the endpoint $\tilde{T}_1$ of the larger-$x$ branch. This branch-end discontinuity is established numerically; the sampled competing endpoint states are locally unstable by the Hessian test, so the discontinuity is not a transition between two locally stable phases.

The orbit radii characterize the geometry and its electromagnetic propagation cones. Relating them to a shadow requires a specified polarization sector, observer and emission model, together with treatment of the AdS boundary \cite{ref083,ref084,refOptNovello}.

\appendix
\section{Hessian test of local thermodynamic stability}\label{app:hessian}

A positive heat capacity tests thermal stability only against entropy fluctuations at fixed ensemble controls. Full local stability requires the Hessian of the CFT energy with respect to the extensive variables allowed to fluctuate in the chosen ensemble to be positive semidefinite. In the canonical ensemble the only fluctuating extensive variable is $S$, so the criterion reduces to $C_Q>0$. In the grand-canonical ensemble the fluctuating variables are $(S,\tilde Q)$, whereas in the mixed ensemble they are $(S,C)$. The corresponding Hessians are
\begin{equation}
H_{\rm gc}=\frac{\partial^2 E}{\partial(S,\tilde Q)^2},
\qquad
H_{\rm mix}=\frac{\partial^2 E}{\partial(S,C)^2},
\end{equation}
where the variables held fixed are those specified by each ensemble, and the physical coupling $\tilde\beta$ is held fixed throughout. For a symmetric $2\times2$ Hessian, positive definiteness requires a positive relevant diagonal element and a positive determinant. In particular,
\begin{equation}
C_{\Phi}=\frac{\tilde T E_{QQ}}{\det H_{\rm gc}},
\qquad
C_{\mu}=\frac{\tilde T E_{CC}}{\det H_{\rm mix}}.
\end{equation}
Thus $C_X>0$ alone is not a sufficient condition for full local stability in the grand-canonical and mixed ensembles.

We evaluated the Hessian derivatives from the closed-form holographic energy while holding the physical coupling $\tilde\beta$ fixed. For the grand-canonical ensemble, $E_{QQ}$ remains positive over the numerical branch examined and tends to zero from above as $x\to\infty$; its infimum over the full large-black-hole branch is therefore zero, not the finite-interval value at a chosen cutoff. In the mixed ensemble, $C_\mu>0$ occurs on both a lower-$x$ interval and a high-$x$ interval. The minimum of $E_{CC}$ over all such intervals lies on the lower-$x$ interval and is negative for all five cases considered, including $\sigma=1.2$. Since $C_\mu=\tilde T E_{CC}/\det H_{\rm mix}$ and $\tilde T>0$, this gives $\det H_{\rm mix}<0$ there: positive heat capacity does not imply full local stability. Table~\ref{tab:hess} compares the determinant roots with the heat-capacity-divergence locations and reports the corresponding diagonal bound. These numerical results apply to the parameter choices stated in the caption and do not establish stability for arbitrary parameters.

\begin{table*}[!t]
\caption{Hessian diagonal test for the grand-canonical and mixed ensembles, with $R=1$ and $\tilde\beta=0.1$. The value $x_{\det}$ is the root of $\det H$; $x_{\rm cap}$ is the corresponding heat-capacity-divergence location in Table~\ref{tab:capdiv}. In the grand-canonical ensemble, $C=5/6$ and $\tilde\Phi R=0.3$; the last column gives $\inf E_{QQ}=0$, approached as $x\to\infty$ on the positive-heat-capacity branch. In the mixed ensemble, $\tilde Q=1$ and $\mu R=0.1$; the last column gives the minimum of $E_{CC}$ over all allowed intervals with $C_\mu>0$. Its location $x_\star$ lies on the lower-$x$ interval.}\label{tab:hess}
\begin{ruledtabular}
\begin{tabular}{llcccc}
Ensemble & Case & $x_{\det}$ & $x_{\rm cap}$ & $x_\star$ & Diagonal bound\\
\hline
Grand-can. & RN & $0.55076$ & $0.5508$ & $x\to\infty$ & $\inf E_{QQ}=0$\\
 & $\sigma=0.12$ & $0.55077$ & $0.5508$ & $x\to\infty$ & $\inf E_{QQ}=0$\\
 & $\sigma=0.45$ & $0.55082$ & $0.5508$ & $x\to\infty$ & $\inf E_{QQ}=0$\\
 & $\sigma=0.9$ & $0.55089$ & $0.5509$ & $x\to\infty$ & $\inf E_{QQ}=0$\\
 & $\sigma=1.2$ & $0.55093$ & $0.5509$ & $x\to\infty$ & $\inf E_{QQ}=0$\\
Mixed & RN & $0.36840$ & $0.3684$ & $0.118846$ & $\min E_{CC}=-0.005629$\\
 & $\sigma=0.12$ & $0.36988$ & $0.3699$ & $0.118984$ & $\min E_{CC}=-0.005648$\\
 & $\sigma=0.45$ & $0.37430$ & $0.3743$ & $0.119390$ & $\min E_{CC}=-0.005704$\\
 & $\sigma=0.9$ & $0.38125$ & $0.3813$ & $0.120015$ & $\min E_{CC}=-0.005786$\\
 & $\sigma=1.2$ & $0.38659$ & $0.3866$ & $0.120481$ & $\min E_{CC}=-0.005847$\\
\end{tabular}
\end{ruledtabular}
\end{table*}

To test the proposed zeroth-order interpretation, we also evaluated the lower-entropy solution at the same temperature $T_1$ where the high-entropy branch terminates. At this endpoint, the high-entropy branch has $GR\to0$, while the lower-entropy solution has a finite potential. Table~\ref{tab:endpointH} reports the resulting branch-valued jump together with $C_\mu$, the relevant Hessian entries, and the smallest Hessian eigenvalue for representative values of $\mu R$ and $\sigma$. The numerical scan shows that, even where $C_\mu$ becomes positive, the determinant remains negative and the Hessian has a negative eigenvalue at the endpoint. Hence all sixteen sampled lower-entropy endpoint states are locally unstable, with $\det H<0$ and $\lambda_{\min}<0$, and the finite branch-end discontinuity does not connect two locally stable phases at these parameter values.

\begin{table*}[!t]
\caption{Numerical endpoint test in the mixed ensemble. $T_1$ is the endpoint temperature of the high-entropy branch, $G_{\rm low}$ is the potential of the lower-entropy solution at the same temperature, and $\Delta(GR)=G_{\rm low}R$ because $G_{\rm high}R\to0$. Negative $\det H$ indicates an indefinite Hessian.}\label{tab:endpointH}
\begin{ruledtabular}
\begin{tabular}{cccccccc}
$\mu R$ & $\sigma$ & $T_1R$ & $\Delta(GR)$ & $C_\mu/\tilde Q$ & $E_{CC}$ & $\det H$ & $\lambda_{\min}(H)$\\
\hline
0.10 & 0.12 & 0.309908 & 0.634685 & -0.229801 & 0.012266 & -0.016541 & -0.093378 \\
0.10 & 0.45 & 0.309908 & 0.631657 & -0.234120 & 0.012345 & -0.016342 & -0.100171 \\
0.10 & 0.90 & 0.309908 & 0.627317 & -0.241985 & 0.012507 & -0.016017 & -0.111730 \\
0.10 & 1.20 & 0.309908 & 0.624271 & -0.249012 & 0.012664 & -0.015761 & -0.121357 \\
0.15 & 0.12 & 0.305309 & 0.534063 & -0.017207 & 0.000521 & -0.009251 & -0.095879 \\
0.15 & 0.45 & 0.305309 & 0.532873 & -0.012255 & 0.000369 & -0.009182 & -0.097457 \\
0.15 & 0.90 & 0.305309 & 0.531210 & -0.005038 & 0.000150 & -0.009078 & -0.099782 \\
0.16 & 0.12 & 0.304350 & 0.514829 & 0.082178 & -0.002184 & -0.008090 & -0.093772 \\
0.16 & 0.45 & 0.304350 & 0.513843 & 0.088685 & -0.002342 & -0.008037 & -0.094947 \\
0.16 & 0.90 & 0.304350 & 0.512469 & 0.098171 & -0.002567 & -0.007957 & -0.096656 \\
0.20 & 0.12 & 0.300362 & 0.439172 & 0.885868 & -0.012970 & -0.004397 & -0.081737 \\
0.20 & 0.45 & 0.300362 & 0.438722 & 0.898022 & -0.013097 & -0.004380 & -0.082107 \\
0.20 & 0.90 & 0.300362 & 0.438100 & 0.915347 & -0.013275 & -0.004356 & -0.082626 \\
0.30 & 0.12 & 0.288940 & 0.244581 & 21.115579 & -0.027427 & -0.000375 & -0.044597 \\
0.30 & 0.45 & 0.288940 & 0.244550 & 21.140075 & -0.027444 & -0.000375 & -0.044617 \\
0.30 & 0.90 & 0.288940 & 0.244507 & 21.173696 & -0.027468 & -0.000375 & -0.044645 \\
\end{tabular}
\end{ruledtabular}
\end{table*}

The scan covers the representative parameters listed in Table~\ref{tab:endpointH}; this result does not exclude locally stable endpoint states elsewhere in parameter space. The mixed-ensemble results are therefore reported as a finite branch-end discontinuity whose lower-entropy side is locally unstable at the sampled points.

\section{Supplementary tables}\label{app:tables}
The tables below collect numerical material referred to in Secs.~\ref{sec:holo} and \ref{sec:cap}.

\begin{table*}[!t]
\caption{CFT quantities for representative bulk states with $\beta=0.1$ in units $R=1$, with $C=l^{2}/4$. The last column is $|\partial E/\partial C-\mu|$, with the derivative at fixed $(S,\mathcal{V},\tilde{Q},\tilde{\beta})$ evaluated numerically and $\mu$ obtained from Eq.~\eqref{eq:pmu}.}\label{tab:dict}
\begin{ruledtabular}
\begin{tabular}{ccccccc}
$q$ & $\sigma$ & $r_{h}$ & $P$ & $C$ & $\tilde{Q}$ & $E$\\
\hline
$0.5$ & $0.12$ & $1.5$ & $0.001$ & $29.842$ & $5.4627$ & $9.2589$\\
$0.5$ & $0.9$ & $3.0$ & $0.001$ & $29.842$ & $5.4627$ & $18.079$\\
$1.2$ & $0.45$ & $2.5$ & $0.001$ & $29.842$ & $13.111$ & $17.517$\\
$0.5$ & $0.9$ & $0.4$ & $0.02$ & $1.4921$ & $1.2215$ & $1.1736$\\
$0.5$ & $1.2$ & $1.5$ & $0.001$ & $29.842$ & $5.4627$ & $9.2579$ \\
\end{tabular}
\end{ruledtabular}
\begin{ruledtabular}
\begin{tabular}{cccc}
$\tilde{T}$ & $\tilde{\Phi}$ & $\mu$ & $|\partial_{C}E-\mu|$\\
\hline
$0.54803$ & $0.33325$ & $0.11946$ & $<10^{-30}$\\
$0.34731$ & $0.16665$ & $0.24626$ & $<10^{-30}$\\
$0.3224$ & $0.47968$ & $0.16417$ & $<10^{-30}$\\
$0.11381$ & $0.99436$ & $-0.022517$ & $<10^{-30}$\\
$0.54837$ & $0.33255$ & $0.11950$ & $<10^{-12}$ \\
\end{tabular}
\end{ruledtabular}
\end{table*}

\begin{table*}[!t]
\caption{Ensembles of the CFT with $\tilde{\beta}$ fixed. The potential is the Legendre transform of $E$ with respect to the variables not held fixed.}\label{tab:ens}
\begin{ruledtabular}
\begin{tabular}{ll}
Fixed variables & Potential\\
\hline
$(\tilde{Q},\mathcal{V},C)$ & $F=E-\tilde{T}S=\tilde{\Phi}\tilde{Q}+\mu C+\tilde{\Psi}\tilde{\beta}$\\
$(\tilde{\Phi},\mathcal{V},C)$ & $W=E-\tilde{T}S-\tilde{\Phi}\tilde{Q}=\mu C+\tilde{\Psi}\tilde{\beta}$\\
$(\tilde{Q},\mathcal{V},\mu)$ & $G=E-\tilde{T}S-\mu C=\tilde{\Phi}\tilde{Q}+\tilde{\Psi}\tilde{\beta}$\\
$(\tilde{\Phi},\mathcal{V},\mu)$ & $E-\tilde{T}S-\tilde{\Phi}\tilde{Q}-\mu C=\tilde{\Psi}\tilde{\beta}$\\
$(\tilde{Q},p,C)$ & $F_{1}=E-\tilde{T}S+p\mathcal{V}$\\
$(\tilde{Q},p,\mu)$ & $F_{2}=E-\tilde{T}S-\mu C+p\mathcal{V}=\tilde{\Phi}\tilde{Q}+\tilde{\Psi}\tilde{\beta}+p\mathcal{V}$\\
$(\tilde{\Phi},p,\mu)$ & $F_{3}=p\mathcal{V}+\tilde{\Psi}\tilde{\beta}$\\
$(\tilde{\Phi},p,C)$ & $F_{4}=\mu C+\tilde{\Psi}\tilde{\beta}+p\mathcal{V}$\\
\end{tabular}
\end{ruledtabular}
\end{table*}

\begin{table*}[!t]
\caption{Divergences of the heat capacity, given by the extrema of $\tilde{T}(x)$. In the canonical and grand-canonical cases, $\tilde{\beta}=0.1$ and $C=5/6$ are fixed, so $\hat{\beta}=0.03$; the canonical case uses $y=0.12$ and the grand-canonical case $\tilde{\Phi}R=0.3$. In the mixed ensemble, $\tilde{\beta}=0.1$, $\tilde Q=1$ and $\mu R=0.1$ are fixed, so $\kappa=0.1$ and $\hat{\beta}=\kappa y$ varies. In the canonical ensemble the branches $x<x_{1}$ and $x_{1}<x<x_{2}$ are stable and unstable, respectively, and the branch $x>x_{2}$ is stable.}\label{tab:capdiv}
\begin{ruledtabular}
\begin{tabular}{llcc}
Ensemble & Case & $x$ & $\tilde{T}R$\\
\hline
Canonical & RN & $0.2258, 0.5313$ & $0.30679, 0.26898$\\
 & $\sigma=0.12$ & $0.2197, 0.5314$ & $0.30878, 0.26898$\\
 & $\sigma=0.45$ & $0.1909, 0.5317$ & $0.31732, 0.26900$\\
 & $\sigma=0.9$ & $0.5320$ & $0.26901$\\
 & $\sigma=1.2$ & $0.5322$ & $0.26903$\\
Grand-can. & RN & $0.5508$ & $0.26297$\\
 & $\sigma=0.12$ & $0.5508$ & $0.26297$\\
 & $\sigma=0.45$ & $0.5508$ & $0.26298$\\
 & $\sigma=0.9$ & $0.5509$ & $0.26299$\\
 & $\sigma=1.2$ & $0.5509$ & $0.26300$\\
Mixed & RN & $0.3684$ & $0.17590$\\
 & $\sigma=0.12$ & $0.3699$ & $0.17662$\\
 & $\sigma=0.45$ & $0.3743$ & $0.17866$\\
 & $\sigma=0.9$ & $0.3813$ & $0.18161$\\
 & $\sigma=1.2$ & $0.3866$ & $0.18368$\\
\end{tabular}
\end{ruledtabular}
\end{table*}
\begin{center}
\textbf{Declaration on the Use of Artificial Intelligence Technologies}
\end{center}
In the preparation of this manuscript, the authors utilized artificial intelligence-based technologies solely for optimizing and refining the linguistic structure, ensuring compliance with academic journal publication standards, and improving graphics, language quality, grammar and \LaTeX{} typesetting. 
\textbf{All ideas, statements, calculations, and interpretations of the results are entirely original and the direct outcome of the authors' research.} 
It is emphasized that the text refined by these tools was critically and meticulously reviewed multiple times to ensure its alignment with the authors' original intent and scientific findings.\\ \textit{The authors assume full responsibility for the originality, accuracy, and integrity of the intellectual content, including all analyses, interpretations, and conclusions.}

\end{document}